# Spectral Geometry and Bosonic–Bloch Probes: Explorations in Quantum Learning

Santanu Ganguly[†], Xing Liang[*] and Dimitrios Makris
*Quantum AI Research Group,*
*School of Computer Science and Mathematics,*
*Kingston University London,*
*Penrhyn Road, Kingston upon Thames, KT1 2EE, United Kingdom*
[†]*e-mail address: ganguly.santanu@kingston.ac.uk*
[*]*e-mail address: x.liang@kingston.ac.uk*

This paper investigates how spectral and state-space geometry can be used as diagnostic information in two complementary quantum-learning settings. In a graph-regularized quantum network, an experimental run with 15 nodes (10 supervised and 5 unsupervised) shows a small increase in algebraic connectivity and effective spectral rank after training. A synchronized five-seed regularization sweep reveals a non-monotonic finite-system reorganization of representation variance and graph connectivity: the strongest contraction occurs at intermediate regularization, but none of the tested conditions satisfies the prespecified joint criterion for representation concentration and graph fragmentation. On the learned 15-node graph, a systematic two-boson analysis over all 105 distinct input pairs and 31 propagation times yields a small but reproducible association between interference magnitude and the Fiedler edge split (primary statistic 0.108, 95% bootstrap interval [0.084, 0.140], node-permutation $p=4.998\times10^{-4}$). The existing IBM-hardware study is retained as a separate implementation-level benchmark of a three-qubit Z-diagonal energy landscape and is not interpreted as a measurement of the bosonic interference observable. In a complementary anomaly-detection study on CIC-DDoS2019 SSDP traffic, dimension-matched CAE-6D and QAE-6Q models are compared over five initialization seeds. The classical baseline attains higher reconstruction and latent-distance discrimination on this dataset, while the QAE provides additional Bloch-space and parameter-space observables that expose internal quantum-state geometry. Together, these findings establish a unified spectral–geometric framework for diagnosing quantum learning systems – using physically interpretable geometry to interrogate learned representations – without identifying the two representation spaces as the same manifold.

# I. INTRODUCTION

A quantum network connected via suitable nodes can be represented by graphs and used to implement quantum computations, quantum-to-classical interfaces for measurement, quantum-to-quantum interfaces between physical systems, and quantum sensing [1,2]. Graph structure is also natural for describing relational costs based on quantum quantities such as fidelity, purity, or entanglement [2-5]. In this work we study two complementary, but distinct, quantum-learning settings. The first uses graph regularization and Laplacian diagnostics to characterize a learned similarity graph and then probes that learned structure with two-particle interference. The second uses a hybrid quantum autoencoder for denial-of-service anomaly detection and analyzes its latent quantum states using Bloch-space observables and parameter-space QFI. The common theme is diagnostic rather than computational: we do not claim a quantum speedup or a shared physical manifold between the two studies.

## A. Spectral structure in graph learning

Learning over structured data is fundamentally geometric [12, 18]. In graph-based models, relational inductive biases govern information propagation and representation organization [7, 57]. Spectral graph theory [19, 21, 52] provides a natural language for this structure: the Laplacian eigenvalues and eigenvectors encode connectivity, mixing, and partitioning, with algebraic connectivity and the Fiedler vector [24] characterizing global coherence and principal graph cuts [11, 13, 19, 24, 52]. In classical machine learning, this spectral viewpoint underlies graph convolutional networks and message-passing architectures [15, 21, 67], where propagation can be interpreted as filtering in the Laplacian eigenbasis [7, 10, 12]. Graph neural networks (GNNs) formalize this paradigm through permutation invariance and local aggregation rules [24, 50, 52, 59]. Beyond task-level performance, collective behavior and internal organization in quantum neural-network models have themselves emerged as physically relevant objects of study [124].

### B. Bosonic probes as structural diagnostics

Multiphoton interference in linear optical systems – originally studied in boson sampling [5, 28, 55, 93] – is sensitive to global unitary structure. Recent work has explored photonic protocols such as Gaussian boson sampling for graph-aware feature generation and combinatorial tasks [8, 28, 43]. These developments suggest that interference can serve not only as a computational primitive but also as a diagnostic of structural organization [108-111]. Here we adopt the same general philosophy without making a complexity-theoretic claim: *bosonic interference* is used as a *physically grounded probe of learned structure*. Two-boson interference is used as a physically interpretable probe of the learned 15-node output-similarity graph. This analysis uses the same combinatorial Laplacian to define both the Fiedler structure and the optical propagator, scans all unordered input pairs and a prespecified propagation-time grid, and evaluates the observed association against bootstrap and node-permutation controls. We demonstrate that training modifies spectral entropy and effective spectral dimension, and that the bosonic enhancement observable $\Delta P_{uv}$ correlates with Laplacian-derived partitions.

### C. Quantum machine learning and geometry

This geometric viewpoint is especially natural QML, a domain that extends this perspective into Hilbert space, where variational circuits encode information in amplitudes and phases [6, 9, 65, 67, 68]. Quantum graph neural networks (QGNNs) and related relational variational architectures embed graph structure directly into quantum circuits [39, 55, 60, 90]. Yet most existing analyses remain either architecture-centric or optimization-centric. Much less attention has been paid to simple, model-agnostic observables that diagnose how learning reshapes the geometry of the representations themselves.

### D. Bloch-space geometry for anomaly detection

Quantum autoencoders extend the same geometric picture into latent quantum state space [75-78]. Originally introduced as variational compression models [45], they have only recently been explored for anomaly detection [82, 103, 104, 114, 115], despite the broader importance of anomaly detection across machine learning, cybersecurity, and cyber–physical systems [94, 95]. Classical autoencoders identify anomalies through reconstruction error [36], while quantum variants have been studied through variational and optimization-based approaches [82, 95, 103].

Anomaly detection [72, 73, 82, 93-95] is central to cybersecurity and cyber–physical systems, where rare events must be identified without labelled attack data [36, 37, 94]. Recent surveys further show that graph anomaly detection and graph-based intrusion detection remain methodologically fragmented and sensitive to graph construction, benchmark choice, and deployment assumptions [15, 36, 46, 72, 94, 95].

In this setting, a hybrid quantum autoencoder defines a manifold of reduced quantum states in its latent layer, making Bloch-space drift a natural local probe of how samples move relative to the benign manifold representations [9, 32, 48]. Parameterized (variational) quantum circuits that provide an alternative route to expressive latent representations been explored for supervised learning and quantum-enhanced feature mappings, commonly using hardware-efficient ansätze suited to near-term devices [48, 65, 68]. This is useful because hybrid quantum–classical models based on parameterized circuits can generate expressive latent representations [9, 32, 48, 105]; yet most studies still emphasize aggregate metrics such as ROC-AUC [ 6, 16, 17, 48] rather than the geometry of learned states.

Trainability of variational circuits can be constrained by flat optimization landscapes and barren-plateau phenomena [87,88]. The quantum Fisher information (QFI) supplies a local Riemannian metric in circuit-parameter space and is useful for characterizing parameter sensitivity and identifiability [17,96,97]. In the present work QFI is therefore used only as a local parameter-space diagnostic of the trained latent circuit. We do not infer input-space Lipschitz continuity, anomaly separability, or the absence of barren plateaus from the QFI spectrum alone.

These considerations motivate the use of **Bloch-space geometry** [36, 96, 98, 103-107] as a diagnostic tool to probe how anomalies are encoded in quantum latent representations. We, therefore, *in contrast to previous works*, analyze Pauli expectation values to construct Bloch-space drift metrics that distinguish persistent geometric displacement from transient fluctuations, clarifying how anomalies deform the learned quantum manifold [77-79, 114, 115]. The intrusion detection task is used as a structured, high-dimensional benchmark to probe the behavior of quantum latent bottlenecks under realistic thresholding constraints.

### E. Contributions

This work makes three major contributions grounded in spectral geometry and physically interpretable diagnostics.

(1) Spectral reorganization and bosonic probing of a learned graph: We characterize a fixed 15-node, 10-SL/5-UL graph-learning configuration using synchronized spectral diagnostics and a five-seed regularization sweep. Rather than inferring a phase transition, we report representation concentration and graph fragmentation as separate finite-system diagnostics. We then evaluate two-boson interference systematically on the learned 15-node graph and quantify its association with the Fiedler edge split using bootstrap uncertainty and a node-permutation null.

(2) Bloch-space diagnostics for a matched hybrid quantum autoencoder: We compare a 6-dimensional classical autoencoder bottleneck with a 6-qubit quantum latent bottleneck under the same preprocessing, benign training subset, optimization settings, and five initialization seeds. Reconstruction performance is reported alongside a classical latent-centroid distance control and quantum Bloch-space drift. Consecutive drift is treated as an order-sensitivity control rather than a temporal observable.

(3) Reproducible implementation-level diagnostics: We report exact graph splits, seeds, fixed-kernel procedures, permutation/bootstrap settings, CIC-DDoS2019 data provenance, and matched CAE/QAE sampling. Existing IBM-hardware results are retained as a separate three-qubit energy-landscape consistency benchmark; they are not used as hardware validation of the bosonic interference calculation.

The two experimental sections are linked by a common methodological principle rather than by an assertion that they occupy the same manifold. In the graph-regularized study, geometry is encoded globally by the spectrum and Fiedler structure of a learned similarity graph. In the autoencoder study, geometry is observed locally through distances and Bloch-vector displacement in a latent state space. Both analyses ask whether internal geometric observables reveal organization that is hidden by aggregate task metrics, but the observables and representation spaces remain distinct.

Across both settings, we therefore use spectral and state-space geometry as complementary diagnostic languages. No computational, statistical, or practical quantum advantage is claimed by the experiments reported here.

### F. Structure of the Paper

The paper is organized as follows. Section II introduces the graph-regularized quantum learning framework, spectral diagnostics, two-boson probe, hardware energy benchmark, and the dimension-matched CAE/QAE methodology. Section III presents the synchronized multi-seed spectral results, systematic boson-Fiedler analysis, retained hardware benchmark, and matched anomaly-detection results with classical and quantum latent-space controls. Section IV discusses the shared diagnostic perspective while keeping the two representation spaces conceptually distinct. Section V concludes, followed by Appendices A, B, and C containing additional spectral, hardware, QFI, and latent-geometry diagnostics.

## II. METHODS

This section describes the methodologies used in two distinct studies: graph-regularized quantum learning with spectral and two-particle diagnostics, and matched classical/quantum autoencoding with latent-space diagnostics. The studies share a geometric diagnostic perspective but are not assumed to realize the same learned manifold.

### A. Spectral structure in graph learning

Dataset and graph configuration: All primary spectral experiments use the 15-node graph shown schematically in Fig. 1. For the experimental analysis, 10 nodes are assigned to the supervised-learning (SL) subset and 5 to the unsupervised-learning (UL) subset. SL nodes enter both supervised and graph terms, whereas UL nodes are updated through graph coupling and regularization. The fixed split is recorded in the reproducibility manifest and is held constant for the primary pre/post analysis and the synchronized multi-seed regularization study.

All spectral graph experiments are synthetic and use explicitly generated quantum training states and graph construction rules; no external dataset enters Study I. The primary configuration uses architecture [3,1], 1000 training rounds, and a fixed 15-node 10-SL/5-UL split. The synchronized Fig. 6 study uses five paired initialization seeds and a fixed RBF bandwidth across all regularization strengths so that changes in variance and Laplacian quantities are not confounded by adaptive rescaling of the output graph.

Graph construction and node roles: The graph in Fig. 1 contains 15 learned representations, with 10 supervised and 5 unsupervised nodes in the primary experimental configuration. The supervised nodes are constrained by the task term and the graph penalty, while the unsupervised nodes are influenced through relational coupling. To avoid any sign ambiguity, the paper reports a non-negative graph-regularization strength $\lambda_{\mathrm{reg}}$ and writes the optimized objective in the maximization convention as task fidelity minus $\lambda_{\mathrm{reg}}$ times graph disagreement. The legacy implementation variable satisfies $\gamma = -\lambda_{\mathrm{reg}}$.

$$C[s] = C_{SL}[s] + \lambda_G \sum_{(u,v)\in E} A_{uv} \| \mathbf{h}_u(s) - \mathbf{h}_v(s) \|_2^2 \tag{1}$$

with
$$C_{\mathrm{SL}}[s] = \sum_{u\in \mathcal{V}_{\mathrm{SL}}} \ell(\hat{y}_u(s), y_u) \tag{2}$$

and,
$$C_{\mathrm{graph}}(\theta) = \sum_{(i,j)\in E} w_{ij} \| f_\theta(x_i) - f_\theta(x_j) \|^2 \tag{3}$$

Here A denotes the fixed adjacency matrix and the graph term measures disagreement between representations connected by the graph. Throughout this analysis, $\lambda_{\mathrm{reg}} \geq 0$ denotes the physical regularization strength. The legacy optimizer used the internal variable $\gamma = -\lambda_{\mathrm{reg}}$; therefore increasingly negative $\gamma$ in the code corresponds to increasing non-negative regularization strength. This convention is used consistently in all figures and tables.

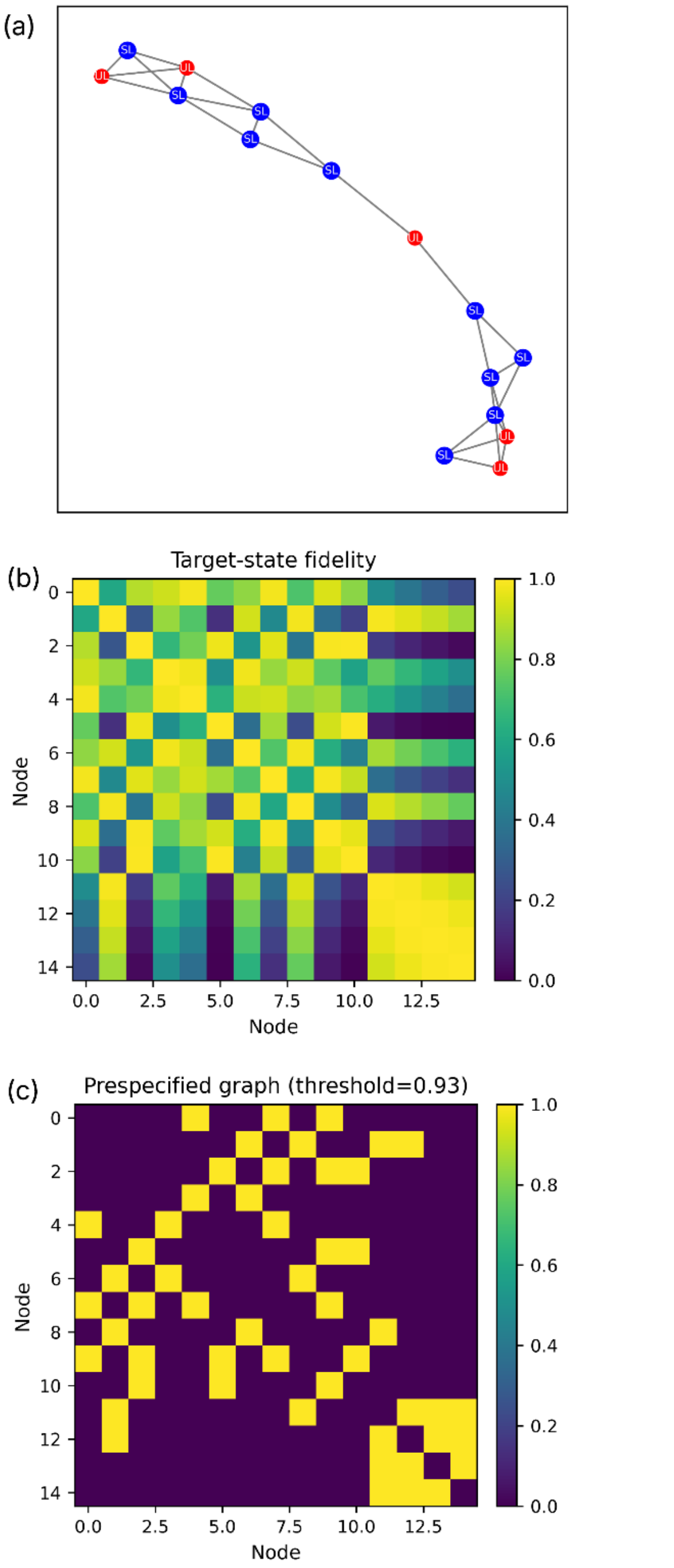

FIG. 1. (Color online) (a) 15-node Graph used for the primary spectral experiment**.** Ten nodes belong to the supervised-learning (SL) subset (blue) and five to the unsupervised-learning (UL) subset (red). Edges encode the fixed relational structure entering the graph-regularized training objective; the learned output-similarity graph used for spectral and bosonic diagnostics is constructed from the trained quantum outputs. (b) target-state fidelity matrix for the 15 training nodes. (c) prespecified adjacency obtained using the 0.93 threshold. The primary split contains 10 supervised and 5 unsupervised nodes.

The graph comprises nodes $u \in \mathcal{V}$ representing learned outputs (or latent representations) $\mathbf{h}_u(s)$, with a partition $\mathcal{V} = \mathcal{V}_{\text{SL}} \cup \mathcal{V}_{\text{UL}}$ into supervised (SL) and unsupervised (UL) nodes. SL nodes $u \in \mathcal{V}_{\text{SL}}$ are directly constrained by labels through $C_{\text{SL}} = \sum_{u \in \mathcal{V}_{\text{SL}}} \ell(\hat{y}_u, y_u)$, whereas UL nodes have no label term and are updated only via graph coupling and regularization. Graph structure enters through a smoothness penalty $\sum_{(u,v)\in E} A_{uv} \| \mathbf{h}_u - \mathbf{h}_v \|_2^2$, (see Appendix A for more details) which propagates information across the SL/UL partition and yields the Laplacian spectra analyzed below.

We consider a quantum learning model in which relational structure is incorporated through graph (Fig. 1) regularization. Let $G = (V, E)$ denote a similarity graph constructed from the outputs of a parametrized quantum circuit. The nodes $V$ correspond to data samples, and weighted edges encode pairwise similarity. The geometry of this graph is characterized by its Laplacian

$$L = D - A \tag{4}$$

where $A$ is the adjacency matrix and $D$ the degree matrix (more detailed calculations are shown in Appendix A). The Laplacian spectrum $\{\lambda_k\}$ provides a basis for diagnosing connectivity, mixing, and partition structure of the learned representation. An example of the similarity graph construction and the corresponding Laplacian spectrum is shown in Figs. 9 and 21.

**Empirical observation 1: Graph-regularized training reshapes spectral geometry**: We augment the task objective with a non-negative graph-disagreement penalty, graphically depicted in Fig. 2. Let $\theta$ denote the trainable parameters of the quantum network and let $f_\theta(x_i)$ denote the learned output representation associated with node $i$. The graph-disagreement term is,

$$C_{\text{graph}}(\theta) = \sum_{(i,j)\in E} w_{ij} \| f_\theta(x_i) - f_\theta(x_j) \|^2$$

where $E$ is the edge set and $w_{ij} \geq 0$ is the weight associated with edge $(i, j)$. This is the same graph-smoothness contribution introduced in Eq. (3). We write the training objective as,

$$\mathcal{J}(\theta) = \mathcal{J}_{\text{task}}(\theta) - \lambda_{\text{reg}} C_{\text{graph}}(\theta), \quad \lambda_{\text{reg}} \geq 0 \tag{5}$$

where $\mathcal{J}(\theta)$ is the scalar objective maximized during training and $\mathcal{J}_{\text{task}}(\theta)$ denotes the task-dependent part of that objective, including the supervised contribution and any terms held fixed across the graph-regularization comparison.

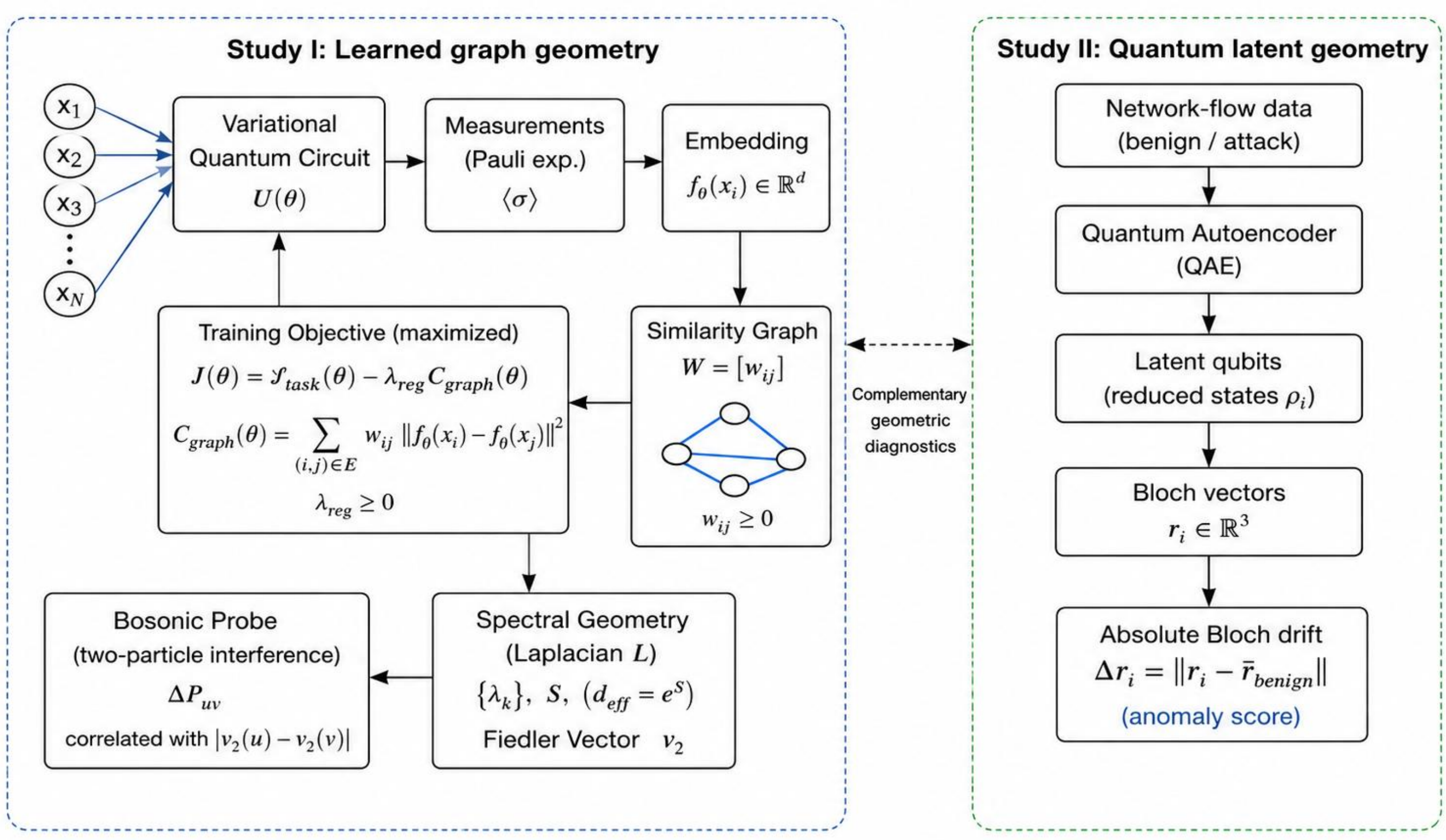

FIG 2. (Colour online) **(a) Overview of the complementary geometric diagnostics considered in this work.** In Study I, a variational quantum model produces output embeddings $f_\theta(x_i)$ from which a weighted similarity graph and its Laplacian spectral observables are constructed. Training maximizes $\mathcal{J}(\theta) = \mathcal{J}_{\text{task}}(\theta) - \lambda_{\text{reg}} C_{\text{graph}}(\theta)$, with $\lambda_{\text{reg}} \geq 0$. The learned graph is subsequently interrogated through its spectrum, Fiedler coordinate, and a two-particle interference probe. Study II independently analyzes a quantum-autoencoder latent representation through qubit-resolved Bloch-vector displacement. The two branches represent complementary geometric diagnostics and are not assumed to constitute the same learned state space.

Because $C_{\text{graph}}(\theta) \geq 0$, increasing $\lambda_{\text{reg}}$ penalizes disagreement between representations connected by graph edges.

For comparison with the conventional **minimization-loss convention**, we equivalently define,

$$\mathcal{L}(\theta) \equiv -\mathcal{J}(\theta) = \mathcal{L}_{task}(\theta) + \lambda_{\text{reg}} C_{\text{graph}}(\theta), \qquad \mathcal{L}_{task} = -\mathcal{J}_{task} \tag{6}$$

Therefore, $\mathcal{J}(\theta)$ and $\mathcal{L}(\theta)$ describe the same optimization problem under maximization and minimization conventions, respectively. In the legacy implementation the graph term enters the maximized objective as $+\gamma C_{\text{graph}}$; consequently,

$$\gamma = -\lambda_{\text{reg}}$$

The paper reports the physically transparent non-negative quantity, except when the legacy code mapping is explicitly documented. The analysis treats the resulting changes in representation spread and Laplacian structure as empirical finite-system responses to regularization rather than as a theorem requiring monotonic increases in connectivity or effective spectral dimension.

The non-negative parameter $\lambda_{\text{reg}}$ controls the strength of graph regularization. In the implementation used for the experimental runs, $\gamma = -\lambda_{\text{reg}}$; the paper reports only $\lambda_{\text{reg}}$ except when explicitly documenting the code mapping.

Eqns. (1)–(3) define the task and graph contributions to the training criterion. Eqn. (5) expresses the same optimization in the maximization convention used by the implementation, while Eq. (6) gives the equivalent minimization-loss convention.

Under this graph-regularized objective, training can reorganize the learned output-similarity graph. We therefore measure, rather than assume, the response of algebraic connectivity, spectral entropy, effective spectral rank, and representation variance. The synchronized multi-seed analysis in Fig. 6 is the primary evidence for the regularization-dependent response. The algebraic connectivity $\lambda_2$, together with its associated Fiedler vector, provides a network-level characterization of connectivity and identifies structurally important edge contributions to collective dynamics [120, 121].

Training proceeds via gradient-based updates with step size $\epsilon$. As parameters evolve, the induced similarity graph and its Laplacian spectrum change dynamically, allowing geometric diagnostics to track the emergence or degradation of structure. Fig. 7 shows a similarity Laplacian graph generated with the following graph-regularization terms: Weighted similarity graph $W$ constructed from embeddings $f_\theta(x_i)$. Edge weights $w_{ij}$ define the adjacency matrix and degree matrix $D_{ii} = \sum_j w_{ij}$; Eigenvalue spectrum of the corresponding graph Laplacian $L = D - A$. The spectral quantities $\{\lambda_k\}$, algebraic connectivity $\lambda_2$, and spectral entropy $S$introduced below are computed from this operator.

### 1. *Spectral Diagnostics*

To quantify geometric organization, we monitor three spectral quantities. First, the algebraic connectivity $\lambda_2$ measures global coherence and signals representational fragmentation when it approaches zero. Second, we define the spectral entropy

$$S = -\sum_k p_k \log p_k, \quad p_k = \frac{\lambda_k}{\sum_j \lambda_j} \tag{7}$$

which captures the distribution of spectral weight across modes. The associated effective spectral dimension, $d_{\text{eff}} = e^S$ provides a compact measure of representational spread.

Finally, the Fiedler vector $v_2$ [24], corresponding to $\lambda_2$, encodes the principal partition of the graph. Edges with large spectral split $|v_2(i) - v_2(j)|$ connect nodes that lie on opposite sides of the dominant structural cut and thus play a central role in characterizing learned geometry.

These diagnostics are interpreted separately. Representation concentration is monitored through relative Hilbert-Schmidt variance, while graph fragmentation is monitored through relative algebraic connectivity and connected-component count. The effective spectral rank is $d_{eff} = e^S$ and therefore obeys the mathematical bound $1 \le d_{eff} \le N$. A joint-collapse label is used only if both the prespecified concentration and fragmentation criteria are sustained; none of the experimental runs meets that joint criterion.

We quantify representation spread by the Hilbert–Schmidt (HS) variance $V(s) = N^{-1} \sum_i \| \rho_i(s) - \bar{\rho}(s) \|_{HS}^2$, where $\bar{\rho}(s)$ is the mean output density matrix at training round $s$. To compare runs with different initial spreads, we report the normalized quantity $V(s)/V(0)$, such that values below unity indicate contraction relative to initialization (ref. Fig. 6).

Fig. 3 shows an example of defining a Laplacian spectral entropy as followed in our methods. $S = -\sum_k p_k \ln p_k$ with $p_k = \lambda_k / \sum_k \lambda_k$ (normalized Laplacian eigenvalue distribution); the effective dimension $d_{\text{eff}}$ is shown for each panel. This conceptual figure illustrates how an increase in $S$ corresponds to a delocalization of spectral weight and motivates the use of $d_{\text{eff}}$ as a compact scalar diagnostic of spectral geometry [123].

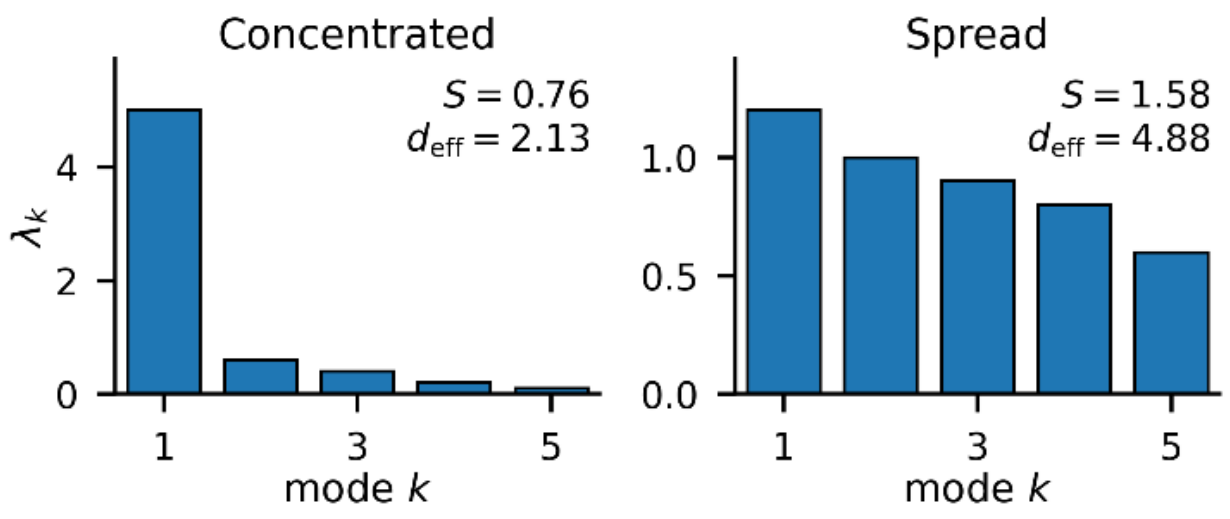

FIG 3. (Colour online) *Conceptual Laplacian spectra and spectral entropy.* Left: a concentrated Laplacian spectrum with most weight on a few modes (low spectral entropy $S$ and small effective dimension $d_{\text{eff}} = e^S$). Right: a spread spectrum with weight distributed across modes (high $S$ and larger $d_{\text{eff}}$).

#### a. *Graph Geometry to Physical Observables.*

For the physical probe, the learned undirected output-similarity graph is mapped to a single-particle propagator using the same combinatorial Laplacian that defines the spectral diagnostics:

$$U(\tau) = \exp\left[-i\tau \frac{L}{\|L\|_2}\right]$$

Here $L = D - W$ is the combinatorial Laplacian of the learned similarity graph, $\|L\|_2$ is its spectral norm and $\tau$ is a dimensionless propagation-time parameter controlling the evolution of the continuous-time quantum walk.. Normalization by $\|L\|_2$ removes the overall spectral scale of the learned graph, allowing propagation conditions to be compared on a common dimensionless scale. Using $L$ in both the Fiedler analysis and the propagator removes any operator mismatch and makes the optical diagnostic directly comparable with the reported spectral quantities. Laplacian-generated quantum walks provide a natural framework for coherent propagation on network structures [122]. We therefore use the learned graph Laplacian itself as the generator of the single-particle propagator.

*b. Two-*photon *edge probabilities*

**Identity 1**: **Edge-level two-boson interference**

For distinct input modes s and t and distinct output modes u and v, we associate each output edge with a coincidence event. For indistinguishable bosons [105-108], the coincidence probability is

$$P_{\text{bos}} = |U_{us}U_{vt} + U_{ut}U_{vs}|^2$$

For distinguishable particles, the corresponding baseline is

$$P_{\text{dis}} = |U_{us}U_{vt}|^2 + |U_{ut}U_{vs}|^2$$

The edge-resolved interference contribution is defined by the difference between indistinguishable and distinguishable coincidence probabilities,

$$I_{uv} = P_{\text{bos}} - P_{\text{dis}} \tag{8}$$

which isolates the two-path exchange-interference contribution for the selected input and output modes.

$$\Delta P_{uv} = \mathrm{I}_{uv} = 2\,\mathrm{Re}\left[\mathrm{U}_{us}\mathrm{U}_{vt}\,\mathrm{U}_{ut}\overline{\mathrm{U}}_{vs}\right]$$

Equation (8) is an exact two-particle interference identity for the chosen propagator; it is not a theorem asserting universal alignment with a graph partition. Positive and negative values correspond to constructive and destructive exchange interference, respectively. Whether its magnitude is structured by the learned Fiedler mode is tested empirically below. The analysis therefore treats Fiedler alignment as a statistical property of the learned graph, input-pair ensemble, and propagation-time interval rather than as a universal consequence of bosonic statistics.

*c. Spectral* correlation *analysis.*

We compute the Fiedler vector $v_2$ of the learned combinatorial Laplacian and compare $|I_{uv}|$ with the edge split $|v_2(u)-v_2(v)|$. The primary analysis uses all positive learned-similarity edges, all 105 unordered distinct input pairs of the 15-node graph, and 31 prespecified propagation times from 0.10 to 6.00. The primary statistic is the median across input/time conditions of the absolute Spearman association, with a 2000-replicate cluster bootstrap interval and a 2000-permutation node-label null. Signed interference is reported as a secondary diagnostic.

## 2. *Quantum Network Model*

We model a quantum communication network [27] as a time-dependent graph $G_t = (V, E)$ where vertices represent quantum nodes with finite memory and edges represent physical channels capable of probabilistic entanglement generation. Successful generation on an elementary link produces a Werner state

$$\rho_e = F_{0,e}\left|\Phi^+\right\rangle\left\langle\Phi^+\right| + \left(1 - F_{0,e}\right)\mathbb{I}/4 \ ,$$

with initial fidelity $F_{0,e}$ determined by channel loss and noise [5, 6]. denotes the **maximally entangled Bell state** used as the reference state in Werner states. Entangled pairs stored in memory undergo decoherence, which we model as depolarizing noise. After a storage time $a$, the fidelity evolves as

$$F(a) = \frac{1}{4} + \left(F_0 - \frac{1}{4}\right)e^{-a/T_2}$$

where $T_2$ denotes the memory coherence time. Entanglement swapping at repeater nodes is performed via Bell-state measurements, yielding a Werner state with fidelity,

$$F_{\text{swap}} = \eta_{\text{sw}}\left[F_1F_2 + \frac{(1-F_1)(1-F_2)}{9}\right]$$

where $\eta_{\text{sw}}$ captures imperfections in local operations [6, 9].

*a. Conceptual extension: Bosonic sampling as a quantum feature map*

To generate expressive representations of local network structure, selected nodes perform Gaussian boson sampling on induced subgraphs corresponding to their local neighborhoods. For a node $v$, the adjacency matrix of its neighborhood $A_{\mathcal{N}(v)}$ is embedded into the covariance matrix of a Gaussian state, and photon-number samples, $s_v^{(k)} \sim \mathcal{P}_{\text{GBS}}\left(A_{\mathcal{N}(v)}\right)$ are generated, where $\mathcal{P}_{\text{GBS}}$ is the probability distribution over photon-number patterns generated by a Gaussian Boson Sampling device. Statistical moments or histograms of these samples define a quantum-enhanced feature vector $z_v$. Owing to the relationship between GBS output probabilities and matrix hafnians, these features encode higher-order connectivity and correlation structure beyond pairwise interactions [16–18].

The Gaussian-boson-sampling construction introduced here is retained as a broader architectural extension and literature connection. It is not part of the numerical pipeline used to generate the spectral, boson-Fiedler, hardware, or QAE results. Accordingly, no performance or quantum-advantage claim in this paper depends on evaluating hafnians or GBS sampling hardness.

### b. Conceptual extension: GNN architecture

An extension was considered to combine neighborhood-level photonic sampling statistics with classical node and edge information in a message-passing graph neural network. To generate expressive representations of local network structure, selected nodes perform Gaussian boson sampling on induced neighborhood subgraphs. The resulting photon-number samples define quantum-enhanced feature maps whose statistics are governed by hafnians $h$, encoding higher-order connectivity beyond pairwise correlations [13–15]. These features are combined with classical node and edge information to produce a global network state that is processed by a *graph neural network* via message passing,

$$h_v^{(l+1)} = \psi\left( h_v^{(l)} \sum_{u \in \mathcal{N}(v)} h_u^{(l)}, z_v \right) \tag{9}$$

In this formulation, the GNN parameterizes control and inference policies for entanglement generation, swapping, optional purification, optimized using a delivery-dependent objective favoring high-fidelity, low-latency end-to-end entanglement. This architecture provides context for future work but is not evaluated in the experimental experiments reported in Sec. III.. The GNN control-policy formulation is therefore retained as an outlook-level design connection rather than an executed component of the present experiments.

### c. The Learning Objective

The network control problem is formulated as a sequential decision process with training augmented by the loss function defined equation (5). The GNN parameterizes policies for entanglement generation, swapping, and optional purification. A delivery-dependent reward is issued as

$$r = 1\left[F_{\mathrm{e2e}} \geq F_{\min}\right] R_{\mathrm{succ}} - \lambda_T T - \lambda_W W \tag{10}$$

where $F_{\mathrm{e2e}}$ is the end-to-end fidelity, $T$ is delivery latency, and $W$ quantifies wasted entanglement resources. From a theoretical perspective, the combined bosonic-sampling–GNN architecture can be interpreted as a **variational approximation** to the optimal control policy over *quantum network states*, with the bosonic sampler defining a nonclassical feature map and the GNN acting as a structured variational ansatz. The broader bosonic-sampling/GNN construction may be viewed as a candidate structured feature architecture, but no quantum computational advantage is inferred from it.

## 3. *Diagnostics of representation structure*

We quantify representation spread using Hilbert-Schmidt variance of the quantum outputs, graph connectivity through the second eigenvalue $\lambda_2$ of the learned combinatorial Laplacian, and spectral organization through entropy $S$ and $d_{eff} = e^S$. For the synchronized regularization sweep, the operational concentration criterion is $V(s)/V(0) \leq 0.10$ and the fragmentation criterion is $\lambda_2(S)/\lambda_2(0) \leq 0.01$ (or graph disconnection). These criteria are reported separately.

### a. Operational concentration and fragmentation criteria

Representation concentration is defined operationally as a sustained reduction of the relative output variance $V(s)/V(0)$ to $0.10$ or below over the specified final training tail. Graph fragmentation is defined independently as a sustained relative algebraic connectivity $\lambda_2(S)/\lambda_2(0)$ of $0.01$ or below, or the appearance of more than one connected component. The same fixed output-graph bandwidth is used across seeds and regularization strengths.

A joint-collapse label requires both criteria to be sustained in the same run. The effective spectral rank is reported continuously as a third diagnostic and is not used as an additional arbitrary collapse threshold.

For each input pair $(s \mid t)$ and propagation time $\tau$, we quantify the monotonic association between the edgewise interference magnitude and the Fiedler-coordinate separation using Spearman's rank correlation coefficient,

$$\rho_S = \mathrm{Corr}[\mathrm{rank}|\left(|I_{uv}|\right)\mathrm{rank}\left(|v_2(u) - v_2(v)|\right)],$$

where the correlation is evaluated across all retained learned-graph edges $(u \mid v)$. Here, $\rho_S = 1$ denotes a perfectly increasing monotonic relation, $\rho_S = -1$ a perfectly decreasing monotonic relation, and $\rho_S \approx 0$ little or no monotonic rank association. Because the direction of the edgewise association can vary with the input modes and propagation time, the primary condition-level statistic is $|\rho_S|$. The global statistic reported below is the median of $|\rho_S|$ over all prespecified input-pair and propagation-time conditions. To this end, we have, over all prespecified input-pair and propagation-time conditions, of the absolute Spearman rank association between the bosonic interference magnitude and the Fiedler-edge separation:

$$T_{\mathrm{BF}} = \mathrm{median}_{s<t,\tau} \left| \rho_S\left( \left| I_{uv}^{(s|t|\tau)} \right|, \left| v_2(u) - v_2(v) \right| \right) \right|,$$

where $T_{\mathrm{BF}}$ is the global Boson–Fiedler association statistic.

### 4. *Bosonic probe of learned structure*

**Empirical test 2: Bosonic probe of spectral partitions**

In the learned 15-node graph, we test whether edge-resolved two-boson interference is statistically structured by the Fiedler edge split. This is formulated as an empirical association test; no universal proportionality or theorem is assumed. The graph is encoded through the normalized Laplacian propagator, indistinguishable and distinguishable coincidence probabilities are evaluated for all input pairs and propagation times, and the exchange-interference contribution is compared with Fiedler edge separation using bootstrap and permutation controls.

### 5. *Finite-system regularization sweep*

The Fig. 6 experiment uses $\lambda_{\text{reg}} \in \{0, 0.5, 1.5, 3.0\}$, five paired initialization seeds ($11, 22, 33, 44, 55$), 1000 training rounds, synchronized checkpoints every $20$ rounds, a fixed RBF bandwidth across all runs, and 2000 bootstrap replicates. This design measures a finite-system crossover/reorganization and does not constitute finite-size scaling or a thermodynamic phase-transition analysis.

#### *a. Heuristic model for representation spread*

To provide qualitative context for the evolution of representation spread during training, we consider a coarse-grained stochastic description of fluctuations about the learned representation configuration. In this picture, graph regularization contributes an effective restoring tendency that can reduce the variance of the output representations. Such a linearized stochastic description motivates a stationary-variance scale, but it is not used here to define a critical point or to establish a phase transition. In particular, the synchronized multi-seed analysis reported below is treated as the primary evidence for the regularization-dependent finite-system response. Under a linearized Ornstein–Uhlenbeck-type approximation,

$$V_\infty = \frac{D}{\kappa_{eff}},$$

where $D$ represents an effective fluctuation scale, and $\kappa_{eff}$ an effective contraction rate. This relation is used only as a qualitative model of variance suppression and is not interpreted as evidence of a critical transition.

### 6. *Hardware Experiments*

Selected three-qubit energy observables are evaluated on IBM hardware and compared with ideal and noisy simulators. These experiments benchmark implementation-level preservation of a Z-diagonal energy landscape under finite shots and device noise. They do not measure the two-boson coincidence observable $I_{uv}$ and are not presented as hardware validation of the boson-Fiedler association.

#### *b. Architecture Choice*

We consider a quantum neural network with a $[3, 1]$ architecture, mapping three-qubit input states to a single-qubit output. This choice deliberately imposes a strong representational bottleneck, reducing the output Hilbert space dimension from $2^3 = 8$ to $2$. Such compression makes the interaction between supervision, graph regularization, and quantum dynamics explicitly observable.

In this setting, graph regularization constrains how three-qubit inputs are compressed to a single-qubit output. The bottleneck makes changes in output variance and graph geometry readily measurable. The analysis uses explicit operational criteria to distinguish representation concentration from graph fragmentation rather than assuming that either phenomenon necessarily occurs for strong coupling.

The architecture is small enough for reproducible statevector and density-matrix simulation while retaining nontrivial graph structure. The results are therefore interpreted as finite-system behavior of this specified architecture; no thermodynamic phase transition or universal scaling law is inferred.

#### *c. The Hamiltonian Model*

We consider a parameterized quantum circuit acting on $n = 3$ qubits and evaluate the expectation value of a 3-qubit diagonal Pauli Hamiltonian

$$H = Z_0 + 0.5\,Z_1 + 0.25\,Z_2 \qquad (12)$$

represented as a `SparsePauliOp` in Qiskit [4, 14, 70, 71], where $Z_i$ represents the Pauli-$Z$ operator acting on qubit $i$. The diagonal structure of $H$ enables direct estimation from computational-basis measurements without basis rotations. Throughout, Pauli operators are indexed according to Qiskit's convention, with qubit indices increasing from left to right in the Pauli string representation.

The variational ansatz is constructed using a layered, *hardware-agnostic circuit template* consisting of single-qubit rotations and entangling gates, instantiated here using the `TwoLocal` pattern and subsequently decomposed into elementary gates. The decomposition yields standard rotation and entangling primitives and does not affect the compiled circuit or results reported here. A *model pipeline for the Hamiltonian generation* is shown in Fig. 29 of Appendix C.

#### *d. Classical simulation baselines*

All algorithmic development, training, and regression testing are performed using classical simulation prior to hardware execution. Two complementary simulator baselines were used.

### *e. Ideal (noiseless) simulator*

An ideal baseline is obtained by sampling the logical three-qubit circuit using a noiseless QASM simulator. This baseline isolates finite-shot effects while excluding hardware noise. Energies are estimated from measurement counts using the relation

$$\langle H \rangle = \sum_{i=0}^{2} c_i \langle Z_i \rangle$$

with $\langle Z_i \rangle$ computed from bitstring frequencies. Shot-noise uncertainties are estimated assuming independent binomial statistics. In order to evaluate the expectation value of the Hamiltonian on the hardware, we use a parameter $\alpha$ as a *scan coordinate* to probe the hardware-vs-simulator agreement along one chosen parameter direction.

### *f. Hardware noise modeling*

We simulate an approximate device noise baseline by importing the target backend's noise snapshot into Qiskit Aer and applying the resulting `NoiseModel` to the logical (3-qubit) circuits. This Aer-based noisy simulation is intended as a *diagnostic* tool that reproduces first-order error channels (readout error, decoherence and native gate infidelities) under the backend calibration used when the snapshot was retrieved. We do not treat the noisy-Aer curve as a predictive forecast of the instantaneous device output: differences can arise from temporal calibration drift, mapping/transpilation artifacts, and open-plan execution constraints.

To contextualize hardware results, we compare experimental energies against two simulator baselines. An *ideal* baseline is obtained from noiseless sampling of the same logical circuit, isolating finite-shot effects. A *noisy* baseline is generated using the Aer simulator with a noise model constructed from device calibration data via `NoiseModel.from_backend`. This model incorporates gate-dependent depolarizing noise, asymmetric readout errors, and $T_1 / T_2$ relaxation during idle periods, but neglects correlated, non-Markovian, and leakage effects. As such, the noisy-Aer baseline provides a lower-bound estimate of hardware noise and is not expected to fully reproduce experimental behavior. A breakdown of noise sources and budget included in the simulator and present on hardware is provided in Table AI in Appendix A.

### *g. Hardware execution details*

Table I summarizes the hardware resources used for the retained implementation benchmark. Hardware circuits were executed on `ibm_fez` using Qiskit Runtime SamplerV2 with 2000 shots per circuit evaluation. In the 13-point parameter-space landscape, each point was repeated ten times and the sample standard deviation across repetitions is reported. A separate shot-matched hardware/Aer comparison also used 2000 shots per evaluation. No post-selection was applied.

TABLE I. Quantum hardware resources

| Backend | Qubits | Shots | Noise | Queue | Notes |
|---|---|---|---|---|---|
| ibm_fez | 3 | 2,000 | Hardware | 2–5 min | SamplerV2; 10 repeats/point in Fig. 11 |
| AerSimulator | 3 | 2,000 | Cal.-derived | 0 s | Shot-matched noisy reference |
| Statevector/Aer | 3 | Exact | None | 0 s | Noiseless reference |

Dense statevector simulation justification: the three-qubit Hamiltonian benchmark lies comfortably in the exact statevector regime, so noiseless expectations can be obtained at negligible cost relative to hardware scheduling and finite-shot execution. Hardware and shot-matched noisy simulations use SamplerV2/count-based expectation estimation with 2000 shots per evaluation. The hardware benchmark is retained to assess implementation-level energy-landscape consistency, not as evidence for the bosonic interference observable used elsewhere in Study I.

### *h. End-to-end training efficiency*

Assessment of the practical impact of different expectation-value backends on training throughput, was done by measuring the average wall-clock time per training epoch (mean ± standard deviation) using two implementations: (i) a QuTiP-based dense statevector workflow and (ii) a Qiskit-based dense expectation evaluation. As shown in Fig. 20, the Qiskit dense implementation yields a lower and more stable per-epoch runtime than the QuTiP-based approach for the problem sizes considered.

### *i. Circuit compilation and measurement*

Logical three-qubit circuits are transpiled to the target backend using Qiskit's standard transpilation pipeline, yielding circuits expressed entirely in the backend's instruction set architecture (ISA), consisting of native single-qubit rotations and entangling gates. During transpilation, logical qubits are mapped onto a subset of the device's physical qubits; the resulting circuits may span the full physical register of the backend.

To ensure unambiguous measurement interpretation, final measurements are removed from the transpiled circuit and reintroduced such that *only the three physical qubits corresponding to the original logical qubits are measured into a three-bit classical register*. The logical-to-physical qubit mapping is extracted from the transpiler layout metadata, guaranteeing that each measured classical bit corresponds to a specific logical qubit. All reported hardware energies are computed exclusively from these three measured bits.

*j. Energy estimation on hardware*

For a given parameter vector $\boldsymbol{\theta}$, the transpiled and measurement-cleaned circuit is executed using `SamplerV2` with 2000 shots per evaluation. Expectation values $\langle Z_i \rangle$ are computed from measurement outcomes and combined linearly to obtain $\langle H \rangle$. Shot-noise uncertainties are estimated via standard error propagation assuming independent sampling of Pauli observables. To assess variability, hardware evaluations are repeated multiple times for selected parameter points, and empirical standard deviations are reported where appropriate.

*k. Comparison methodology*

Hardware results are compared against both ideal and noisy simulator baselines using identical logical parameter points. To facilitate direct comparison despite differing internal parameter orderings introduced by transpilation, simulator parameter vectors are mapped onto the transpiled circuit's parameter ordering prior to hardware execution. This ensures that all three curves—ideal simulator, noisy Aer, and hardware—correspond to the same logical circuit parameters.

Differences $\Delta E = E - E_{\text{ideal}}$ are reported to highlight the contribution of calibrated noise and residual hardware effects. Agreement between hardware and noisy-Aer within error bars indicates that dominant error mechanisms are captured by the calibration-derived noise model, while systematic deviations beyond this baseline are attributed to effects not included in the simulator.

Relation to quantum kernels and bosonic sampling hardness: the hardware experiment validates only qualitative trends and noise sensitivity of the three-qubit Z-diagonal energy benchmark. It is separate from the learned-graph two-boson calculation and from the conceptual GBS feature-map extension. No claim of sampling hardness or hardware-confirmed boson-Fiedler alignment is made.

### 7. *Implicit quantum kernel*

If $p_v(s)$ denotes the GBS output distribution over photon-count patterns defined by $s$ generated by encoding the local neighborhood subgraph around node $v$, then, an implicit kernel between two neighborhoods $v$ and $v'$ can be written as:

$$K_{\text{GBS}}(v,v') = \sum_{s\in\mathcal{S}} \sqrt{p_v(s)\,p_{v'}(s)} = \left\langle \sqrt{p_v}, \sqrt{p_{v'}} \right\rangle \quad (13)$$

i.e., the inner product between "square-rooted" sampling distributions (a Bhattacharyya/Hellinger-type [98, 99] kernel). In our experiments, $p_v$ was accessed through samples, so the model learns on an **implicit quantum feature map** without explicitly evaluating hafnians.

## B. Bloch-space geometry for anomaly detection

Dataset: The anomaly-detection study uses flow records derived from the CIC-DDoS2019 release [98,99,118]. The analysis is restricted to exact labels BENIGN and DrDoS_SSDP. The benign working CSV was assembled manually from benign-labeled SSDP-related records; because that intermediate file was not assumed to be class-pure, the analysis code normalized the label field and retained only the 21,854 rows labeled exactly BENIGN from 24,292 raw rows. The attack source DrDoS_SSDP.csv contains 2,611,374 records, including 2,610,611 rows labeled exactly DrDoS_SSDP. A two-pass uniform sample without replacement selected 100,000 eligible attack rows with seed 42, producing DrDoS_SSDP_uniform_100k.csv, which was verified to contain only the target attack class. Non-numeric identifiers were removed; infinities were treated as missing values; imputation, scaling, and PCA were fitted from benign training data only.

### 1. *Classical and Hybrid Quantum Autoencoder*

The model consists of a classical encoder, a quantum latent layer, and a classical decoder. Classical features are mapped to rotation angles and embedded via angle encoding. The quantum latent layer is implemented using a shallow, hardware-efficient variational circuit with single-qubit rotations and ring entanglement [100]. Measurements of Pauli operators provide the quantum latent representation.

*a. Classical Autoencoder (CAE)*

The classical autoencoder uses a fully connected encoder-decoder architecture with ReLU activations and a 6-dimensional latent bottleneck, matched to the six-qubit quantum latent dimension. It is trained using mean squared reconstruction error and Adam optimization on the same 5000 benign training rows used by the QAE.

*b. Hybrid Quantum Autoencoder (QAE)*

The hybrid quantum autoencoder (QAE) replaces the classical latent bottleneck with a **variational quantum circuit (VQC)**. A classical preprocessing network maps the reduced feature vector to qubit rotation angles. The quantum circuit employs: Angle embedding, trainable rotation gates, ring entanglement via CNOT operations, measurement of Pauli-Z expectation values as latent features.

For final comparison, CAE-6D and QAE-6Q use the same 16-dimensional PCA input, the same 5000 benign training rows, 20 epochs, batch size 16, learning rate $10^{-3}$, and five initialization seeds (11,22,33,44,55). The benign/attack validation and test partitions are fixed across models. The notation **CAE-6D** denotes the classical autoencoder with a six-dimensional latent bottleneck, whereas **QAE-6Q** denotes the hybrid quantum autoencoder with a six-qubit latent representation. The latent dimensions were matched to reduce representational-capacity differences as a confounding factor and to provide a more direct classical–quantum comparison; both models also used the same benign training rows, preprocessing, optimizer, learning rate, batch size, epoch count, and fixed validation/test partitions. Five prespecified initialization seeds $(\{11,22,33,44,55\})$ were used to quantify sensitivity to model initialization and to report variability across independent training runs as mean $\pm$ sample standard deviation. The numerical seed values have no special physical significance; they provide fixed, reproducible independent initializations rather than results selected from a single favorable run.

### 2. *Model Architecture*

We employ a hybrid quantum–classical autoencoder consisting of:

1. A classical encoder mapping $x \in \mathbb{R}^d$ to rotation angles $\phi(x)$,
2. A variational quantum latent layer,
3. A classical decoder reconstructing $\hat{x}$

The quantum latent layer prepares the state

$$|\psi(x)\rangle = U(\theta,x)|0\rangle^{\otimes n} \tag{14}$$

where $U(\theta,x)$ consists of angle embedding followed by two layers of trainable $R_Y$ and $R_Z$ rotations and ring entanglement [97]. Pauli expectation values define the latent representation

$$\mathbf{z}(x) = \{\langle X\rangle_q, \langle Y\rangle_q, \langle Z_q\rangle\}_{q=1}^{n} \tag{15}$$

The anomaly score is the reconstruction error $s(x) = \| x - \hat{x} \|_2^2$. Fig. 4 below shows the high-level model diagram, with the full model pipeline shared in Fig. 30 of Appendix C.

In Fig. 4, input traffic-flow features are first preprocessed and mapped by a classical front end to a low-dimensional angle representation, which is embedded into a variational quantum latent encoder.

### 3. *Training & Quantum Circuit*

The QAE model was trained end-to-end by minimizing mean squared reconstruction error,

$$\mathcal{L} = \frac{1}{N}\sum_{i=1}^{N} \| x_i - \hat{x}_i \|_2^2 \tag{16}$$

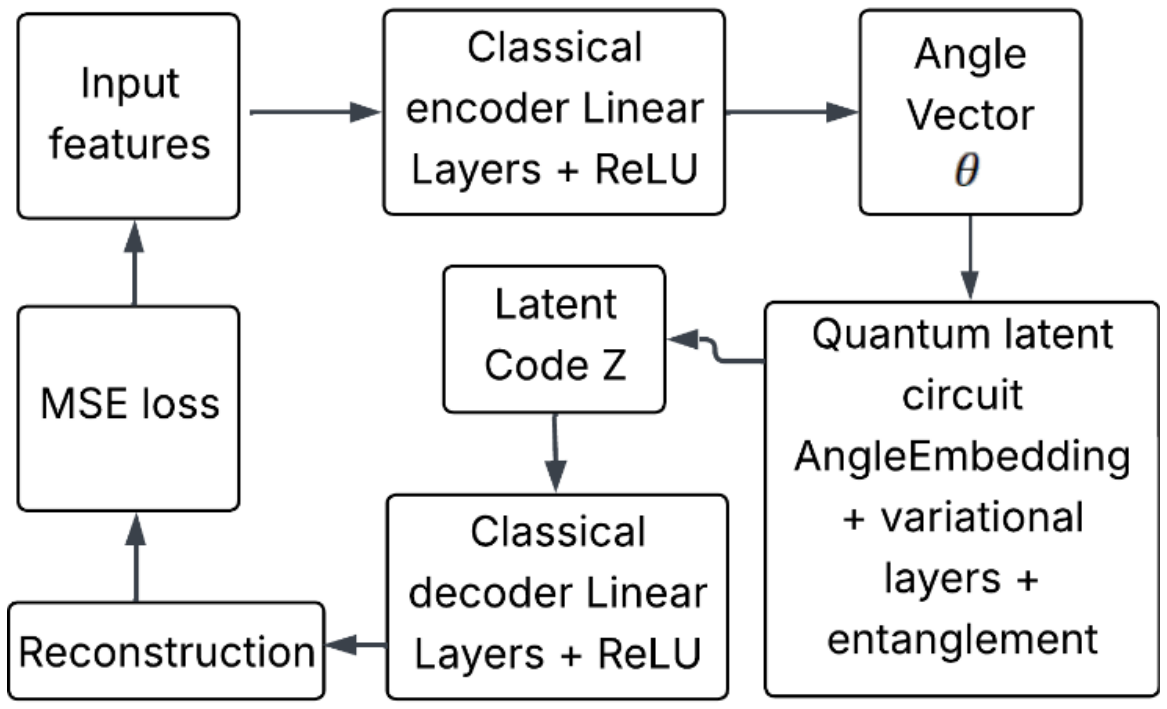

FIG 4. High-level schematic of the extended hybrid classical–quantum autoencoder used for DDoS anomaly detection. Classical preprocessing maps flow features to circuit parameters for a variational quantum latent encoder, whose learned latent state is probed through reconstruction, latent-energy, Bloch-geometric, fidelity/entropy, and QFI-based diagnostic branches. This design allows anomaly detection to be interpreted simultaneously in terms of reconstruction quality, latent-state geometry, effective energy landscape, and trainability. A full model pipeline is shown in Fig. 30 of Appendix C.

Gradients of quantum parameters were computed using the parameter-shift rule [65, 119]. Optimization was performed using Adam. The anomaly threshold $\tau$ was defined as the 95$^{\text{th}}$ percentile of benign validation reconstruction errors and fixed during testing. All experiments were conducted using state-vector simulation without noise modeling to isolate representational effects. **Bloch-Space Representation**: For each qubit $q$, Pauli expectation values define a Bloch vector

$$\mathbf{r}^{(q)}(x) = (\langle X\rangle_q, \langle Y_q\rangle, \langle Z_q\rangle) \tag{17}$$

The **absolute Bloch drift** is defined as

$$d^{(q)}(x) = \| \mathbf{r}^{(q)}(x) - \boldsymbol{\mu}_B^{(q)} \|_2 \tag{18}$$

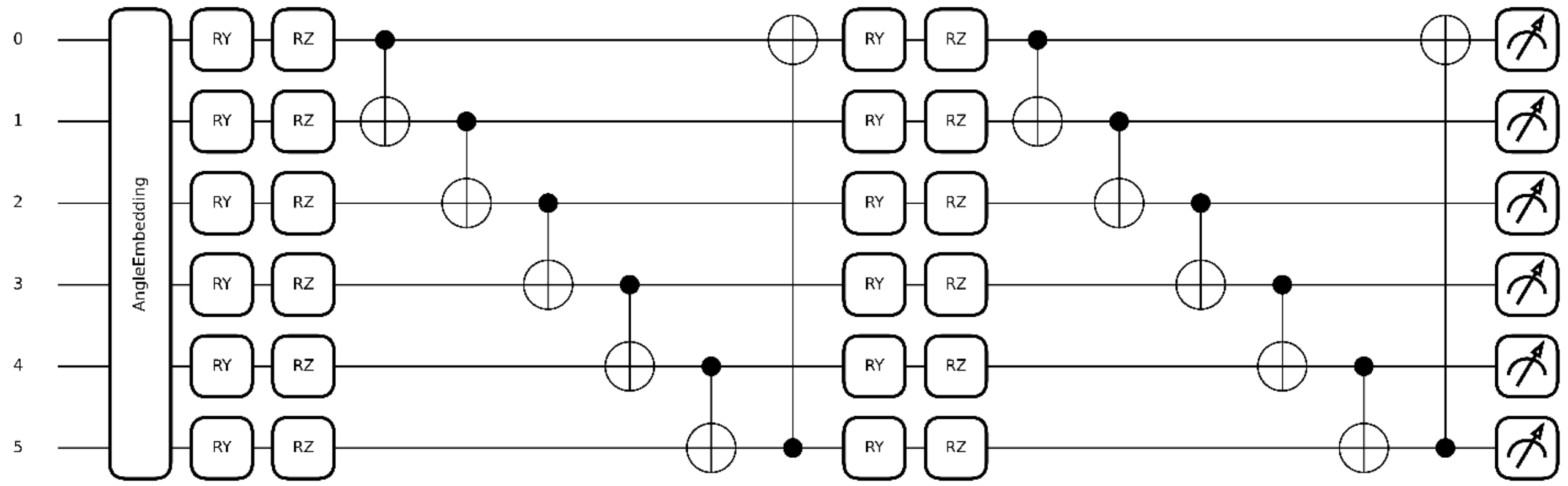

FIG 5. Variational quantum circuit employed as the quantum latent layer of the hybrid quantum autoencoder. Classical input features are encoded via angle embedding, followed by two layers of trainable single-qubit rotations and ring-structured entanglement. Final measurements yield the quantum latent representation, which is passed to the classical decoder. This hardware-efficient ansatz induces structured quantum latent geometry while remaining shallow enough to ensure stable embeddings and avoid spurious dynamical effects.

where $\boldsymbol{\mu}_B^{(q)}$ is the benign mean Bloch vector. The **consecutive Bloch drift** is defined as

$$\Delta^{(q)}(x_t) = \| \mathbf{r}^{(q)}(x_{t+1}) - \mathbf{r}^{(q)}(x_t) \|_2 . \tag{19}$$

Fig. 5 shows the quantum circuit generated during the code implementation as the VQC. The VQC is a *hybrid quantum autoencoder latent circuit* with 6 qubits (indexed 0–5), two variational layers, followed by Pauli measurements. The *AngleEmbedding* block applies a data-dependent rotation to each qubit where each input feature $x_i$ is encoded as $R_Y(x_i)$ that maps classical features into quantum embeddings. This defines the input-dependent quantum state $|\psi(x)\rangle$. In *first variational layer formed by the $RY-RZ$ blocks*, each qubit undergoes a trainable local unitary $U(\theta) = R_Y(\theta_{i,1}) R_Z(\theta_{i,2})$, where parameters are learned jointly with the classical encoder/decoder. This layer shapes the local geometry of the quantum latent space. Following that, the CNOT gates form a ring topology:

$$0 \to 1 \to 2 \to 3 \to 4 \to 5 \to 0$$

The ring of CNOT operations introduces nonlocal correlations while keeping the circuit shallow and hardware-efficient. The circuit geometry motivates Bloch-space analysis of the reduced single-qubit states, but the entangling pattern alone does not establish an advantage over the classical latent representation.

A second trainable layer increases circuit expressivity while preserving a shallow ansatz. Measurements of Pauli operators define the quantum latent observables passed to the classical decoder and used for Bloch analysis. The observed QFI spectrum is reported separately as a local parameter-space diagnostic; neither circuit depth nor the order-sensitivity results are taken as proof that barren plateaus are absent.

The variational quantum circuit shown in Fig. 5 defines the quantum latent mapping $\mathrm{x} \mapsto |\psi(\mathrm{x})\rangle$ used throughout this work. For each input sample, classical features are embedded via angle encoding, followed by trainable single-qubit rotations and ring-structured entanglement. Measurements of Pauli operators yield Bloch vectors $\mathbf{r}^{(q)}(\mathrm{x}) = (\langle X_q \rangle, \langle Y_q \rangle, \langle Z_q \rangle)$ for each qubit $q$, which form the quantum latent representation. These Bloch vectors are subsequently used to define geometric drift metrics that quantify anomaly-induced displacement in quantum state space.

### 4. *Anomaly Scoring and Threshold Selection*

For both models, **reconstruction error** is used as the anomaly score. Higher reconstruction error indicates greater deviation from learned benign patterns. To maintain a fully unsupervised evaluation protocol, the decision threshold was set to the 95$^{th}$ percentile of benign validation reconstruction errors. This approach avoids label leakage and reflects realistic intrusion detection scenarios. Thresholds optimized directly for F1-score were evaluated but found to produce degenerate classifiers and were therefore excluded from final reporting.

Model performance is assessed using ROC-AUC, average precision, fixed-threshold false-positive and false-negative rates, matched classical and quantum latent-distance AUCs, and order-sensitivity controls. For the classical model, latent drift is the distance from the benign latent centroid. For the QAE, Bloch-space observables are aggregated into an absolute latent drift score. Consecutive/source-order drift is compared with 1000 within-class permutations and is not interpreted as physical time evolution.

#### 5. *Quantum Fisher Information (QFI)*

QFI is a local measure of sensitivity of the quantum latent state to circuit-parameter perturbations. We use it only as a parameter-space information-geometry diagnostic, closely related to the Bures metric [113]. The spectrum can reveal active and anisotropic parameter directions, but by itself it does not establish input-space Lipschitz continuity, anomaly separability, or the absence of barren plateaus. The *mean QFI matrix* is computed on 64 benign validation samples for a representative trained seed, with 1000 bootstrap replicates for uncertainty summaries. The interpretation is deliberately restricted to local parameter-space sensitivity. Stronger trainability claims would require additional gradient-variance or scaling analyses that are outside the scope of the present work.

Taken together, these diagnostics make the hybrid quantum autoencoder useful not only as a detector but also *as an analysis instrument*. *Relative to a classical autoencoder*, its main benefit is the ability to derive multiple anomaly views from one latent representation: reconstruction error, energy, geodesic drift, fidelity, purity, entropy, and trainability structure. For *cybersecurity, that richer latent description is often more valuable* than a single scalar anomaly score, particularly when *attacks are expressed through coordinated shifts* across many flow features [45, 112].

### C. Experimental Setup

The experimental runs were executed in documented software environments rather than assumed to share one identical machine configuration. For the primary spectral analysis the reproducibility manifest records Python 3.11.9, NumPy 2.4.0, SciPy 1.16.3, pandas 2.3.3, QuTiP 5.2.2, NetworkX 3.6.1, and Matplotlib 3.10.8. OS used was Linux Ubuntu 24.04 LTS Hardware execution details are reported explicitly below. This separation avoids attributing platform-dependent timing or numerical behavior to a single undocumented environment. Timing results are reported as *mean ± standard deviation* over multiple trials; Hardware tests were conducted on `IBM_fez` .

#### 1. *Bloch-Geometry probed QAE for IDS*

Performance is quantified with matched five-seed CAE/QAE comparisons using identical preprocessing and test partitions. Bloch-space drift is derived from Pauli expectation values of the QAE latent circuit, while the classical control uses distance from the benign CAE latent centroid. Consecutive drift is evaluated as an order-sensitivity diagnostic: source ordering is compared with 1000 within-class permutations, without assigning temporal meaning to the CSV row order. QFI is reported separately as a local parameter-space diagnostic.

## III. RESULTS

### A. Spectral structure in graph learning & Boson Sampling

Fig. 6 reports the synchronized five-seed regularization sweep for the 15-node, 10-SL/5-UL configuration using three diagnostics evaluated from the same checkpoints: relative Hilbert-Schmidt representation variance, relative algebraic connectivity, and effective spectral rank. The RBF bandwidth used to construct the learned output graph is fixed across all regularization strengths and seeds.

The response is non-monotonic. At the final checkpoint, mean relative variance $V/V_0$ is $2.500 \pm 0.323$ for $\lambda_{\text{reg}} = 0, 0.580 \pm 0.071$ for $\lambda_{\text{reg}} = 0.5$, $0.148 \pm 0.021$ for $\lambda_{\text{reg}} = 1.5$, and $1.306 \pm 0.245$ for $\lambda_{\text{reg}} = 3.0$. The corresponding relative algebraic connectivity values are $0.227 \pm 0.033$, $1.713 \pm 0.281$, $4.177 \pm 0.770$, and 0.700±0.148, respectively. Thus the strongest representation contraction occurs at $\lambda_{\text{reg}} = 1.5$ while graph connectivity is simultaneously strengthened rather than fragmented.

None of the 20 runs across the five initialization seeds and four regularization strengths satisfied either the prespecified representation-concentration criterion, $V/V_0 \leq 0.10$ , or the graph-fragmentation criterion, $\lambda_2 / \lambda_{2,0} \leq 0.01$. As such, no run satisfied the joint criterion. Accordingly, Fig. 6 demonstrates a finite-system, regularization-dependent reorganization in which representation spread and graph connectivity respond differently to coupling strength, and does not support a collapsed phase or a geometric phase transition.

The effective spectral rank remains mathematically and numerically well above its lower bound of one. Final mean $d_{eff}$ values are $12.665 \pm 0.060$, $13.800 \pm 0.009$, $13.979 \pm 0.002$, and $13.020 \pm 0.275$ for $\lambda_{\text{reg}} = 0, 0.5, 1.5, 3.0$, respectively. These values show broad spectral support rather than low-dimensional spectral collapse. Taken together, the synchronized diagnostics separate representation concentration from graph fragmentation and remove the earlier inference that one necessarily implies the other. The spectral response is therefore described as a finite-system crossover/reorganization. The learned Fiedler structure is subsequently probed independently by the two-boson calculation of Fig. 7.

Fig. 7 applies the systematic boson-Fiedler analysis to the learned 15-node output graph rather than to the earlier three-edge illustration. The graph contains 105 positive learned-similarity edges. All 105 unordered input pairs and 31 propagation times are evaluated, yielding 3255 input/time conditions. The primary statistic—the median across conditions of the absolute Spearman association between $|I_{uv}|$ and the Fiedler edge split is $0.1085$ with a 95% cluster-bootstrap interval $[0.0845, 0.1397]$. A 2000-permutation node-label null gives $p = 4.998\times10^{-4}$. The effect is therefore small but reproducibly structured under the prespecified analysis. For each input-pair/time condition, Spearman's rank coefficient $\rho_S$ measures the monotonic edgewise association between interference magnitude $|I_{uv}|$ and Fiedler separation $\Delta v_{2,uv} = |v_2(u) - v_2(v)|$. Across all 3255 prespecified conditions, the median absolute association is $T_{BF} = \mathrm{median}|\rho_S| = 0.1085$, with a 95% cluster-bootstrap interval of $[0.0845 \,|\, 0.1397]$. Although the effect size is small, the node-label permutation test gives $p = 4.998\times10^{-4}$, indicating that the observed association is reproducibly structured relative to the specified null model.

Fig. 8 provides an independent-seed audit of the joint criterion used in Fig. 6. Each point represents the final relative variance and relative algebraic connectivity for one seed and one regularization strength. The dashed lines show the prespecified concentration and fragmentation thresholds.

No point lies in the lower-left joint-criterion region. The audit therefore confirms at the run level—not only after averaging—that none of the 20 tested seed/regularization combinations satisfies the operational joint-collapse criterion. The conclusion is limited to a non-monotonic finite-system reorganization over the tested regularization strengths.

The primary pre/post analysis in Fig. 9 provides a complementary view at $\lambda_{\mathrm{reg}} = 0.5$. The learned graph remains connected and dense (15 nodes, 105 positive learned-similarity edges). Algebraic connectivity increases from $\lambda_2 = 5.5791$ before training to $5.8263$ after training, while $\lambda_3$ increases from $6.1566$ to $9.7324$ and the Fiedler eigengap widens from $0.5775$ to $3.9061$. Spectral entropy increases modestly from $S = 2.61537$ to $2.62962$ and $d_{\mathrm{eff}}$ from $13.6723$ to $13.8685$. These changes indicate spectral reorganization with broad support, not collapse.

Having established how training dynamics control representation diversity and graph structure, we now examine how these learned representations manifest in experimentally accessible observables on quantum hardware. Fig. 10(a) compares hardware energies with ideal and noisy Aer baselines across representative parameter settings.

Fig. 10 compares the retained three-qubit Z-diagonal energy benchmark on ibm_fez with ideal and calibration-derived noisy simulations. Hardware deviations contain both positive and negative shifts, while the noisy simulation captures part of the systematic bias. This experiment is used only to assess implementation-level robustness of the energy observable under finite shots and device noise.

The noisy-Aer curve is a diagnostic reference rather than a predictive model of instantaneous device behavior. Residual differences can reflect calibration drift, mapping/transpilation details, correlated errors, and other effects not represented in the calibration-derived model. These hardware results are independent of the boson-Fiedler calculation and are not interpreted as a hardware measurement of $I_{uv}$.

Hardware experiments were conducted using session-less SamplerV2 execution with 2000 shots per circuit evaluation. The 13-point landscape in Fig. 11 uses ten independent hardware repetitions per point; the reported spread is the sample standard deviation across repetitions.

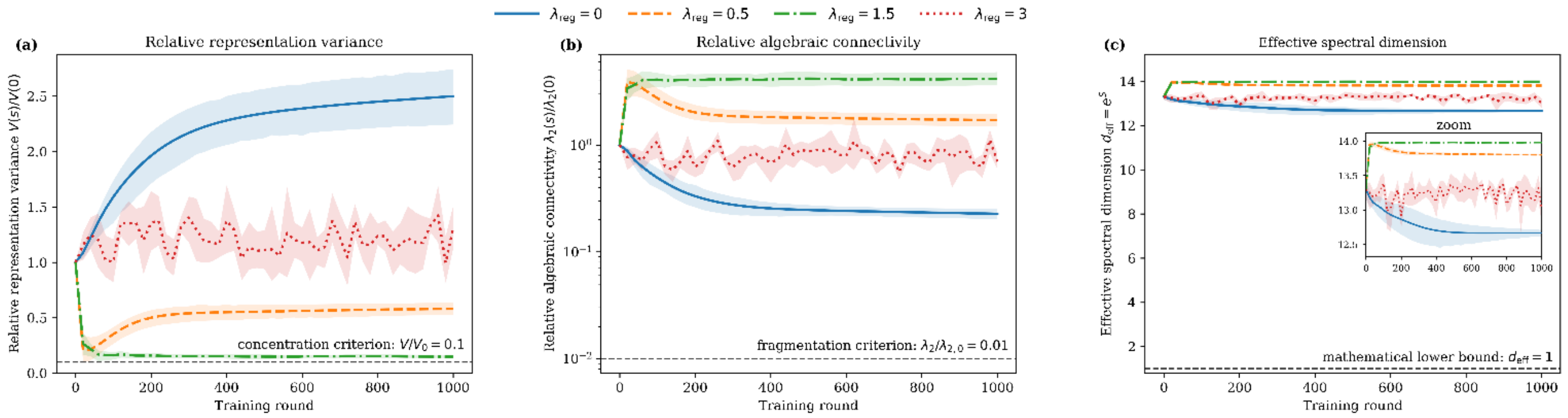

FIG. 6. (Color online) Synchronized multi-seed diagnostics for the 15-node, 10-SL/5-UL graph model. Curves show means across five paired initialization seeds with bootstrap 95% confidence bands. (a) Relative Hilbert-Schmidt representation variance $V(s)/V(0)$; the dashed line marks the prespecified concentration criterion $V/V_0 = 0.10$. (b) Relative algebraic connectivity $\lambda_2(s)/\lambda_2(0)$ on a logarithmic scale; the dashed line marks the fragmentation criterion 0.01. (c) Effective spectral rank $d_{\mathrm{eff}} = e^S$; the dashed line marks the mathematical lower bound $d_{\mathrm{eff}} = 1$. No tested seed satisfies the joint oncentration/fragmentation criterion.

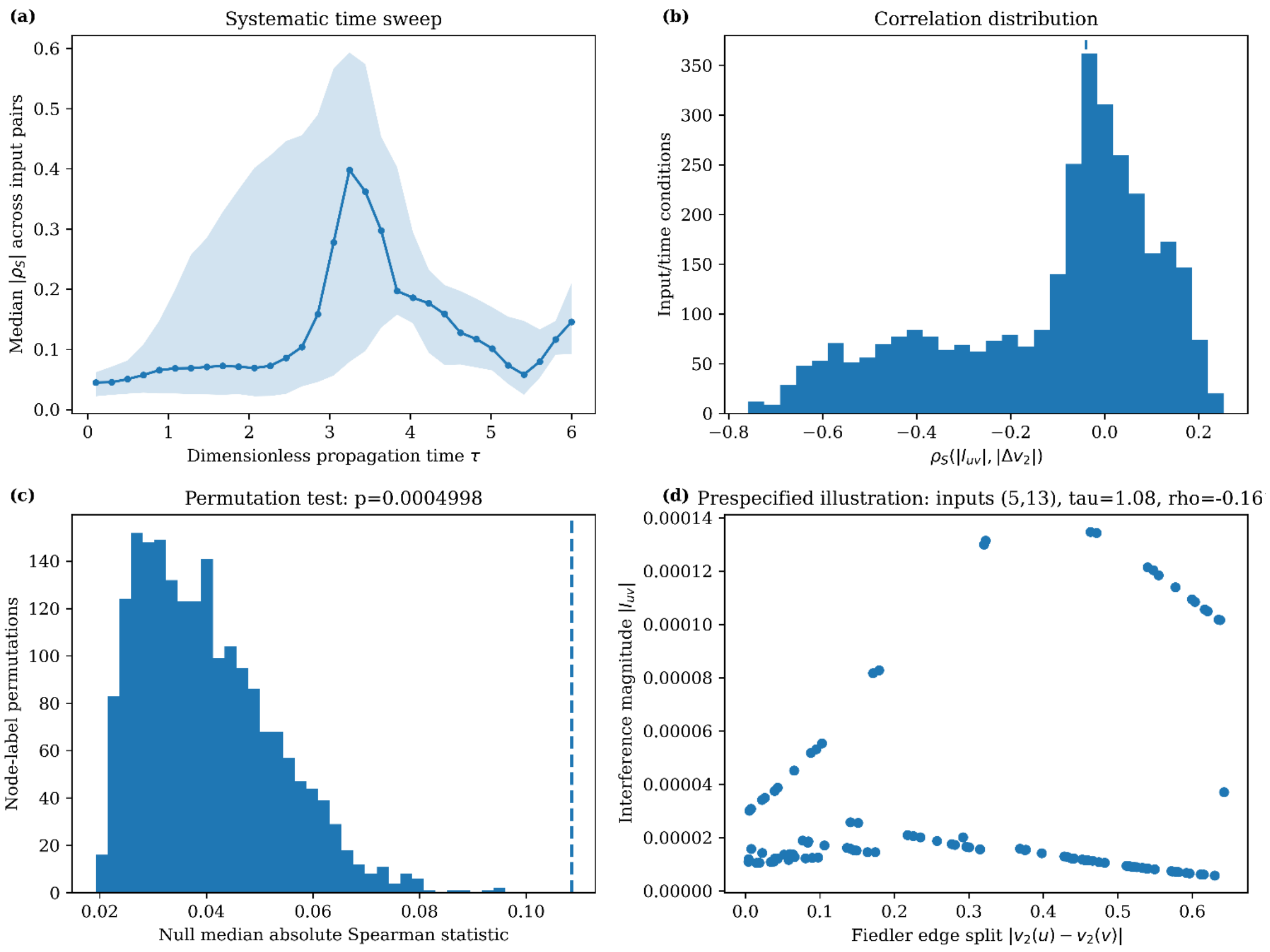

FIG. 7. (Color online) Systematic two-boson/Fiedler analysis on the learned 15-node graph using the combinatorial-Laplacian propagator. (a) Median absolute Spearman association across input pairs versus dimensionless propagation time; shading denotes the interquartile range. (b) Distribution of the condition-level association statistic over all 105 unordered input pairs and 31 propagation times (3255 conditions). (c) Node-label permutation null distribution; the dashed line marks the observed primary statistic 0.1085, with $p = 4.998\times10^{-4}$ and a 95% cluster-bootstrap interval $[0.0845, 0.1397]$. (d) Representative edge-resolved $|I_{uv}|$ versus $|v2(u)-v2(v)|$ for one prespecified input/time condition.

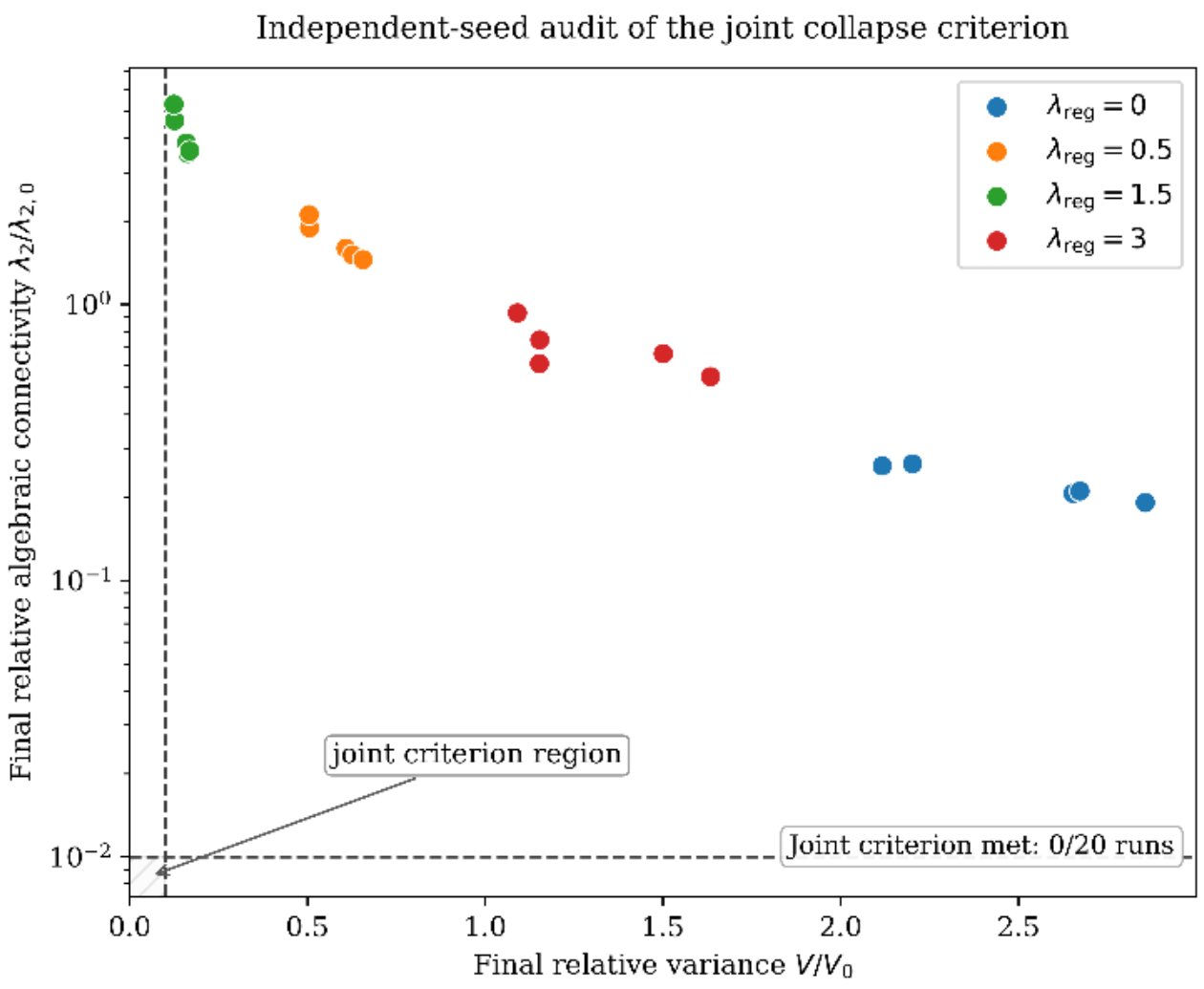

FIG. 8. (Color online) Independent-seed audit of the Fig. 6 operational criteria.

In Fig. 8, each marker represents one of 20 seed/regularization runs. The vertical line marks $V/V_0 = 0.10$ and the horizontal line marks $\lambda_2/\lambda_{2,0} = 0.01$. No run enters the joint-criterion region. This panel is an audit of the classification rule, not evidence for a thermodynamic phase boundary.

Fig. 11 probes the same three-qubit energy expectation along a one-dimensional interpolation in parameter space. Despite finite noise and systematic offsets, the hardware mean follows the qualitative curvature of the statevector reference over the sampled interval.

The interpolation therefore provides an implementation-level consistency check of the local energy landscape rather than a validation of the graph-interference analysis. The scan uses a fixed base parameter vector and a fixed direction in circuit-parameter space.

$$E(\alpha) = \langle H \rangle_{\theta(\alpha)}$$

Fig. 11 compares the noiseless statevector reference with hardware values obtained at the same logical parameter points. Error bars summarize ten independent 2000-shot hardware repetitions at each point.

The hardware section ends with this energy-landscape comparison. The learned similarity graph and bosonic interference results discussed in Figs. 7, 9, and 12 are obtained from the separate systematic simulation workflow described above.

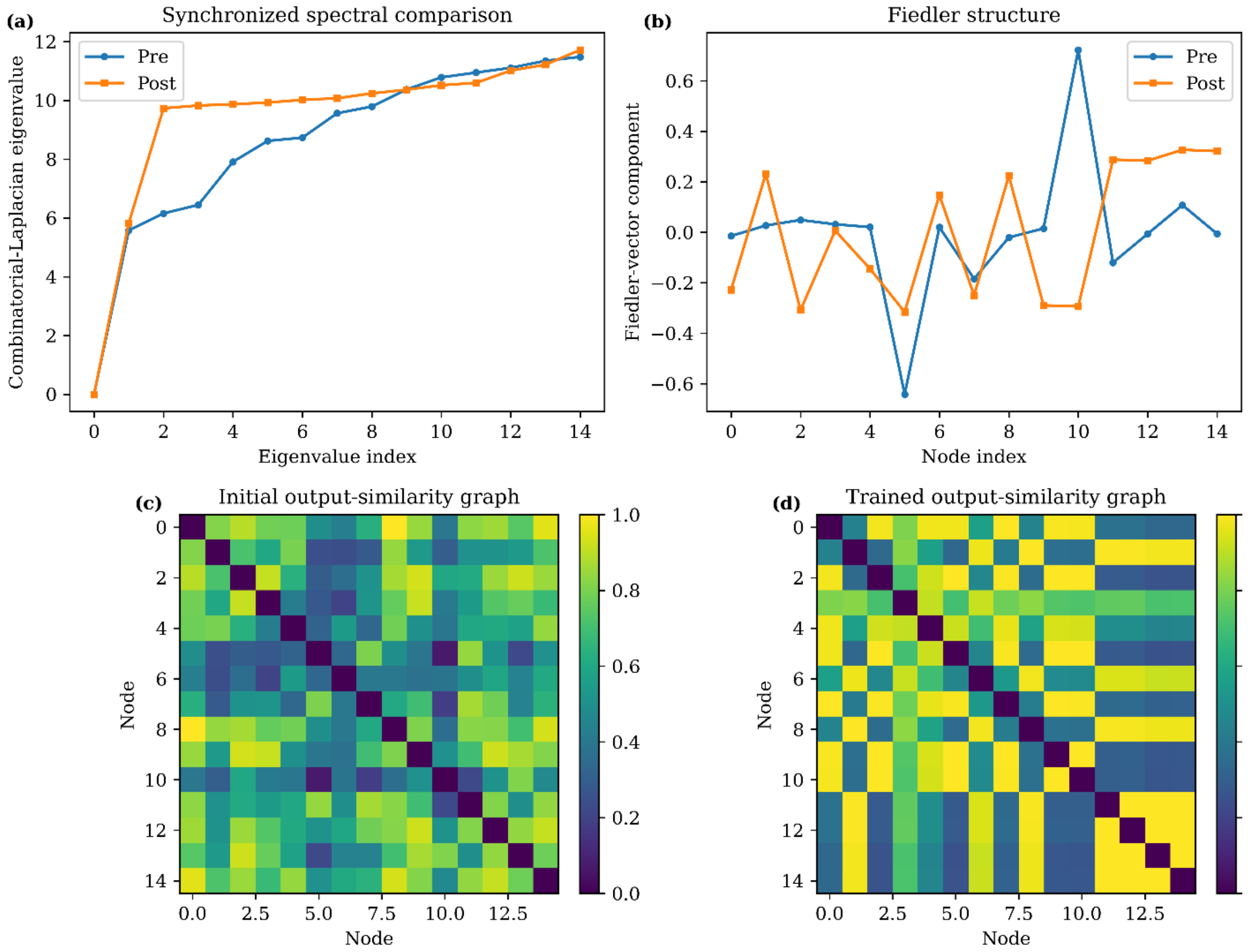

FIG. 9. (Color online) Primary pre/post spectral analysis for the 15-node, 10-SL/5-UL run at $\lambda_{reg} = 0.5$ using a fixed RBF bandwidth. The learned graph remains connected; $\lambda_2$ changes from 5.5791 to 5.8263, spectral entropy from 2.61537 to 2.62962, and effective spectral rank from 13.6723 to 13.8685. The comparison documents modest spectral reorganization while maintaining high effective rank.

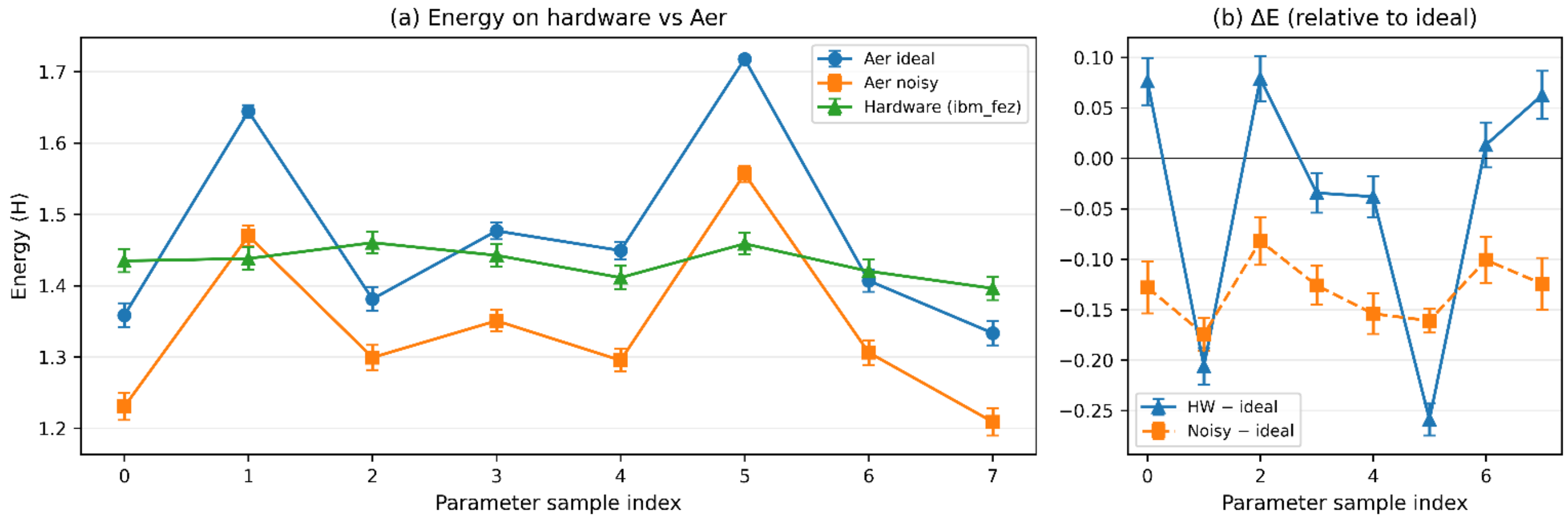

FIG. 10. (Color online) Three-qubit Z-diagonal energy benchmark on `ibm_fez` compared with ideal and calibration-derived noisy Aer references. The comparison diagnoses finite-shot and device-noise effects in the energy observable. This figure does not measure the two-boson interference observable used in Fig. 7.

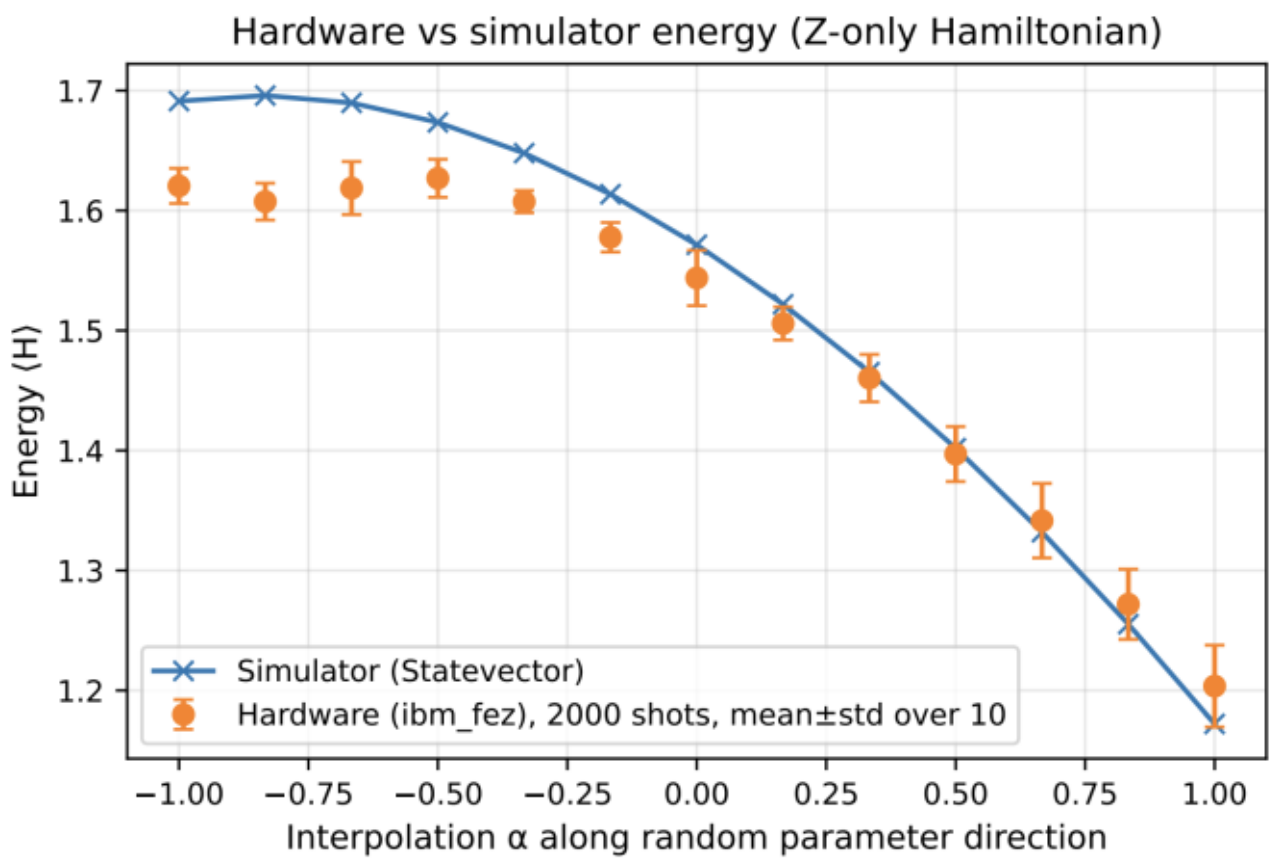

FIG. 11. (Color online) Energy expectation value along a one-dimensional interpolation in parameter space for the retained three-qubit Z-diagonal benchmark. Hardware results on ibm_fez are shown as mean±one sample standard deviation over ten independent runs with 2000 shots each. The statevector curve is a noiseless reference. The experiment is not a measurement of the two-boson coincidence observable.

Fig. 12 shows the pairwise fidelity/similarity structure of the graph data used to motivate the learned relational organization. It is interpreted as a graph-structure visualization rather than as direct evidence of bosonic enhancement.

The block structure in Fig. 12 indicates heterogeneous pairwise similarity and motivates the use of Laplacian spectral diagnostics. Quantitative statements about bosonic interference are based on the systematic learned-graph analysis of Fig. 7 rather than on visual inspection of this matrix.

The pairwise similarity matrix is symmetric with a bright diagonal and heterogeneous off-diagonal structure. Any relation between that structure and the Fiedler/interference observables is evaluated quantitatively through the fixed learned graph and the permutation-controlled analysis, not inferred from an assumed collapse or near-disconnection.

Benchmarks demonstrate numerical equivalence between operator-based and circuit-based implementations, with fidelities consistently approaching unity. Dense Qiskit execution matches QuTiP performance (Fig.21 in Appendix A) for small systems. These results confirm that operator-based training rules can be faithfully reproduced in circuit-based frameworks without prohibitive cost.

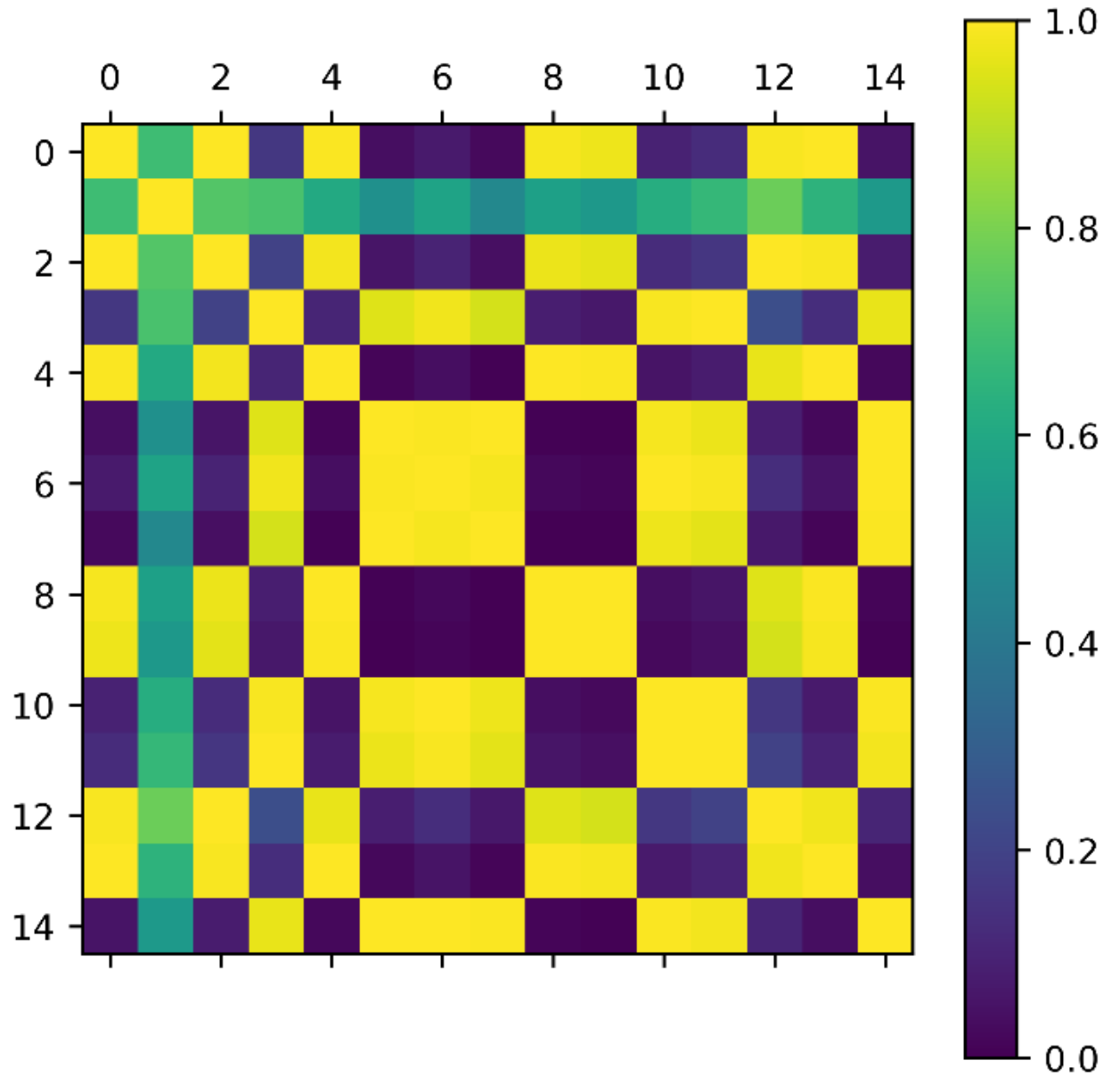

FIG 12. (Colour online) Fidelity matrix. Heatmap of edge-resolved bosonic enhancement, normalized to the maximum value. Each entry corresponds to a two-photon coincidence observable associated with a graph edge.

### B. Bloch-Space diagnostics in Hybrid QAE

This addresses the results of the anomaly detection experiments applying classical autoencoder (CAE) and hybrid QAE to IDS data [98, 99].

#### *1. Reconstruction-Based Anomaly Detection*

Using reconstruction error as the anomaly score, both matched models rank attacks above benign traffic, but the classical baseline performs better on this dataset. Across five seeds, the CAE-6D attains ROC-AUC 0.9827±0.0032 and average precision 0.9705±0.0037, whereas the QAE-6Q attains ROC-AUC 0.9570±0.0157 and average precision 0.9388±0.0171.

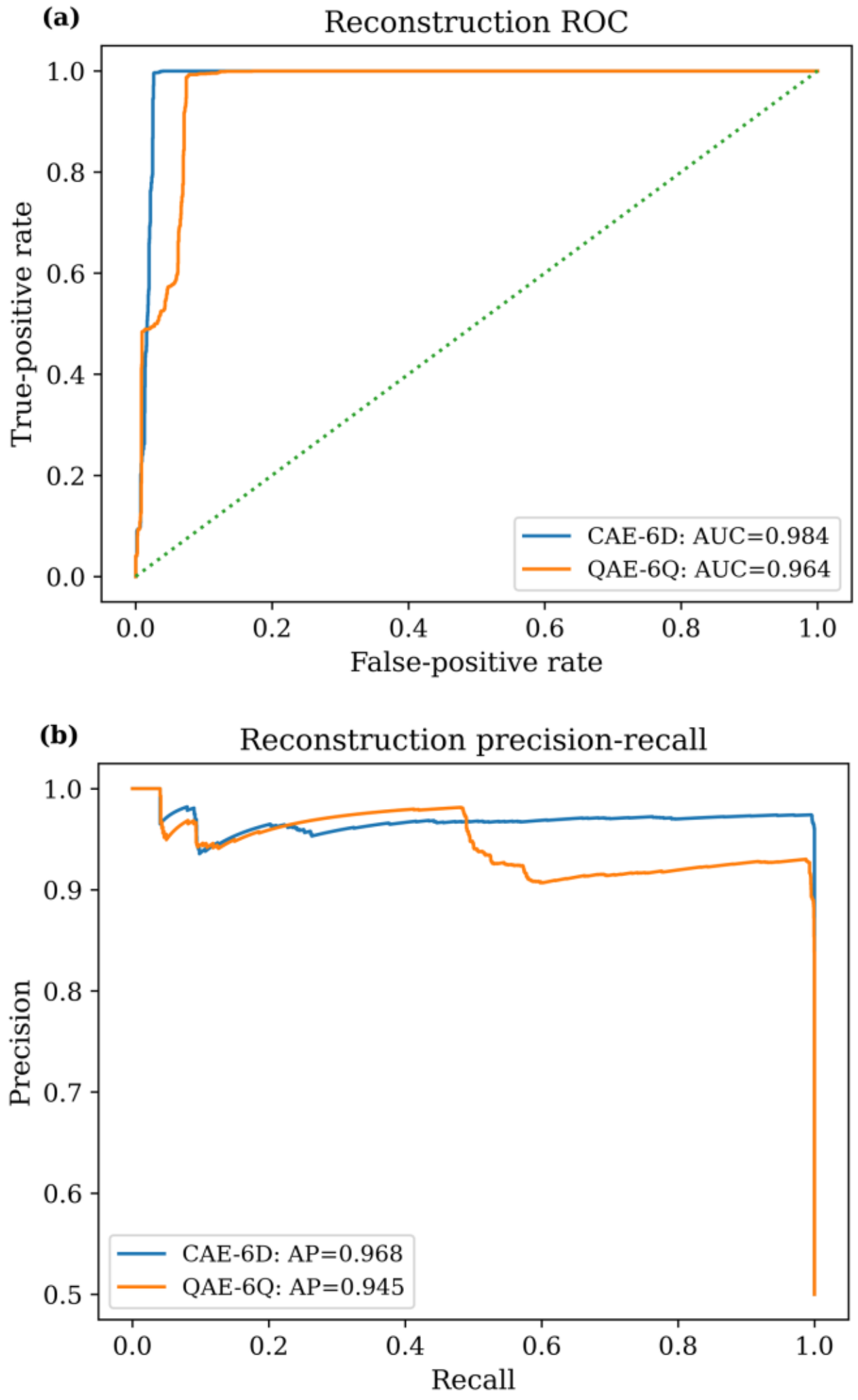

FIG. 13. (Color online) Representative seed-44 reconstruction-error ROC and precision-recall curves for the dimension-matched CAE-6D and QAE-6Q models. Both models use the same preprocessing, 5000 benign training rows, and fixed test partition. Multi-seed mean±standard-deviation metrics are reported in Table II.

Fig. 13 shows the representative seed-44 ROC/precision-recall curves. The multi-seed means in Table II are the primary quantitative comparison; the representative curves are included to display the shape of the score distributions.

The matched experiment therefore does not show a reconstruction-performance advantage for the QAE. Instead, the quantum model is analyzed for the additional state-space observables available from its latent circuit, while a matched classical latent-distance control is retained so that geometric discrimination is not attributed uniquely to quantum mechanics.

#### *2. Bloch-Space Drift as a Geometric Diagnostic*

Absolute latent-distance discrimination is also stronger for the classical control in the five-seed aggregate: CAE centroid distance gives ROC-AUC 0.9401±0.0084, whereas QAE absolute Bloch-space drift gives 0.8520±0.0427. Fig. 14 shows the representative seed-44 comparison (0.9335 versus 0.7992). Thus Bloch drift is a useful internal quantum-state diagnostic, but it is not uniquely superior to a classical latent-distance measure.

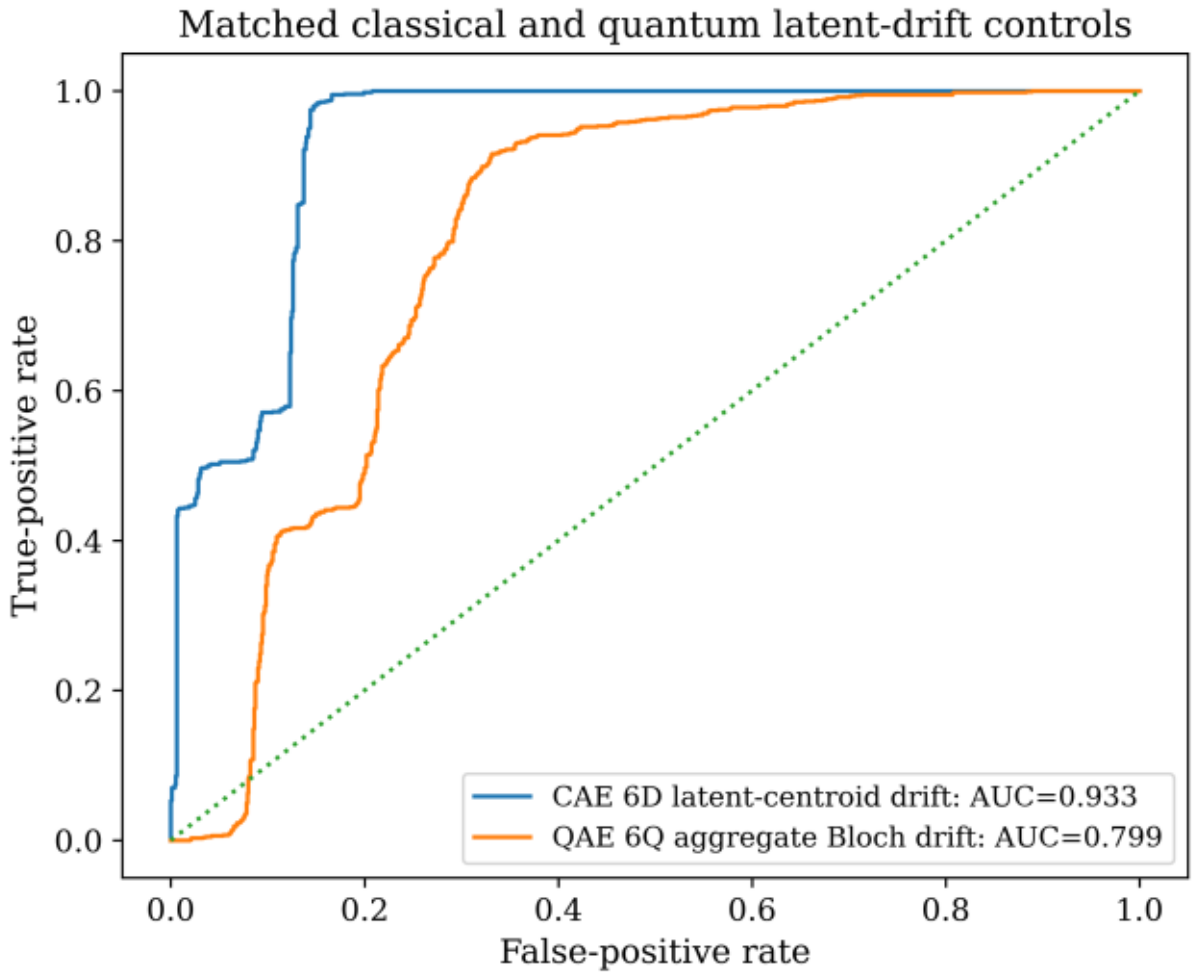

FIG. 14. (Color online) Representative seed-44 comparison of matched absolute latent-distance diagnostics. The CAE score is distance from the benign latent centroid and the QAE score is aggregate absolute Bloch-space drift. The representative AUCs are 0.9335 (CAE-6D) and 0.7992 (QAE-6Q); five-seed means are reported in Table II.

Consecutive/source-order drift is treated as an order-sensitivity control rather than a temporal observable. Across five seeds, the source-order AUC is 0.6869±0.0046 for CAE and 0.6059±0.0218 for QAE, compared with within-class permutation means of 0.6741±0.0086 and 0.5681±0.0337, respectively.

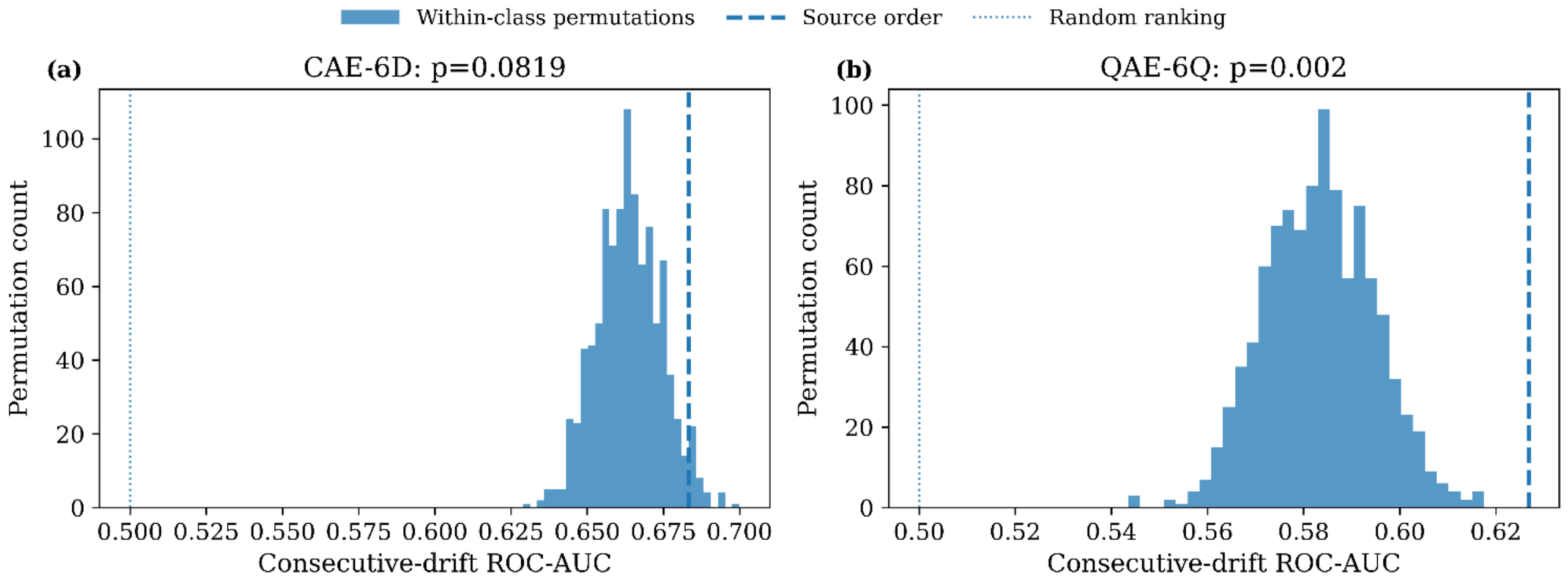

FIG. 15. (Color online) Representative seed-44 order-sensitivity control for consecutive latent drift. Histograms show 1000 within-class permutations; dashed lines mark the source-order statistic. The representative p values are 0.0819 for CAE-6D and 0.0020 for QAE-6Q. The analysis tests sensitivity to sample ordering and is not interpreted as physical time evolution.

Fig. 15 shows the representative seed-44 permutation distributions. For that seed, the QAE source-order statistic exceeds the permutation distribution (p=0.002), whereas the CAE comparison gives p=0.0819. Because the CSV order is not a controlled physical time variable, these differences are interpreted only as order sensitivity and not as dynamical stability, temporal drift, or evidence against barren plateaus.Table II reports mean±standard deviation over five matched initialization seeds. At the fixed 95th-percentile benign-validation threshold, both models have similar false-positive rates (0.0411 on average), but the QAE has a substantially larger false-negative rate (0.4845±0.0307) than the CAE (0.0688±0.1533, with large seed variability). This result therefore favors the CAE in operational detection performance while retaining the QAE as a source of directly measurable quantum-state diagnostics.

The operational threshold remains fully unsupervised: it is defined from benign validation reconstruction errors only and is fixed before attack labels are evaluated. For the matched experimental run, the five-seed mean false-positive rate is 0.0411 for both models. The mean false-negative rate is 0.0688 for CAE-6D and 0.4845 for QAE-6Q. These values show that threshold-free ranking performance and fixed-threshold operating performance must be interpreted separately.

The threshold therefore provides an operational complement to ROC-AUC rather than a geometric proof. It identifies a fixed benign-calibrated decision boundary, while the latent-distance and Bloch-drift analyses separately characterize how representations are displaced. In this experiment, the QAE's larger false-negative rate prevents a claim of improved operational robustness.

TABLE II. Comparison between the classical autoencoder (AE) and the hybrid quantum autoencoder (QAE).

| Metric | Classical AE (CAE-6D) | Hybrid QAE (QAE-6Q) |
|---|---|---|
| ROC–AUC (reconstruction error) | 0.9827 ± 0.0032 | 0.9570 ± 0.0157 |
| Average Precision (AP) | 0.9705 ± 0.0037 | 0.9388 ± 0.0171 |
| False Positive Rate | 0.0411 ± 0.0032 | 0.0411 ± 0.0018 |
| False Negative Rate | 0.0688 ± 0.1533 | 0.4845 ± 0.0307 |
| Absolute latent-drift ROC–AUC | 0.9401 ± 0.0084 | 0.8520 ± 0.0427 |
| Source-order consecutive-drift ROC–AUC | 0.6869 ± 0.0046 | 0.6059 ± 0.0218 |
| Within-class permutation mean ROC–AUC | 0.6741 ± 0.0086 | 0.5681 ± 0.0337 |
| Latent representation | 6D classical latent vector | 6-qubit Bloch-space latent state |

Fig. 16 summarizes the five-seed reconstruction, latent-drift, and order-sensitivity results. The combined view emphasizes the conclusion: the classical baseline is stronger on reconstruction and latent-distance discrimination, whereas the QAE contributes additional state-space observables that are useful for interpretation but do not constitute a performance advantage.

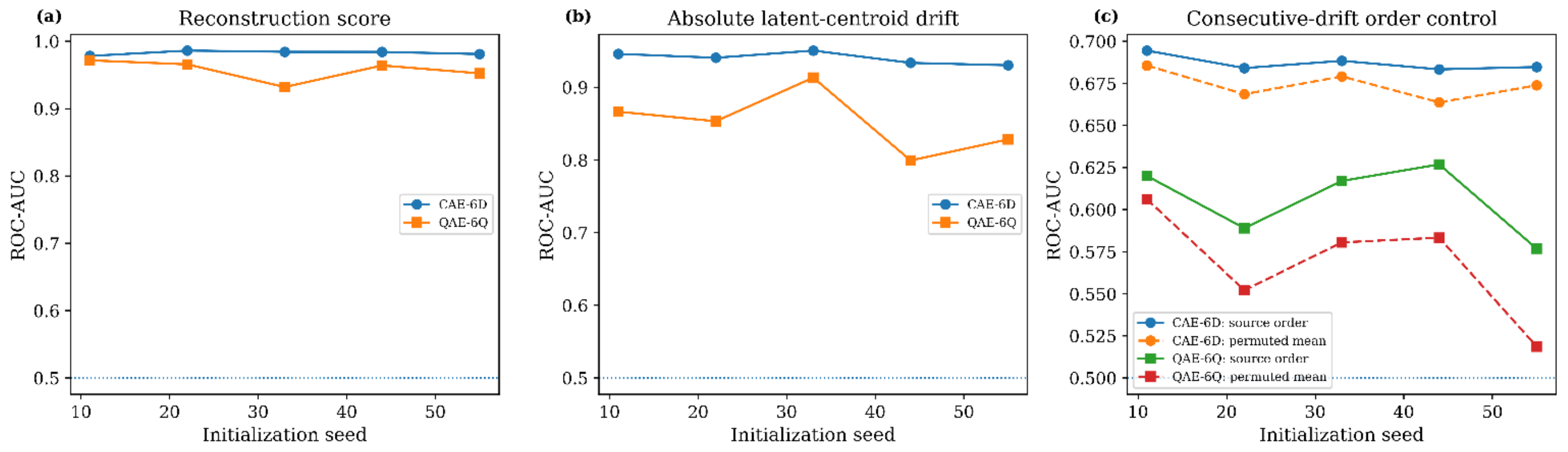

FIG. 16. (Color online) Matched five-seed summary of reconstruction, absolute latent-distance, and order-sensitivity diagnostics for CAE-6D and QAE-6Q. Curves/bars summarize the same fixed data partitions and model dimensions. The classical baseline is stronger on reconstruction and absolute latent-distance discrimination; order-sensitive consecutive drift is reported only as a control.

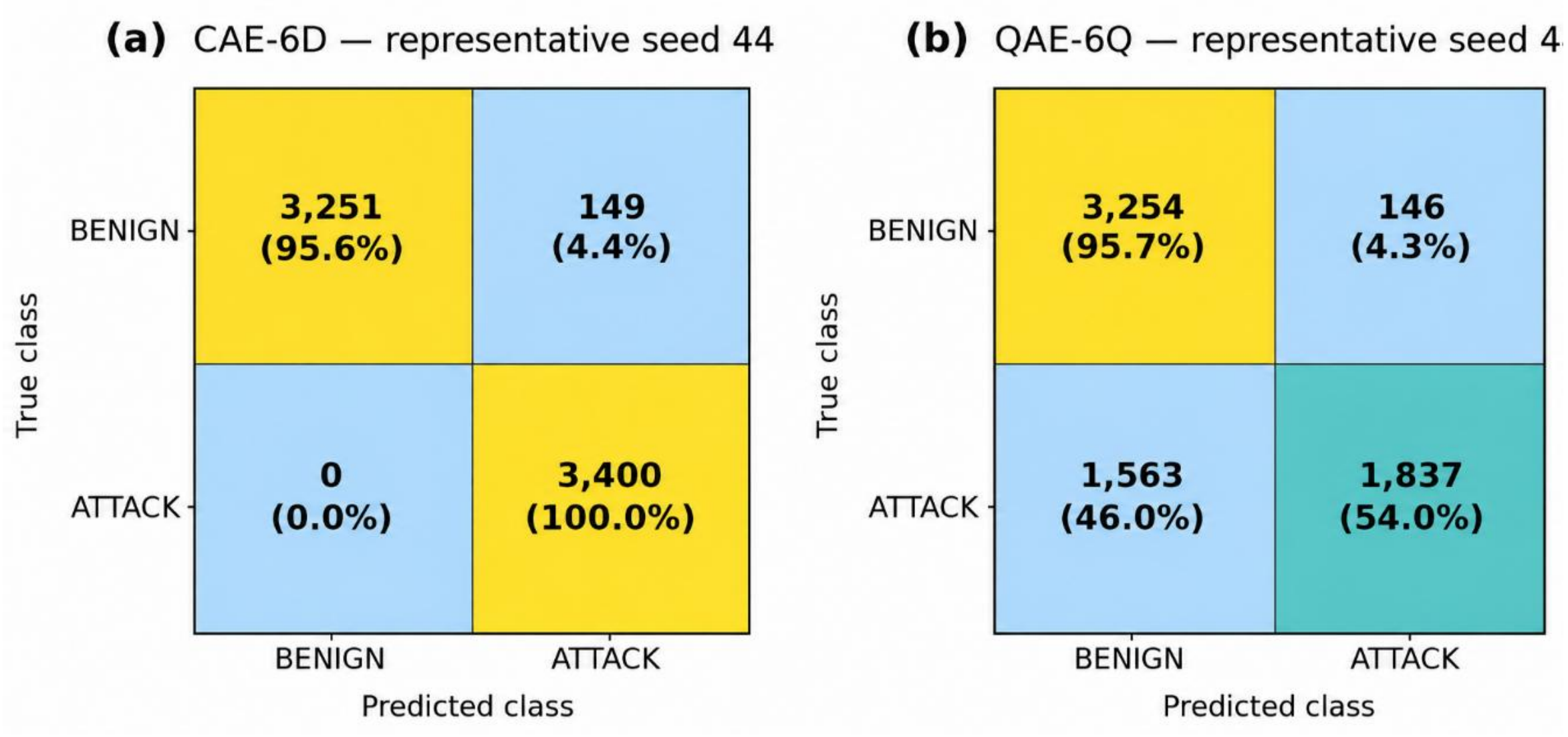

FIG. 17. (Color online) Representative seed-44 confusion matrices for CAE-6D and QAE-6Q at the fixed 95th-percentile benign-validation reconstruction threshold**.** Cell annotations give the number of test samples and the corresponding percentage within each true class. Both models produce similar false-positive rates for BENIGN traffic, but their ATTACK sensitivities differ substantially: the CAE correctly identifies all 3400 attack samples in this representative run, whereas the QAE misclassifies 1563 attacks as benign

Fig. 17 shows representative seed-44 confusion matrices at the fixed benign-validation threshold. The CAE has TN=3251, FP=149, FN=0, TP=3400, whereas the QAE has TN=3254, FP=146, FN=1563, TP=1837. These representative counts illustrate why comparable false-positive rates do not imply comparable false-negative behavior. The QAE is retained because its latent circuit permits direct Bloch-vector and QFI diagnostics of internal quantum-state geometry.

### 3. *Quantum Fisher Index*

The representative QFI matrix for Fig. 18 has trace 17.295, numerical rank 24, active condition number 72.99, participation ratio 13.73, and entropy effective rank 16.54. The bootstrap 95% interval for the trace is [16.840,17.742]. These values indicate a full numerical rank with anisotropic parameter sensitivity for this trained circuit.

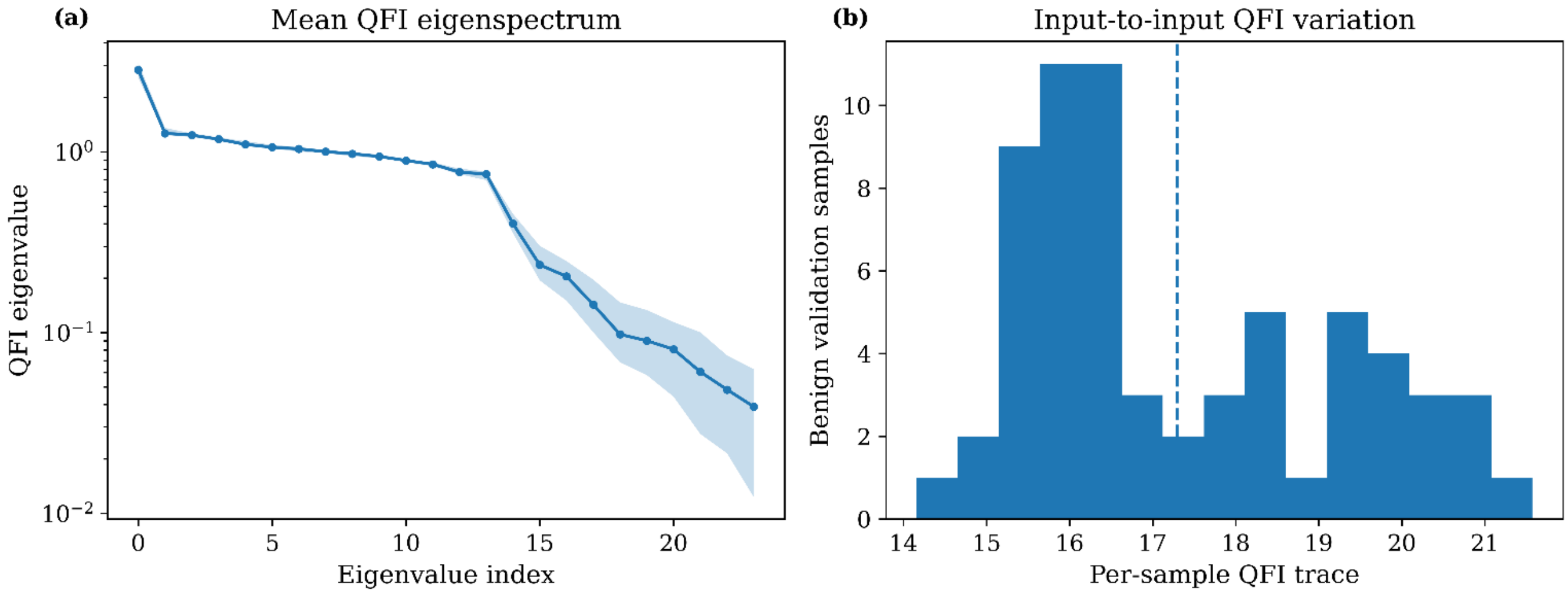

FIG. 18. (Color online) Parameter-space QFI diagnostics for the representative trained QAE seed. The mean QFI is computed on 64 benign validation samples with 1000 bootstrap replicates. The spectrum is interpreted as local circuit-parameter sensitivity only.

The QFI result is not used to infer input-space Lipschitz continuity, anomaly separability, or the absence of barren plateaus. Those stronger claims require additional derivatives or scaling analyses that were not performed.

Within its restricted interpretation, QFI provides a complementary parameter-space view of the trained QAE: the latent circuit retains multiple locally active parameter directions, while the task-level comparison is determined separately by reconstruction and latent-distance metrics. More details have been shared in Section B1 AppendixB

### 4. *Geometry of the Quantum Latent Manifold*

To quantify how Bloch-space discrimination is distributed across the six-qubit register, Fig. 19 reports the per-qubit absolute-drift AUC averaged over the five initialization seeds. The response is heterogeneous across qubits rather than concentrated in a single universal coordinate.

The per-qubit result is used as a quantum-state diagnostic and should be interpreted alongside the matched classical latent-distance control in Fig. 20. The latter shows that the classical centroid-distance score is consistently competitive or stronger across the same seeds.

Fig. 20 presents the paired five-seed comparison of the absolute latent-distance diagnostic. The CAE centroid-distance AUC exceeds the QAE aggregate Bloch-drift AUC for most matched initializations, reinforcing that the value of Bloch observables here is interpretive rather than a demonstrated performance advantage.

Fig. 20 provides the dimension-matched classical control for the geometric anomaly diagnostic. Across the five matched initialization seeds, the CAE-6D benign-centroid distance consistently yields higher ROC-AUC than the aggregate QAE-6Q Bloch-drift score. The five-seed mean decreases from ($0.9401 \pm 0.0084$) for CAE-6D to ($0.8520 \pm 0.0427$) for QAE-6Q, corresponding to a mean difference of approximately (0.088) ROC-AUC. The QAE diagnostic also exhibits substantially greater seed-to-seed variability, as reflected by its larger standard deviation. The paired comparison therefore shows that discrimination based on distance from a benign latent reference is not unique to the quantum representation and, for the present CIC-DDoS2019 SSDP experiment, is stronger and more stable in the classical latent space. Accordingly, Bloch-space drift is interpreted here as an additional physically interpretable diagnostic of the QAE latent state rather than as evidence of improved anomaly-detection performance or a quantum advantage.

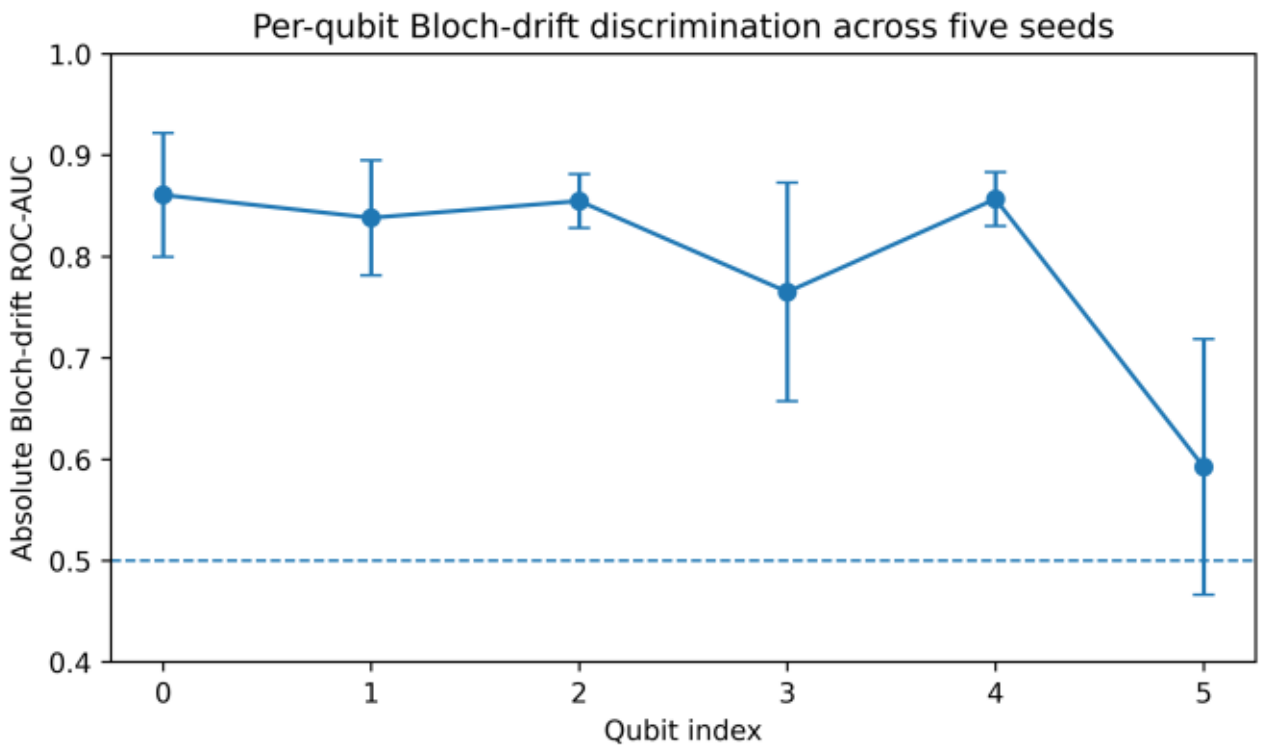

FIG. 19. (Color online) Per-qubit absolute Bloch-drift ROC-AUC across five QAE initialization seeds. Points show the seed mean and error bars the sample standard deviation; the dashed reference marks random ranking (AUC=0.5).

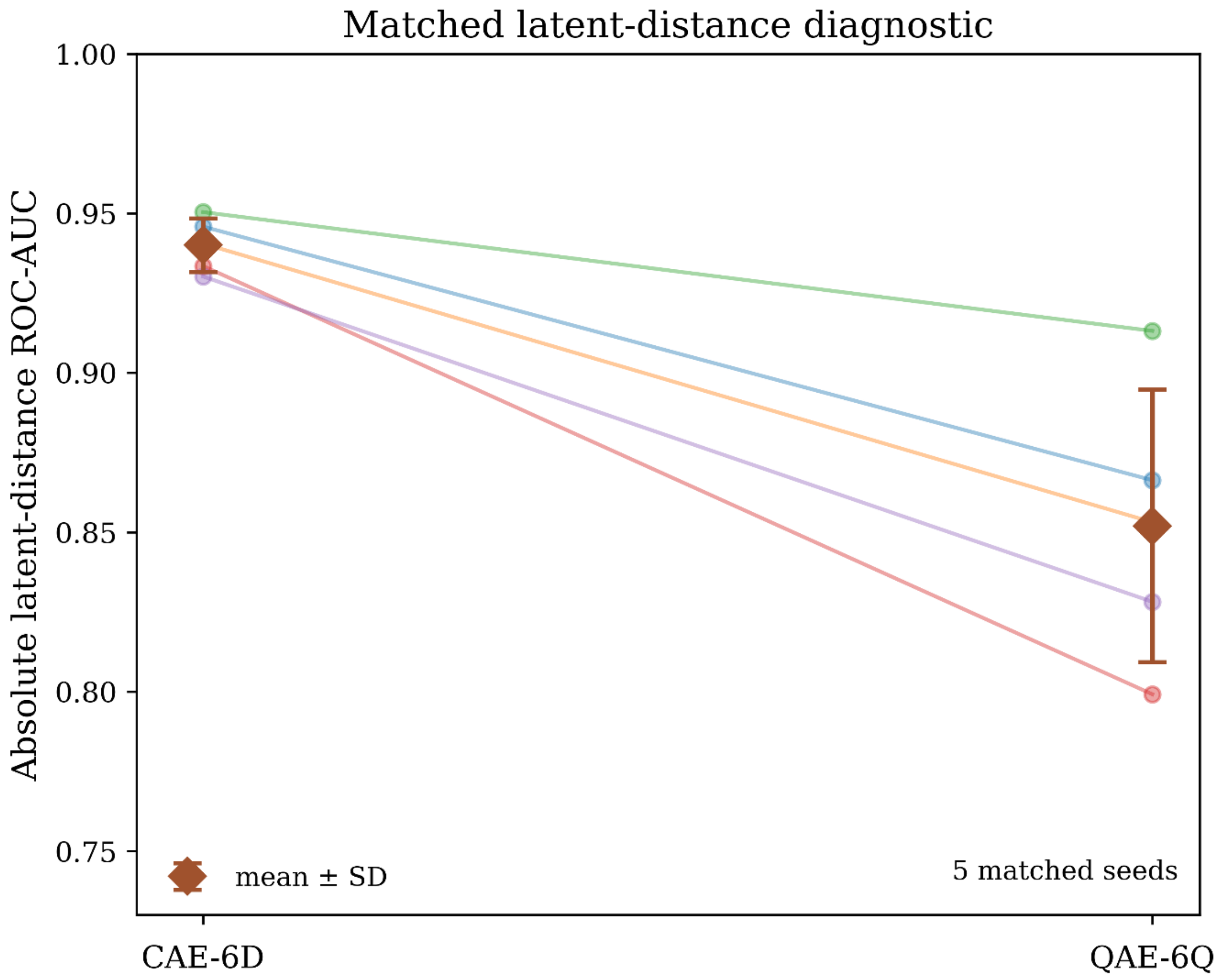

FIG. 20. (Color online) Matched latent-distance discrimination for CAE-6D and QAE-6Q across five initialization seeds. For the classical autoencoder, the anomaly score is the Euclidean distance of the six-dimensional latent representation from the benign latent centroid; for the quantum autoencoder, the corresponding score is the aggregate absolute Bloch-space drift of the six-qubit latent representation. Thin lines connect CAE and QAE results obtained with the same initialization seed and fixed data partition. Diamonds and error bars denote the mean ± sample standard deviation across the five seeds. The CAE-6D latent-distance diagnostic yields (0.9401±0.0084) ROC-AUC, compared with (0.8520±0.0427) for QAE-6Q.

## IV. CONCLUSIONS

This study evaluates a hybrid quantum autoencoder [45] as a representation-learning module within an unsupervised anomaly-detection framework and uses Bloch-space observables to interrogate its latent quantum state. In the matched experiment, the classical autoencoder provides the stronger reconstruction and latent-distance baseline, while the QAE provides additional quantum-state diagnostics rather than a demonstrated performance advantage.

Although the graph-regularized model and the quantum autoencoder operate in different settings, they can be tied together by a common diagnostic principle: learning creates internal representation structure that can be interrogated geometrically. In Study I the relevant object is a learned similarity graph, characterized globally through its Laplacian spectrum and locally probed by two-particle interference. In Study II the relevant object is a quantum latent state, characterized through Bloch-vector displacement and parameter-space QFI. These are not the same manifold and the observables are not interchangeable; rather, they illustrate two levels at which physically interpretable geometry can supplement conventional task metrics.

**1. Quantum learning model:** The PQC or hybrid quantum model maps inputs to learned quantum representations:

$$x_i \mapsto h_i = f_\theta(x_i)$$

or, as in the autoencoder case,

$$x_i \mapsto \rho_i^{\mathrm{latent}}$$

Hence, the models studied here do more than produce scalar predictions or reconstruction scores: they generate structured internal representation spaces whose organization can be inspected with model-appropriate geometric observables.

**2. Spectral geometry:** Representations $h_i$ gives a similarity graph:

$$W_{ij} = K\left(h_i, h_j\right), L = D - W$$

The learned similarity graph supplies a global graph geometry. Its Laplacian spectrum reports connectivity and spectral organization, while the Fiedler vector identifies a principal low-frequency partition. The results show regularization-dependent reorganization without satisfying the operational joint-collapse criterion.

- Algebraic connectivity reports global graph coherence and possible fragmentation;
- spectral entropy and $d_{eff} = \exp(S)$ report how broadly Laplacian spectral weight is distributed;
- and the Fiedler vector provides the partition coordinate used in the systematic two-boson association test.

Thus spectral geometry is the global description used for the learned graph representation in Study I.

**3. Quantum autoencoder:** The quantum autoencoder is one concrete quantum learning model in which the latent layer itself is quantum. That means the learned manifold is not only an abstract embedding space – it is a family of quantum states:

$$\varrho_i^{\text{latent}}$$

The central question—how internal structure is encoded—therefore appears in two different representation spaces: globally through a Laplacian graph in Study I and locally through Bloch coordinates in the QAE study. The connection is methodological, not an identification of the two state spaces.

**4. Bloch-space drift:** Bloch drift is then the **local geometric probe** of the latent manifold. If the latent state of a sample is represented by a Bloch vector $r_i$, then drift relative to the benign reference state measures how far that sample moves in quantum state space:

$$\Delta r_i = \| r_i - \bar{r}_{\text{benign}} \|$$

This is an analogy between diagnostic levels rather than a claim that Bloch displacement is literally the same quantity as graph displacement.

*Spectral geometry* describes how a set of learned outputs is organized collectively, whereas *Bloch drift* describes how an individual quantum latent state is displaced relative to a benign reference. Both can expose internal organization that aggregate performance metrics alone do not reveal.

That shared diagnostic role is the bridge between the two studies.

## A. Spectral Geometry and Bosonic Probes in Quantum Learning

The spectral results reveal a non-monotonic finite-system response to graph regularization rather than a verified collapse transition. Across the tested $\lambda_{reg}$ values, no seed satisfies both the prespecified representation-concentration and graph-fragmentation criteria. The primary pre/post run nevertheless shows measurable spectral reorganization, including a modest increase in algebraic connectivity and effective spectral rank.

The learned structure also produces a small but statistically structured two-boson/Fiedler association when evaluated systematically over all input pairs and propagation times. The IBM-hardware measurements are retained as a separate benchmark of a three-qubit energy landscape under finite shots and device noise; they are not measurements of the bosonic coincidence observable.

Study I therefore establishes a reproducible connection between learning-induced graph reorganization and a physically interpretable two-particle probe within the specified finite system, while avoiding a claim of thermodynamic phase transition, universal Fiedler law, hardware-confirmed bosonic interference, or quantum computational advantage.

### *1. Outlook*

The edge-resolved bosonic observables introduced here suggest a natural interface between quantum interference and graph-structured learning. In particular, interference-induced modulation of edge weights parallels message weighting in classical GNNs, but arises intrinsically from unitary dynamics rather than trained parameters. This points toward *quantum graph neural network architectures* in which learning is implemented through state preparation, phase control, or measurement selection, with interference patterns serving as quantum features. While no quantum advantage is claimed, our results identify a concrete, experimentally accessible primitive that may inform future designs of quantum graph learning models.

## B. QAE for Anomaly Detection in IDS Data

The matched anomaly-detection results show that the QAE can encode attack-related displacement in Bloch space, but the CAE provides the stronger reconstruction and latent-distance discrimination on CIC-DDoS2019 SSDP. Bloch drift is therefore interpreted as an additional quantum-state diagnostic rather than as evidence of superior anomaly-detection performance.

The consecutive-drift analysis is recast as an order-sensitivity control. Its difference from within-class permutations can reveal sensitivity to sample ordering for particular seeds, but the CSV row order is not treated as physical time and the result is not used as evidence for dynamical stability or against barren plateaus.

The QFI spectrum adds a complementary parameter-space description of the trained QAE. Because the classical baseline remains stronger on the principal task and latent-distance metrics, the value of the quantum model in this study lies in access to additional physically interpretable observables, not in a demonstrated performance or quantum advantage.

#### *1. Outlook*

Future work may extend Bloch-space diagnostics to other tasks such as fault detection and process monitoring, explore their relationship to quantum Fisher information and trainability [94], and evaluate robustness under realistic noise models and hardware execution. More broadly, Bloch-space analysis offers a principled framework for interpreting quantum machine-learning models in applied settings.

Summary: The two studies support a common geometric diagnostic philosophy at two distinct levels. In Study I, graph regularization reorganizes a learned similarity graph and spectral observables quantify its collective structure; two-particle interference supplies a local physical probe whose weak but reproducible association with the Fiedler split is established statistically. In Study II, a quantum latent circuit produces sample-dependent states whose Bloch displacement and parameter-space QFI expose local state and circuit geometry, while matched classical controls establish the performance baseline. The broader conclusion is therefore not that both experiments realize the same manifold or a common phase transition, but that robustness and interpretability in hybrid quantum learning can benefit from diagnostics tailored to the geometry of the representation actually learned.

# V. REPRODUCIBILITY NOTE AND CODE AVAILABILITY

Spectral geometry: The graph-learning study uses synthetic training states and fully specified generation, split, kernel, optimization, bootstrap, permutation, and propagation-time procedures. The code and processed data supporting the findings of this study are available from the corresponding author upon reasonable request.

Data and code availability: The anomaly-detection study uses the publicly available CIC-DDoS2019 release and the exact BENIGN/DrDoS_SSDP filtering and uniform attack-sampling procedure described in Sec. II and Appendix B. Code, processed data, run manifests, and analysis outputs supporting the reported results are available from the corresponding author upon reasonable request.

# VI. ETHICS AND COMPLIANCE

This study involves synthetic graph simulations and publicly available cybersecurity datasets. No human subjects, personal medical information, or personally identifiable data were used. The work complies with the usage and redistribution policies of the referenced datasets and adheres to applicable research ethics standards.

## VII. APPENDIX A: BOSONIC PROBES IN SPECTRAL GEOMETRY

### A. Cost Function for Spectral Geometry

Learning is governed by a cost functional $C[s]$ given by

$$C[s] = C_{SL}[s] + \lambda_G \sum_{(u,v)\in E} A_{uv} \| \mathbf{h}_u(s) - \mathbf{h}_v(s) \|_2^2 + \lambda_R C_{reg}[s]$$

with

$$C_{\mathrm{SL}}[s] = \sum_{u\in \mathcal{V}_{\mathrm{SL}}} \ell(\hat{y}_u(s), y_u)$$

and,

$$C_{\mathrm{graph}}(\theta) = \sum_{(i,j)\in E} w_{ij} \| f_\theta(x_i) - f_\theta(x_j) \|^2$$

Here, $\| \mathbf{h}_u(s) - \mathbf{h}_v(s) \|_2^2$ is the **squared Euclidean distance** between the representations of nodes $u$and $v$at training time $s$. If each representation is a $d$-dimensional vector,

$$h_u(s), h_v(s) \in \mathbb{R}^d,$$

Then, $\| h_u(s) - h_v(s) \|_2^2 = \sum_{\alpha=1}^{d} (h_{u,\alpha}(s) - h_{v,\alpha}(s))^2.$

Therefore, in the cost,

$$C[s] = C_{\mathrm{SL}}[s] + \lambda_G \sum_{(u,v)\in E} A_{uv} \| h_u(s) - h_v(s) \|_2^2 + \lambda_R C_{\mathrm{reg}}[s],$$

this term measures **how different neighbouring node embeddings are**. This is the standard graph-Laplacian regularization term:

- If $u$ and $v$ are connected and $A_{uv}$ is large, the loss penalizes them for being far apart.
- Minimizing this term encourages **smoothness over the graph**: nearby/related nodes get similar representations.

Overall:

- $h_u(s)$ : learned output for node $u$ at step $s$

- $h_v(s)$ : learned representation/output for node $v$
- $\| h_u - h_v \|_2^2$ : "disagreement" between them

If the representation is scalar, $h_u(s) \in \mathbb{R}$, then it is just $\| h_u(s) - h_v(s) \|_2^2 = (h_u(s) - h_v(s))^2$. If all nodes are represented into a matrix $H(s)$, then this graph term becomes equivalent to

$$\sum_{(u,v)\in E} A_{uv} \| h_u - h_v \|_2^2 \propto \mathrm{Tr}\left(H^\top L H\right),$$

where $L = D - A$ is the graph Laplacian as in Equation 4.

### B. Empirical observation A1: Graph regularization and spectral geometry

For the specified 15-node quantum-learning model, graph regularization changes the learned output-similarity graph and its Laplacian spectrum. The direction and magnitude of those changes are measured empirically and are not asserted as a universal monotonic theorem.

1. The algebraic connectivity satisfies

$$\lambda_2\left(\tilde{L}\right) \geq \lambda_2(L)$$

In the primary $\lambda_{\text{reg}} = 0.5$ run, algebraic connectivity increases from $5.5791$ to $5.8263$; across the synchronized sweep, however, relative $\lambda_2$ responds non-monotonically to $\lambda_{\text{reg}}$.

2. The spectral entropy

$$S = -\sum_k p_k \log p_k, p_k = \frac{\lambda_k}{\sum_j \lambda_j}$$

In the primary run, spectral entropy increases from 2.61537 to 2.62962 and $d_{eff}$ from 13.6723 to 13.8685. The synchronized multi-seed sweep is used to determine whether representation concentration or graph fragmentation occurs under the prespecified operational criteria.

3. No tested seed satisfies the concentration criterion $V/V_0 \leq 0.10$, the fragmentation criterion $\lambda_2 / \lambda_{2,0} \leq 0.01$, or the joint criterion; the result therefore does not identify a collapsed phase. This empirical observation summarizes the spectral results in Figs. 6-9:
   - Intermediate $\lambda_{\text{reg}} = 1.5$ produces the strongest representation contraction while increasing relative algebraic connectivity.
   - The largest tested value $\lambda_{\text{reg}} = 3.0$ does not generate the strongest contraction and does not fragment the learned graph, demonstrating the non-monotonic finite-system response.
   - The interpretation is therefore based on measured synchronized trajectories and seed-level audit data rather than an assumed phase diagram.

### C. Hardware noise modeling

A breakdown of noise sources included in the simulator and present on hardware is provided in Table AI.

Table AI. Components of the Aer noise budget model

| Noise source | Aer | Approx./model | Notes |
|---|---|---|---|
| Finite sampling | ✓ | ✓ | 2000 shots per circuit |
| Single-qubit gate error | ✓ | ✓ | Depolarizing approximation |
| Two-qubit gate error | ✓ | ✓ | Calibrated CZ error rates |
| Readout error | ✓ | ✓ | Asymmetric bit-flip model |
| $T_1/T_2$ re-laxation | ✓ | ✓ | During idle periods |
| Crosstalk | ✗ | ✓ | Not modeled in Aer |
| Non-Markovian drift | ✗ | ✓ | Calibration-time dependent |
| Leakage | ✗ | ✓ | Outside computational subspace |

To contextualize hardware results, we compare experimental energies against two simulator baselines. This model incorporates gate-dependent depolarizing noise, asymmetric readout errors, and $T_1/T_2$ relaxation during idle periods, but neglects correlated, non-Markovian, and leakage effects. As such, the noisy-Aer baseline provides a lower-bound estimate of hardware noise and is not expected to fully reproduce experimental behavior.

An *ideal* baseline is obtained from noiseless sampling of the same logical circuit, isolating finite-shot effects. A *noisy* baseline is generated using the Aer simulator with a noise model constructed from device calibration data via `NoiseModel.from_backend`. While more detailed noise models are possible, they require assumptions about correlated and time-dependent processes that are not independently calibrated. We therefore restrict to a standard calibration-derived noise model.

*Interpretation of noise modeling*: The noisy-Aer baseline is not intended as a faithful predictor of instantaneous hardware behavior. Instead, it serves as a diagnostic reference that reproduces first-order error channels inferred from calibration data. Discrepancies between noisy-Aer and hardware results are expected due to temporal calibration drift, correlated errors, control imperfections, and non-Markovian dynamics. This interpretation is consistent with prior benchmarking studies of NISQ-era devices.

This noise modelling definition, in conjunction with Fig. 22, and the hardware-simulator results in Fig. 10 and Fig. 11, shows that backend choice impacts both micro-level expectation evaluation and macro-level training throughput. Qiskit dense implementation achieves lower per-epoch runtime with reduced variance, enabling more efficient large-scale training and hyperparameter sweeps.

### a. *Pre-Training Laplacian (Fidelity Graph)*

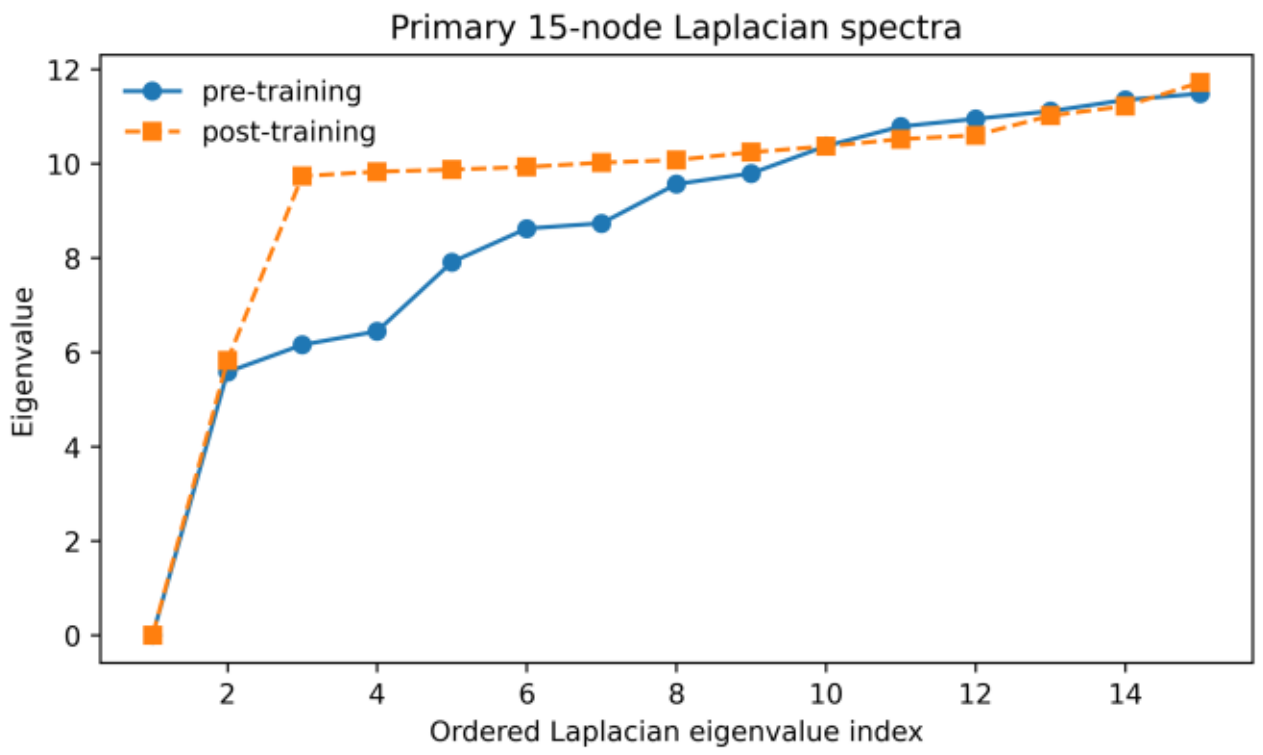

FIG. 21. (Color online) Primary pre/post combinatorial-Laplacian spectra for the 15-node, 10-SL/5-UL run. Training changes the low-frequency spectrum without producing graph fragmentation; $\lambda_2$ changes from 5.5791 to 5.8263 and $\lambda_3$ from 6.1566 to 9.7324.

Fig. 21 reproduces the primary pre/post combinatorial-Laplacian spectra using the same fixed output-graph bandwidth as the main analysis. The graph is connected before and after training, with $\lambda_2 = 5.5791$ pre-training and 5.8263 post-training.

The spectrum therefore shows a finite, connected graph on both sides of training. The widened $\lambda_3 - \lambda_2$ separation is interpreted as spectral reorganization of this particular learned graph rather than evidence for an almost-disconnected pre-training state:

- the pre-training graph is connected;
- the post-training graph is also connected;
- and the principal Fiedler mode remains well defined in both cases.

The comparison does not require a near-zero pre-training $\lambda_2$. Instead, the measured pre/post values provide the baseline for the learned Fiedler vector and the subsequent systematic interference analysis.

Compared with the pre-training graph, the learned output-similarity graph shows a modest increase in $\lambda_2$ and a larger increase in $\lambda_3$, together with small increases in spectral entropy and effective rank. These measured changes supply the spectral context for Fig. 23 and the boson-Fiedler analysis without implying collapse or a thermodynamic phase transition.

### a. *End-to-end efficiency comparison*

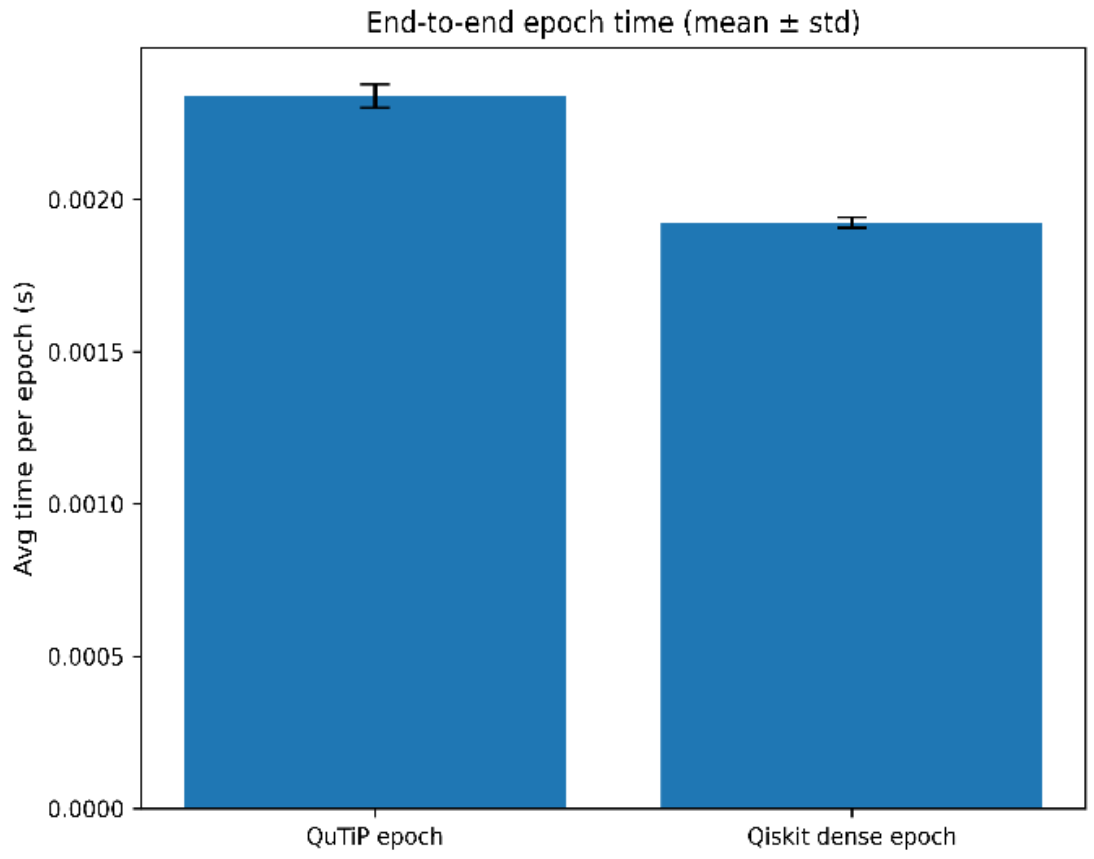

FIG 22. (Colour online) *End-to-end training epoch time for different dense simulators.* Average wall-clock time per training epoch (mean±standard deviation) comparing a QuTiP-based dense statevector implementation and a Qiskit dense expectation workflow.

As mentioned in Methods section, to assess the practical impact of different expectation-value backends on training throughput, we measured the average wall-clock time per training epoch (mean±standard deviation) using two implementations: (i) a QuTiP-based dense statevector workflow and (ii) a Qiskit-based dense expectation evaluation. This is exhibited in Fig. 22.

Fig. 22, along with the hardware-simulator results in Fig. 10 and Fig. 11, shows that backend choice impacts both micro-level expectation evaluation and macro-level training throughput. The Qiskit dense implementation achieves lower per-epoch runtime with reduced variance, enabling more efficient large-scale training and hyperparameter sweeps. This difference reflects reduced Python overhead and tighter

integration of expectation evaluation within the Qiskit workflow, and motivates our choice of backend for large-scale sweeps.

### D. Fiedler vector Output

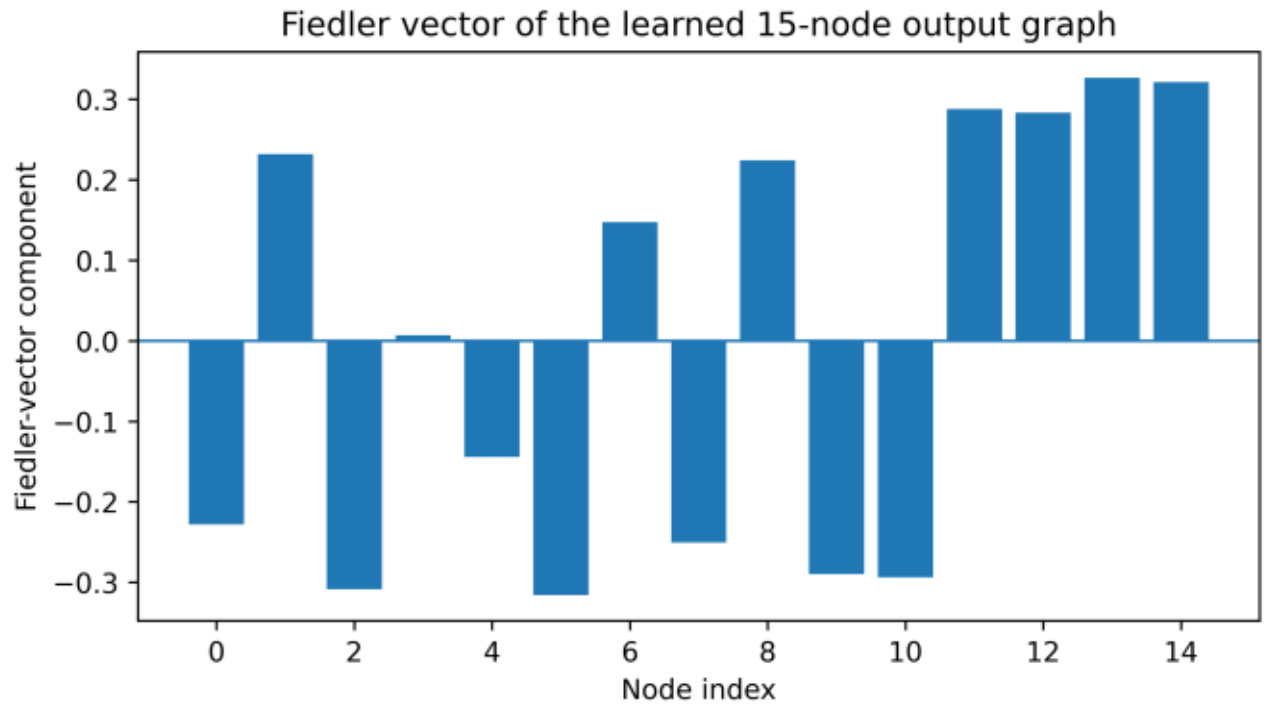

FIG. 23. (Color online) Fiedler vector of the learned 15-node output-similarity Laplacian for the primary 10-SL/5-UL run. The sign and magnitude of $v_2$ define the principal spectral coordinate used to compute the edge split $|v_2(u)-v_2(v)|$ in the systematic interference analysis.

(Ref: empirical spectral observation and systematic boson-Fiedler test.)

Let $v_2$ be the **Fiedler vector** [24] (the eigenvector associated with $\lambda_2$) of the learned output graph Laplacian.

1. The **sign of** $v_2^{(i)}$ assigns node $i$ to one side of the **minimum-energy spectral cut**.
2. For a connected graph, the Fiedler vector provides a low-frequency partition coordinate; a small $\lambda_2$ would indicate a weak cut, but the present learned graph has finite $\lambda_2 = 5.8263$ and is not near disconnected.

Hence, if $v_2$ denotes the Fiedler vector of the graph Laplacian, then in the regime where the learned geometry is dominated by the second eigenmode,

$$\mathrm{corr}\left(\Delta P_{uv}, | v_2(u)-v_2(v)|\right) < 0$$

The Fiedler vector is used as a structural coordinate rather than as a theorem predicting the sign of interference on every edge. Across all 3255 input/time conditions, the primary absolute-association statistic is 0.1085 with 95% bootstrap interval [0.0845,0.1397] and node-permutation $p = 4.998\times10^{-4}$. The effect is therefore small but systematically different from the node-label null.

1. Spectral separation: the learned Fiedler vector defines a reproducible principal coordinate of the connected output graph.
2. Learned organization: the sign structure partitions nodes for descriptive spectral analysis, but it is not interpreted as proof of two universally coherent communities.
3. Bosonic association: the systematic analysis finds a weak but reproducible relationship between interference magnitude and Fiedler edge separation; representative edge plots illustrate, but do not define, the reported statistic.
4. Relation to the empirical structure:
   - Primary pre/post analysis: spectral entropy and effective rank increase modestly while the graph remains connected.
   - Systematic boson-Fiedler test: interference magnitude is weakly but significantly structured relative to the Fiedler edge coordinate under the prespecified permutation analysis.

Fig. 23 visualizes the spectral coordinate used in the learned-graph interference test; it does not by itself establish a universal interference law.

## VIII. APPENDIX B: BLOCH SPACE GEOMETRY FOR ANOMALY DETECTION

QFI is related to local information geometry and circuit-parameter sensitivity [17, 96, 97, 105, 113]. In the analysis it is not used to infer that the embedding map is Lipschitz in its inputs or that barren plateaus are absent. Those questions concern different derivatives and would require additional gradient-variance or scaling evidence. The reported QFI spectrum is therefore interpreted only as a local parameter-space diagnostic of the representative trained QAE.

### A. Definition B1: Bloch representation and drift induced by the quantum latent circuit

Let $x \in \mathbb{R}^d$ denote an input feature vector after classical preprocessing. The quantum circuit in Fig. 5 implements a parameterized unitary

$$U(\theta,x) = U_L(\theta_L)U_E \cdots U_1(\theta_1)U_E U_{\mathrm{enc}}(x)$$

where $U_{\mathrm{enc}}(x)$ denotes angle embedding, $U_\ell(\theta_\ell)$ are trainable single-qubit rotations, and $U_E$ represents the ring entanglement layer. The circuit prepares the quantum state

$$|\psi(x)\rangle = U(\theta,x)|0\rangle^{\otimes n}$$

For each qubit $q$, we define the associated **Bloch vector**

$$\mathbf{r}^{(q)}(x) = \begin{pmatrix} \langle \psi(x)|X_q|\psi(x)\rangle \\ \langle \psi(x)|Y_q|\psi(x)\rangle \\ \langle \psi(x)|Z_q|\psi(x)\rangle \end{pmatrix}$$

which provides a geometric representation of the reduced single-qubit state induced by the full entangled circuit.

**Absolute Bloch Drift (Geometry-Based Anomaly Score):** To quantify anomaly-induced displacement [96, 108-111], we compute the **benign mean Bloch vector** for each qubit,

$$\boldsymbol{\mu}_B^{(q)} = \mathbb{E}_{x\sim \text{BENIGN}}\left[\mathbf{r}^{(q)}(x)\right]$$

and define the **Bloch drift magnitude**

$$d^{(q)}(x) = \| \mathbf{r}^{(q)}(x) - \boldsymbol{\mu}_B^{(q)} \|_2$$

This metric measures absolute geometric displacement of the reduced single-qubit state relative to the benign reference and underlies the Bloch-drift ROC analysis in the main text.

Consecutive Bloch drift (order-sensitivity diagnostic): We additionally compute the distance between latent states assigned to adjacent rows of a chosen sample ordering.

$$\Delta^{(q)}(x_t) = \| \mathbf{r}^{(q)}(x_{t+1}) - \mathbf{r}^{(q)}(x_t) \|_2$$

Because CSV row order is not treated as a physical time variable, consecutive drift is compared with within-class permutations and is interpreted only as an order-sensitivity control. It is not used to infer dynamical stability.

### B. Empirical observation B2: Absolute drift and matched latent-distance controls

Let $U(\theta, x)$ be a parameterized quantum circuit preparing a pure state

$$|\psi(x)\rangle = U(\theta, x)|0\rangle^{\otimes n}$$

and let $\mathbf{r}^{(q)}(x) \in \mathbb{R}^3$ denote the Bloch vector of qubit $q$. We define:

i. The benign mean Bloch vector as:

$$\boldsymbol{\mu}_B^{(q)} = \mathbb{E}_{x\sim \mathcal{D}_B}\left[\mathbf{r}^{(q)}(x)\right]$$

ii. The absolute Bloch drift:

$$d^{(q)}(x) = \| \mathbf{r}^{(q)}(x) - \boldsymbol{\mu}_B^{(q)} \|_2$$

iii. The consecutive Bloch drift:

$$\Delta^{(q)}(x_t) = \| \mathbf{r}^{(q)}(x_{t+1}) - \mathbf{r}^{(q)}(x_t) \|_2$$

Operational interpretation:

1. Absolute Bloch drift is evaluated empirically as an anomaly-ranking score relative to the benign reference state.

$$\| \mathbf{r}^{(q)}(x) - \mathbf{r}^{(q)}(x') \|_2 \le L \| x - x' \|$$

2. A matched classical control uses Euclidean distance from the benign CAE latent centroid, allowing geometric discrimination to be compared without attributing all latent-distance effects uniquely to quantum mechanics.

No theorem connecting QFI to input-space Lipschitz continuity is assumed in this analysis.

$$\| \mathbb{E}_{x\sim \mathcal{D}_A}\left[\mathbf{r}^{(q)}(x)\right] - \boldsymbol{\mu}_B^{(q)} \|_2 > \delta$$

Order-sensitive consecutive drift is compared with 1000 within-class permutations; source-order deviations are reported as order effects rather than temporal dynamics.

$$\mathbb{E}\left[\Delta^{(q)}\right] \le L\, \mathbb{E}\left[\| x_{t+1} - x_t \|\right]$$

Class separability and latent-geometry claims are therefore based on the measured ROC/permutation statistics rather than on the former Lipschitz assumptions.

**Relation to quantum Fisher information** (parameter space only)**:** A nonzero QFI eigenvalue indicates local sensitivity to circuit-parameter perturbations. It does not imply a bound on derivatives with respect to the classical input and does not, by itself, prove the absence of barren plateaus. This paper keeps QFI and input-space/Bloch-drift interpretations separate.

#### *1. Quantum Fisher Information*

This is a summary of the representative parameter-space QFI analysis reported in Fig. 18.The representative QFI statistics are interpreted as follows:

I. trace=17.295 quantifies the total local parameter sensitivity of the mean QFI matrix;
II. numerical rank=24 indicates that all 24 trainable parameter directions are active above the stated numerical tolerance;
III. active condition number=72.99 indicates anisotropic, but nonsingular, local parameter sensitivity;
IV. participation ratio=13.73 and entropy effective rank=16.54 summarize how the local QFI weight is distributed across parameter directions.

The QFI matrix is full rank for the representative trained circuit and has finite anisotropy. This is evidence only about local parameter-space geometry. It is not used as a traffic feature, an input-space smoothness bound, or proof that the training landscape is free of barren plateaus.

The trace, condition number, numerical rank, participation ratio, and entropy effective rank provide complementary summaries of local parameter sensitivity. The bootstrap 95% interval for the trace is [16.840,17.742].

### *a. Evaluation Metrics*

Model performance was evaluated using threshold-free ranking metrics and fixed-threshold operational diagnostics.

**Ranking Metrics:** The receiver operating characteristic (ROC) curve evaluates discrimination across all possible thresholds. The ROC-AUC is defined as

$$\mathrm{AUC} = \Pr(s(x_A) > s(x_B))$$

where $s(x)$ is the anomaly score, $x_A \sim \mathcal{D}_A$ is an attack sample, and $x_B \sim \mathcal{D}_B$ is a benign sample.

**Fixed-Threshold Metrics:** A fixed operational threshold $\tau$was defined as the 95$^{\mathrm{th}}$ percentile of benign validation scores. At this threshold, predictions are given by

$$\hat{y}(x) = \begin{cases} 1 & \text{if } s(x) > \tau, \\ 0 & \text{otherwise.} \end{cases}$$

The false-positive and false-negative rates are

$$\mathrm{FP}(\tau) = \Pr_{\mathcal{D}_B}[s > \tau], \mathrm{FN}(\tau) = \Pr_{\mathcal{D}_A}[s \le \tau].$$

### *b. Multi-qubit Bloch Embeddings*

Fig. 25 shows the per-qubit absolute Bloch-drift AUC for every seed. The response is distributed heterogeneously across the six-qubit register; some qubits consistently separate the classes better than others, while no single qubit dominates every initialization.

This seed-by-qubit audit supports the use of Bloch observables as a distributed diagnostic of the QAE latent state, while the matched CAE latent-distance control remains the appropriate classical comparison for discrimination strength.

### *c. Bloch-Drift rankings based on AUC*

Fig. 26 ranks qubits by their mean absolute Bloch-drift ROC-AUC across the five experimental seeds.

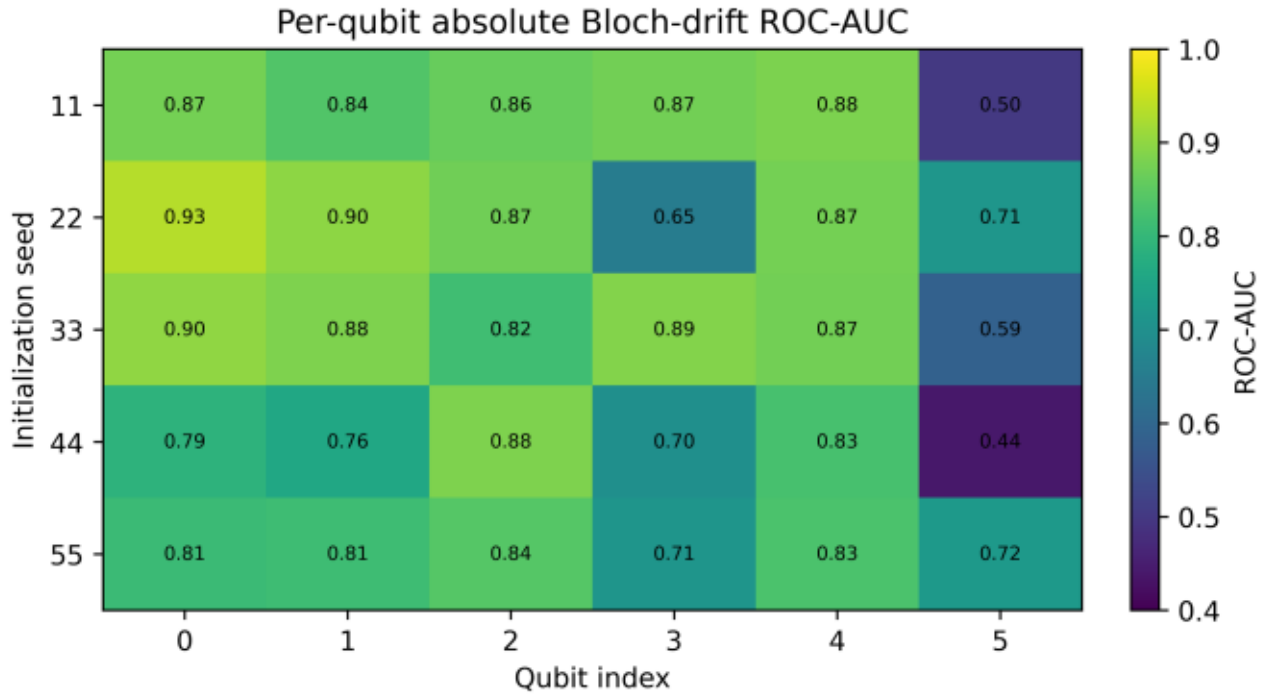

FIG. 25. (Color online) Heat map of per-qubit absolute Bloch-drift ROC-AUC for all five QAE initialization seeds. Values are computed from the fixed test partition; the variation across qubits and seeds quantifies heterogeneity of the quantum latent diagnostic.

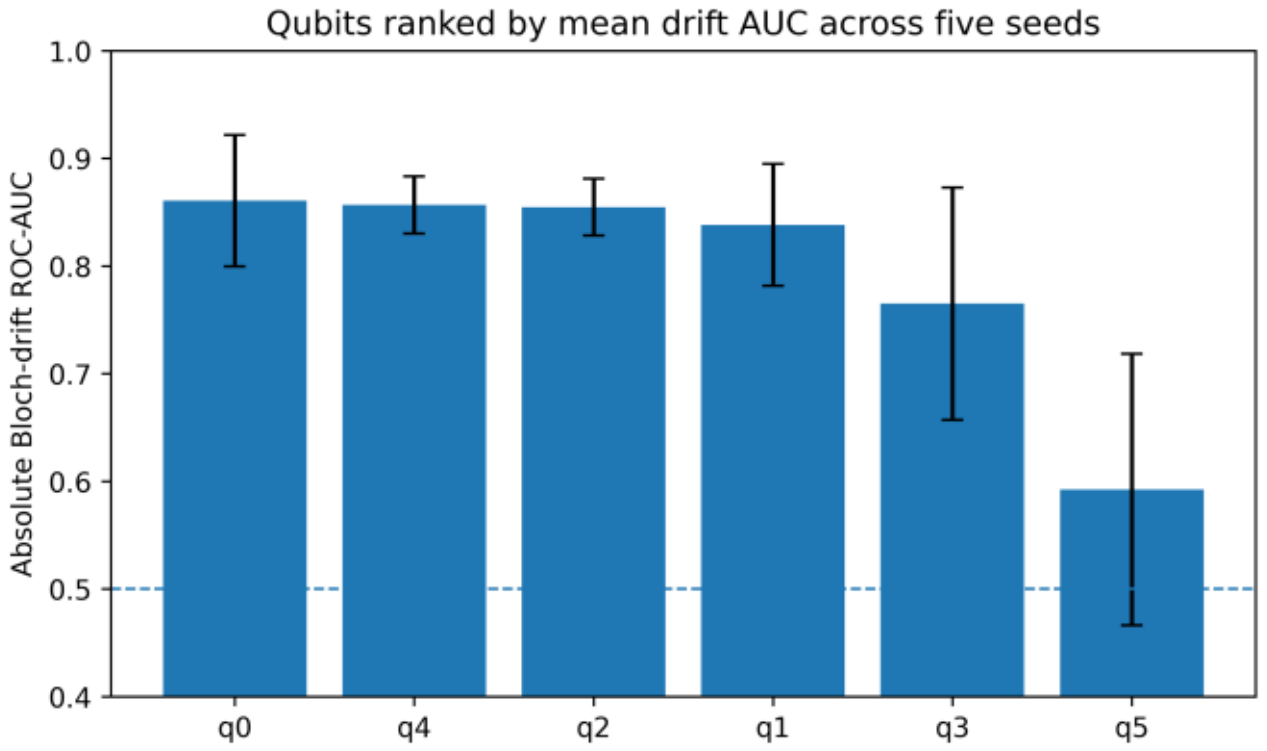

FIG. 26. (Color online) Qubits ranked by mean absolute Bloch-drift ROC-AUC across five QAE initialization seeds; error bars show the sample standard deviation across seeds.

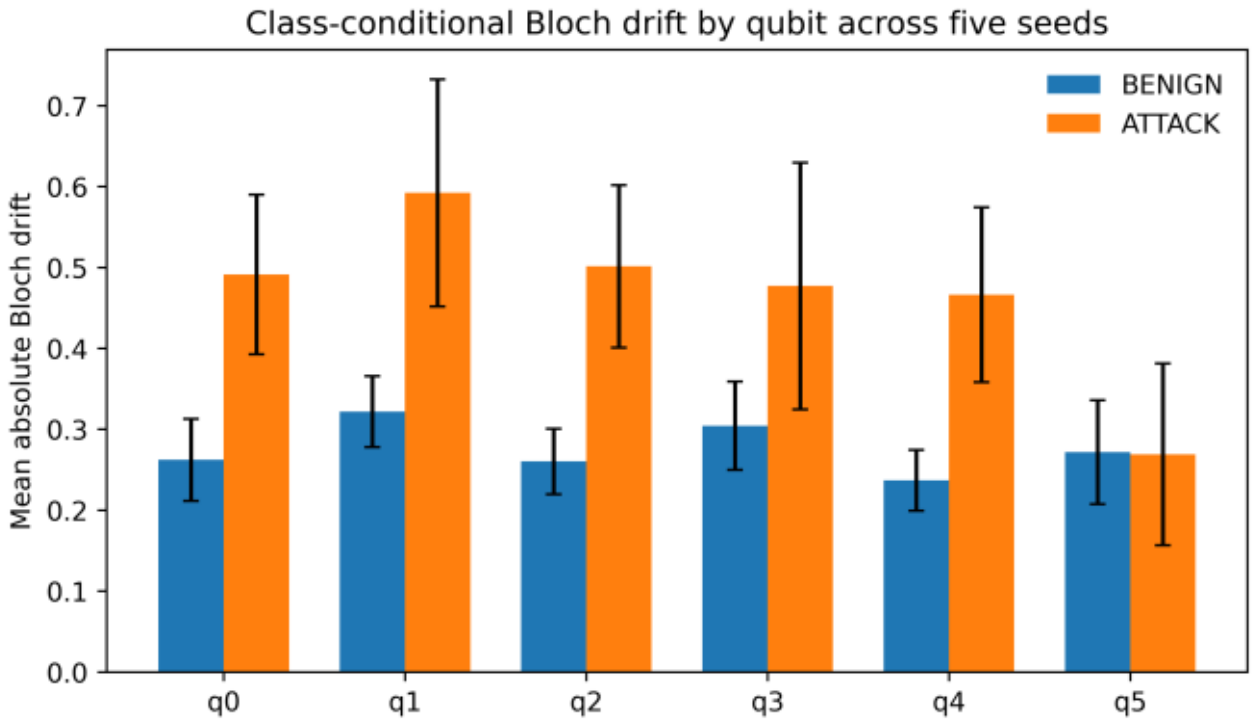

FIG. 27. (Color online) Mean absolute Bloch drift for BENIGN and ATTACK samples by qubit, averaged across five QAE seeds; error bars show the sample standard deviation across seeds. The plot complements Fig. 27 by showing the class-conditional displacement underlying the AUC ranking.

Fig. 27 shows that the class-conditional Bloch response is heterogeneous across the six latent qubits. Averaged across five initialization seeds, ATTACK samples exhibit consistently larger mean absolute Bloch drift than BENIGN samples for qubits $q0-q4$, with the largest separation observed for $q1$ and similarly strong differences for $q0$, $q2$, and $q4$. In contrast, $q5$ shows nearly identical mean drift for the two classes, indicating little class-discriminative displacement on that qubit. The error bars also show appreciable seed-to-seed variability, particularly for the ATTACK class. Overall, the result suggests that anomaly-related Bloch displacement is distributed non-uniformly across the quantum latent register rather than arising uniformly from all qubits.

Next, in Appendix C we share the main model pipelines for the Hamiltonian in the spectral geometry (Fig. 28) and the QAE in the DDoS data analysis (Fig. 29).

# IX. APPENDIX C: MODEL PIPELINES

## A. Hamiltonian model pipeline for spectral analysis

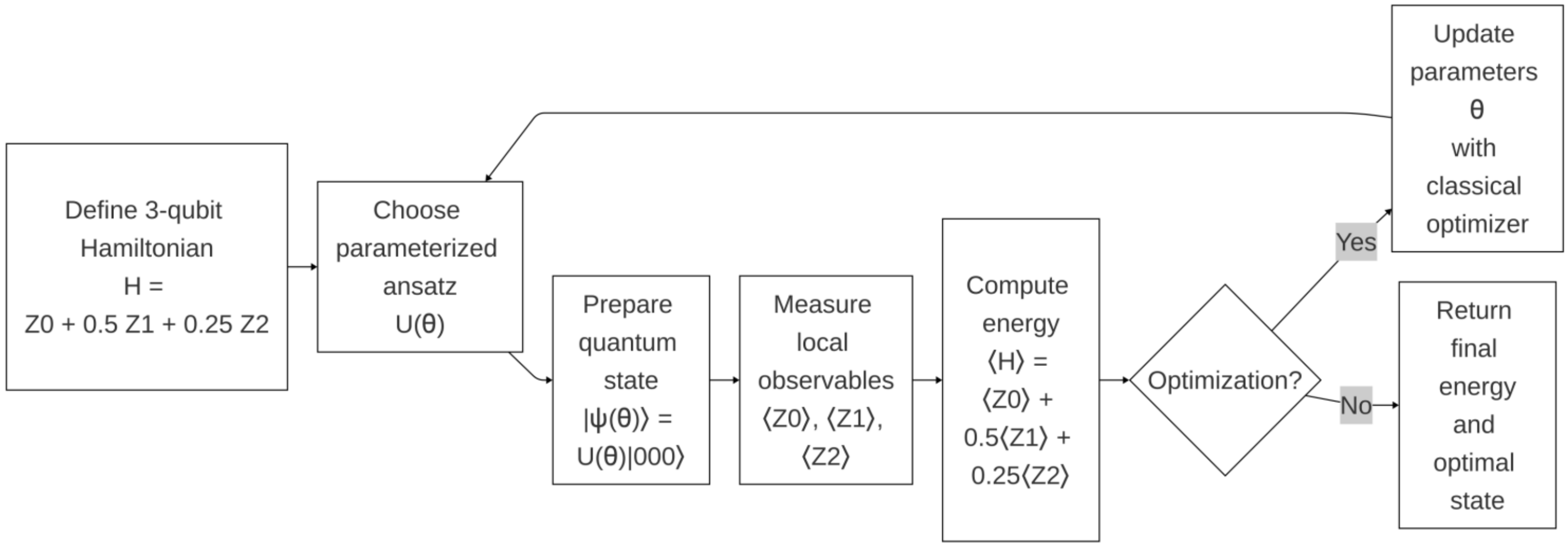

FIG. 28. *Graph-regularized quantum learning architecture.* Input samples are processed by a variational quantum circuit and measured to produce embeddings $f_\theta(x_i)$, from which a similarity graph $W$ is constructed. Training minimizes a supervised objective augmented by a graph-smoothness penalty, while the learned graph is analyzed through Laplacian-based spectral geometry. The resulting representations are probed using bosonic edge interference $\Delta P_{uv}$ and Bloch-space drift $\Delta r$, linking learned structure to physically interpretable observables.

## B. Model pipeline for quantum autoencoder for DDoS data

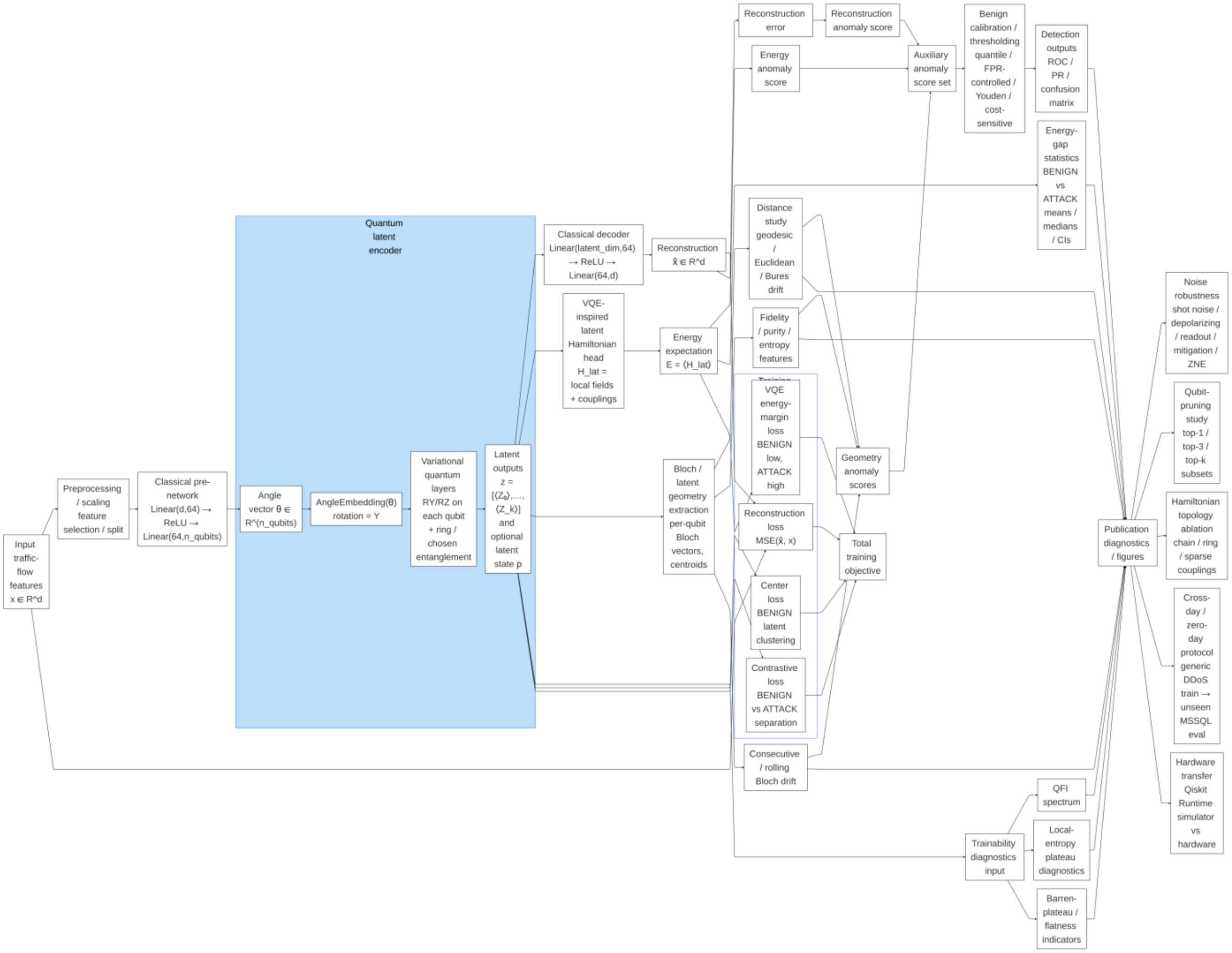

FIG. 29. (Colour online) Full model architecture of the extended hybrid classical–quantum autoencoder for DDoS anomaly detection. Flow-level traffic features are preprocessed and mapped by a classical front end to circuit angles, which are embedded in a variational quantum latent encoder. The resulting latent representation supports reconstruction, a VQE-inspired latent-energy head, Bloch-geometry and drift diagnostics, fidelity/purity/entropy analysis, and QFI/local-entropy trainability diagnostics, thereby enabling reconstruction-, geometry-, energy-, and trainability-based anomaly assessment within a unified framework.

## References

1. C. H. Valahu, M. P. Stafford, Z. Huang, V. G. Matsos, M. J. Millican, T. Chalermpusitarak, N. C. Menicucci, J. Combes, B. Q. Baragiola, and T. R. Tan, Quantum-Enhanced Multiparameter Sensing in a Single Mode, Science Advances 11, (2025).
2. G. S. Hartnett, D. Kielpinski, S. Maity, P. S. Mundada, Y. Baum, and M. R. Hush, *Automated discovery of heralded ballistic graph state generators for fusion-based photonic quantum computation*, arXiv:2508.16505v1 [quant-ph] (2025), last accessed 2026/02/01.
3. H. Yamauchi, S. Kanno, Y. Sato, H. Tezuka, Y.-A. Shimada, E. Kaminishi, and N. Yamamoto, Quantum spectroscopy of topological dynamics via a supersymmetric Hamiltonian, arXiv:2511.23169 [quant-ph] (2025), last accessed 2026/03/25.
4. G. Aleksandrowicz *et al.*, "Qiskit: An Open-source Framework for Quantum Computing," Zenodo (2019), https://doi.org/10.5281/zenodo.2562111.

5. S. Aaronson and A. Arkhipov, "The computational complexity of linear optics," *Theory Comput.* **9**, 143 (2013)
6. A. Abbas *et al*., "The power of quantum neural networks," *Nat. Comput. Sci.* **1**, 403 (2021), https://doi.org/10.1038/s43588-021-00084-1.
7. A. Björklund, J. Mäkelä, and K. Puolamäki, *SLISEMAP: Supervised Dimensionality Reduction through Local Explanations*, Machine Learning **112**, 1 (2022).
8. J. B. Spring et al., *Boson Sampling on a Photonic Chip*, Science **339**, 798 (2013).
9. J. Biamonte *et al*., "Quantum machine learning," *Nature* **549**, 195 (2017), https://doi.org/10.1038/nature23474.
10. A. M. Bishop, *Pattern Recognition and Machine Learning* (Springer, New York, 2006)
11. K. M. Borgwardt, H.-P. Kriegel, S. V. N. Vishwanathan, and N. N. Schraudolph, "Graph Kernels for Disease Outcome Prediction from Protein-Protein Interaction Networks," in *Pacific Symposium on Biocomputing* **12**, 4–15 (2007), https://doi.org/10.1142/9789812772435_0002
12. M. M. Bronstein, J. Bruna, Y. LeCun, A. Szlam, and P. Vandergheynst, "Geometric deep learning: going beyond Euclidean data," *IEEE Signal Process. Mag.* **34**, 18 (2017), https://doi.org/10.1109/MSP.2017.2693418.
13. A. J. Bray, "Theory of phase ordering kinetics," *Adv. Phys.* **43**, 357 (1994)
14. Bauer *et al*., "Qiskit Aer: High-performance quantum circuit simulators," arXiv:2108.08870, https://doi.org/10.48550/arXiv.2108.08870.
15. P. W. Battaglia *et al*., "Relational inductive biases, deep learning, and graph networks," *Nature* **588**, 429 (2020)
16. Soorya Rethinasamy, Saptarshi Roy, Titas Chanda, Aditi Sen(De), and Ujjwal Sen, "Universality in distribution of monogamy scores for random multiqubit pure states," *Phys. Rev. A* **99**, 042302 (2019), https://doi.org/10.1103/PhysRevA.99.042302
17. M. Cerezo *et al*., "Variational quantum algorithms," *Nat. Rev. Phys.* **3**, 625 (2021), https://doi.org/10.1038/s42254-021-00348-9.
18. A. Chicco and G. Jurman, "The advantages of the Matthews correlation coefficient (MCC) over F1 score and accuracy in binary classification evaluation," *BMC Genomics* **21**, 6 (2020), https://doi.org/10.1186/s12864-019-6413-7.
19. F. R. K. Chung, *Spectral Graph Theory* (American Mathematical Society, Providence, RI, 1997).
20. A. W. Cross *et al*., "Open Quantum Assembly Language," arXiv:1707.03429, https://doi.org/10.48550/arXiv.1707.03429.
21. M. Defferrard, X. Bresson, and P. Vandergheynst, "Convolutional neural networks on graphs with fast localized spectral filtering," in *Adv. Neural Inf. Process. Syst.* **29**, 3844 (2016)
22. W. Dür and H.-J. Briegel, "Entanglement purification and quantum error correction," *Rep. Prog. Phys.* **70**, 1381 (2007), https://doi.org/10.1088/0034-4885/70/8/R03.
23. A. Farhi and H. Neven, "Classification with quantum neural networks on near-term processors," arXiv:1802.06002, https://doi.org/10.48550/arXiv.1802.06002.
24. M. Fiedler, "Algebraic connectivity of graphs," *Czechoslovak Math. J.* **23**, 298 (1973).
25. T. Fösel, P. Tighineanu, T. Weiss, and F. Marquardt, *Reinforcement Learning with Neural Networks for Quantum Feedback*, Physical Review X **8**, (2018).
26. W. Gardiner, *Stochastic Methods* (Springer, Berlin, 2009), .
27. L. Gyongyosi and S. Imre, *Networked Quantum Services†*, Quantum Information & Computation **25**, 97 (2025).
28. C. S. Hamilton, R. Kruse, L. Sansoni, S. Barkhofen, C. Silberhorn, and I. Jex, "Gaussian Boson Sampling," *Phys. Rev. Lett.* **119**, 170501 (2017), https://doi.org/10.1103/PhysRevLett.119.170501.
29. W. L. Hamilton, Z. Ying, and J. Leskovec, "Inductive Representation Learning on Large Graphs," in *Advances in Neural Information Processing Systems 30 (NeurIPS 2017)*, 1024–1034 (2017); arXiv:1706.02216.
30. H. Kashima, K. Tsuda, and A. Inokuchi, "Marginalized Kernels Between Labeled Graphs," in *Proc. 20th Int. Conf. Mach. Learn. (ICML 2003)* (AAAI Press, 2003), pp. 321–328.
31. H. J. Kimble, "The quantum internet," *Nature* **453**, 1023 (2008), https://doi.org/10.1038/nature07127.
32. M. Schuld and N. Killoran, "Quantum machine learning in feature Hilbert spaces," *Phys. Rev. Lett.* **122**, 040504 (2019), https://doi.org/10.1103/PhysRevLett.122.040504.

33. T. N. Kipf and M. Welling, “Semi-supervised classification with graph convolutional networks,” in *Proc. ICLR* (2017).
34. J. Wen, X. Kong, S. Wei, B. Wang, T. Xin, and G. Long, *Experimental Realization of Quantum Algorithms for a Linear System Inspired by Adiabatic Quantum Computing*, Physical Review. A/Physical Review, A **99**, (2019).
35. A. Javaid, Q. Niyaz, and W. Sun, “A deep learning approach for network intrusion detection system,” *EAI Endorsed Trans. Security and Safety* (2016), .
36. Lazarevic, L. Ertöz, and V. Kumar, “A comparative study of anomaly detection schemes in network intrusion detection,” in *Proc. SIAM Int. Conf. Data Mining* (2005), .
37. J. Li *et al.*, *npj Quantum Inf.* **7**, 60 (2021), https://doi.org/10.1038/s41534-021-00398-9.
38. S. Mandt *et al.*, “Stochastic Gradient Descent as Approximate Bayesian Inference,” *J. Mach. Learn. Res.* **18**, 1 (2017)
39. M. A. Nielsen and I. L. Chuang, *Quantum Computation and Quantum Information* (Cambridge University Press, Cambridge, 2000).
40. J. S. Otterbach *et al.*, “Unsupervised machine learning on a hybrid quantum computer,” arXiv:1712.05771, https://doi.org/10.48550/arXiv.1712.05771.
41. M. Pant, H. Krovi, D. Towsley, L. Tassiulas, L. Jiang, P. Basu, D. Englund, and S. Guha, “Routing entanglement in the quantum internet,” *npj Quantum Inf.* **5**, 25 (2019), https://doi.org/10.1038/s41534-019-0139-x
42. J. Preskill, “Quantum computing in the NISQ era and beyond,” *Quantum* **2**, 79 (2018), https://doi.org/10.22331/q-2018-08-06-79.
43. N. Quesada, J. M. Arrazola, and N. Killoran, *Gaussian Boson Sampling Using Threshold Detectors*, Physical Review A **98**, (2018).
44. M. Ring *et al.*, “A survey of network-based intrusion detection data sets,” *Comput. Secur.* **86**, 147 (2019), https://doi.org/10.1016/j.cose.2019.06.005.
45. J. Romero *et al.*, “Quantum autoencoders for efficient compression of quantum data,” *Quantum Sci. Technol.* **2**, 045001 (2017), https://doi.org/10.1088/2058-9565/aa8072.
46. M. Sakurada and T. Yairi, “Anomaly detection using autoencoders with nonlinear dimensionality reduction,” in *Proc. MLSDA Workshop* (2014).
47. N. Sangouard, C. Simon, H. de Riedmatten, and N. Gisin, *Quantum Repeaters Based on Atomic Ensembles and Linear Optics*, Reviews of Modern Physics **83**, 33 (2011).
48. M. Schuld *et al.*, “Circuit-centric quantum classifiers,” *Phys. Rev. A* **101**, 032308 (2020), https://doi.org/10.1103/PhysRevA.101.032308.
49. M. Schuld and F. Petruccione, *Supervised Learning with Quantum Computers* (Springer, Cham, 2018), .
50. N. Shervashidze, P. Schweitzer, E. J. van Leeuwen, K. Mehlhorn, and K. M. Borgwardt, “Weisfeiler-Lehman Graph Kernels,” *J. Mach. Learn. Res.* **12**, 2539–2561 (2011)
51. N. Shone, T. Ngoc, V. Phai, and Q. Shi, “A deep learning approach to network intrusion detection,” *IEEE Trans. Emerg. Top. Comput. Intell.* **2**, 41 (2018), https://doi.org/10.1109/TETCI.2017.2772792.
52. A. Spielman, *Spectral and Algebraic Graph Theory* (lecture notes, 2019), .
53. R. Sommer and V. Paxson, “Outside the closed world: On using machine learning for network intrusion detection,” in *IEEE Symp. Security and Privacy* (2010), https://doi.org/10.1109/SP.2010.25.
54. W. Wang *et al.*, “Network intrusion detection using deep learning: A survey,” *IEEE Commun. Surv. Tutor.* **24**, 684 (2022), https://doi.org/10.1109/COMST.2021.3134524.
55. H. Wang, J. Qin, X. Ding, M.-C. Chen, S. Chen, X. You, Y.-M. He, X. Jiang, L. You, Z. Wang, C. Schneider, J. J. Renema, S. Höfling, C.-Y. Lu, and J.-W. Pan, “Boson Sampling with 20 Input Photons and a 60-Mode Interferometer in a $10^{14}$-Dimensional Hilbert Space,” *Phys. Rev. Lett.* **123**, 250503 (2019), https://doi.org/10.1103/PhysRevLett.123.250503
56. S. Wehner, D. Elkouss, and R. Hanson, *Quantum Internet: A Vision for the Road Ahead*, Science **362**, (2018).
57. Z. Wu *et al.*, “A comprehensive survey on graph neural networks,” *IEEE Trans. Neural Netw. Learn. Syst.* **32**, 4 (2021), https://doi.org/10.1109/TNNLS.2020.2978386.
58. R. Verdon *et al.*, “A quantum graph neural network,” arXiv:1909.12264, https://doi.org/10.48550/arXiv.1909.12264.
59. L. G. Valiant, “The complexity of enumeration and reliability problems,” *Theor. Comput. Sci.* **8**, 189 (1979), https://doi.org/10.1016/0304-3975(79)90024-6.

60. J. M. Arrazola *et al.*, "Quantum-inspired algorithms in practice," *Nat. Mach. Intell.* **2**, 393 (2020), https://doi.org/10.1038/s42256-020-00201-3.
61. J. M. Arrazola *et al.*, "Quantum circuits with many photons on a programmable nanophotonic chip," *Nature* **591**, 54 (2021), https://doi.org/10.1038/s41586-021-03202-1.
62. J. Zinn-Justin, *Quantum Field Theory and Critical Phenomena* (Oxford University Press, Oxford, 2002).
63. Yan, A. M. Iliyasu, Z. Liu, A. S. Salama, F. Dong, and K. Hirota, "Bloch Sphere-Based Representation for Quantum Emotion Space," *J. Adv. Comput. Intell. Intell. Inform.* **19**, 134–142 (2015), https://doi.org/10.20965/jaciii.2015.p0134.
64. V. V. Shende and I. L. Markov, "Quantum circuits for incompletely specified two-qubit operators," *Quantum Inf. Comput.* **5**(1), 48–56 (2005), .
65. M. Benedetti, E. Lloyd, S. Sack, and M. Fiorentini, "Parameterized quantum circuits as machine learning models," *Quantum Sci. Technol.* **4**, 043001 (2019), https://doi.org/10.1088/2058-9565/ab4eb5.
66. Huang, M. Broughton, M. Mohseni, R. Babbush, S. Boixo, H. Neven, and J. R. McClean, "Power of data in quantum machine learning," *Nat. Commun.* **12**, 2631 (2021), https://doi.org/10.1038/s41467-021-22539-9.
67. S. Ganguly, *Quantum Machine Learning: An Applied Approach: The Theory and Application of Quantum Machine Learning in Science and Industry* (Apress, 2021).
68. V. Havlíček, A. D. Córcoles, K. Temme, A. W. Harrow, A. Kandala, J. M. Chow, and J. M. Gambetta, "Supervised learning with quantum-enhanced feature spaces," *Nature* **567**, 209–212 (2019), https://doi.org/10.1038/s41586-019-0980-2.
69. M. Arjovsky, A. Shah, and Y. Bengio, "Unitary evolution recurrent neural networks," in *Proc. 33rd Int. Conf. Mach. Learn. (ICML 2016)*, 1120–1128 (2016), .
70. Qiskit API documentation, https://qiskit.org/documentation/ (accessed Dec. 17, 2025)
71. A. Javadi-Abhari *et al.*, "Quantum computing with Qiskit," arXiv:2405.08810 (2024), https://doi.org/10.48550/arXiv.2405.08810.
72. A. Gong *et al.*, "Memorizing Normality to Detect Anomaly: Memory-augmented Deep Autoencoder for Unsupervised Anomaly Detection," arXiv:1904.02639 (2019), https://doi.org/10.48550/arXiv.1904.02639.
73. Z. Chen, C. Yeo, B. Lee, and C. T. Lau, "Autoencoder-based network anomaly detection," in *2018 Wireless Telecommunications Symposium (WTS)*, 1–5 (2018), https://doi.org/10.1109/WTS.2018.8363930.
74. V. Bergholm *et al.*, "PennyLane: Automatic differentiation of hybrid quantum-classical computations," arXiv:1811.04968 (2018), https://doi.org/10.48550/arXiv.1811.04968.
75. J. Basit, D. Hanif, and M. Arshad, "Quantum Variational Autoencoders for Predictive Analytics in High Frequency Trading Enhancing Market Anomaly Detection," *Int. J. Emerg. Multidisciplinaries: Comput. Sci. Artif. Intell.* **3**(1), 21 (2024), https://doi.org/10.54938/ijemdcsai.2024.03.1.319.
76. C. Bravo-Prieto, "Quantum autoencoders with enhanced data encoding," *Mach. Learn.: Sci. Technol.* **2**, 035028 (2021), https://doi.org/10.1088/2632-2153/ac0616.
77. A. Sakhnenko *et al.*, "Hybrid classical-quantum autoencoder for anomaly detection," *Quantum Mach. Intell.* **4**, 2 (2022), https://doi.org/10.1007/s42484-022-00075-z.
78. J. Y. Araz and M. Spannowsky, "The role of data embedding in quantum autoencoders for improved anomaly detection," arXiv:2409.04519 (2024), https://doi.org/10.48550/arXiv.2409.04519 (accessed Nov. 10, 2025).
79. J.-R. Jiang and J.-S. Li, "Applying a Parameterized Quantum Circuit to Anomaly Detection," *Eng. Proc.* **3**, 3 (2025), https://doi.org/10.3390/engproc2025092003.
80. A. Elsharkawy *et al.*, "Integration of Quantum Accelerators with High Performance Computing—A Review of Quantum Programming Tools," *ACM Trans. Quantum Comput.* (2025), https://doi.org/10.1145/3743149.
81. S. Chakraborty, S. H. Shaikh, A. Chakrabarti, and R. Ghosh, "A hybrid quantum feature selection algorithm using a quantum inspired graph theoretic approach," *Appl. Intell.* **50**, 1775–1793 (2020), https://doi.org/10.1007/s10489-019-01604-3.
82. V. S. Ngairangbam, M. Spannowsky, and M. Takeuchi, "Anomaly detection in high-energy physics using a quantum autoencoder," *Phys. Rev. D* **105**, 095004 (2022), https://doi.org/10.1103/PhysRevD.105.095004.
83. Y. Zhu, G. Bai, Y. Wang, T. Li, and G. Chiribella, "Quantum autoencoders for

communication-efficient cloud computing," *Quantum Mach. Intell.* **5**, 2 (2023), https://doi.org/10.1007/s42484-023-00112-5.

84. H. Suryotrisongko and Y. Musashi, "Evaluating hybrid quantum-classical deep learning for cybersecurity botnet DGA detection," *Proc. Comput. Sci.* **197**, 111–118 (2022), https://doi.org/10.1016/j.procs.2021.12.135.
85. E. D. Payares and J. C. Martinez-Santos, "Quantum machine learning for intrusion detection of distributed denial of service attacks: a comparative overview," in *Quantum Computing, Communication, and Simulation*, edited by P. R. Hemmer and A. L. Migdall, Proc. SPIE **11726**, 117260R (2021), https://doi.org/10.1117/12.2593297.
86. C. Bauckhage *et al.*, *Quantum Machine Learning in the Context of IT Security: Security within Quantum Machine Learning & Quantum Machine Learning for Cyber Security* (Federal Office for Information Security (BSI), 2022)
87. J. R. McClean, S. Boixo, V. N. Smelyanskiy, R. Babbush, and H. Neven, *Barren Plateaus in Quantum Neural Network Training Landscapes*, Nature Communications **9**, (2018).
88. M. Cerezo, A. Sone, T. Volkoff, L. Cincio, and P. J. Coles, *Cost Function Dependent Barren Plateaus in Shallow Parametrized Quantum Circuits*, Nature Communications **12**, (2021).
89. N. Lambert, E. Giguère, P. Menczel, B. Li, P. Hopf, G. Suárez, M. Gali, J. Lishman, R. Gadhvi, R. Agarwal, A. Galicia, N. Shammah, P. Nation, J. R. Johansson, S. Ahmed, S. Cross, A. Pitchford, and F. Nori, "QuTiP 5: The Quantum Toolbox in Python," *Phys. Rep.* **1153**, 1–62 (2026), https://doi.org/10.1016/j.physrep.2025.10.001
90. A. Ceschini, F. Mauro, F. D. Falco, A. Sebastianelli, A. Verdone, A. Rosato, B. Le Saux, M. Panella, P. Gamba, and S. L. Ullo, "From Graphs to Qubits: A Critical Review of Quantum Graph Neural Networks," arXiv:2408.06524 (2024), https://doi.org/10.48550/arXiv.2408.06524 , last accessed on 02/25/2026
91. Zhe-Yu Jeff Ou, *Multi-Photon Quantum Interference* (Springer Science & Business Media, 2007).
92. J. Stöhr, *Overcoming the Diffraction Limit by Multi-Photon Interference: A Tutorial*, Advances in Optics and Photonics **11**, 215 (2019).
93. M. Tillmann, B. Dakić, R. Heilmann, S. Nolte, A. Szameit, and P. Walther, *Experimental Boson Sampling*, Nature Photonics **7**, 540 (2013).
94. V. Chandola, A. Banerjee, and V. Kumar, "Anomaly Detection: A Survey," *ACM Comput. Surv.* **41**(3), Article 15 (2009), https://doi.org/10.1145/1541880.1541882
95. R. Chalapathy and S. Chawla, "Deep Learning for Anomaly Detection: A Survey," arXiv:1901.03407 (2019), https://doi.org/10.48550/arXiv.1901.03407
96. S. L. Braunstein and C. M. Caves, "Statistical distance and the geometry of quantum states," *Phys. Rev. Lett.* **72**, 3439–3443 (1994), https://doi.org/10.1103/PhysRevLett.72.3439
97. J. Stokes, J. Izaac, N. Killoran, and G. Carleo, "Quantum Natural Gradient," *Quantum* **4**, 269 (2020), https://doi.org/10.22331/q-2020-05-25-269; arXiv:1909.02108.
98. I. Sharafaldin, A. Habibi Lashkari, S. Hakak, and A. A. Ghorbani, "Developing Realistic Distributed Denial of Service (DDoS) Attack Dataset and Taxonomy," in *2019 International Carnahan Conference on Security Technology (ICCST)*, 1–8 (2019), https://doi.org/10.1109/CCST.2019.8888419
99. I. Sharafaldin, A. Habibi Lashkari, and A. A. Ghorbani, "Toward Generating a New Intrusion Detection Dataset and Intrusion Traffic Characterization," in *Proc. 4th Int. Conf. Inf. Syst. Secur. Privacy (ICISSP 2018)* (SciTePress, 2018), pp. 108–116, https://doi.org/10.5220/0006639801080116
100. A. Kandala, A. Mezzacapo, K. Temme, M. Takita, M. Brink, J. M. Chow, and J. M. Gambetta, *Hardware-Efficient Variational Quantum Eigensolver for Small Molecules and Quantum Magnets*, Nature **549**, 242 (2017).
101. T. Bhattacharyya, M. A. Dritschel, and C. S. Todd, *Completely Bounded Kernels*, Acta Scientiarum Mathematicarum **79**, 191 (2013).
102. W. LI and X. JIA, *Feature Selection Algorithm Based on Hellinger Distance*, Journal of Computer Applications **30**, 1530 (2010).
103. F. Bloch, "Nuclear Induction," *Phys. Rev.* **70**, 460 (1946).
104. I. Bengtsson and K. Życzkowski, *Geometry of Quantum States: An Introduction to Quantum Entanglement*, 2nd ed. (Cambridge University Press, Cambridge, England, 2017).
105. R. Jozsa, "Fidelity for Mixed Quantum States," *J. Mod. Opt.* **41**, 2315 (1994).

106. A. Uhlmann, “The ‘transition probability’ in the state space of a ∗-algebra,” *Rep. Math. Phys.* **9**, 273 (1976).
107. M. Hübner, “Explicit computation of the Bures distance for density matrices,” *Phys. Lett. A* **163**, 239 (1992).
108. M. Bina, F. Grasselli, and M. G. A. Paris, “Continuous-variable quantum probes for structured environments,” *Phys. Rev. A* **97**, 012125 (2018).
109. C. Benedetti, F. Salari Sehdaran, M. H. Zandi, and M. G. A. Paris, “Quantum probes for the cutoff frequency of Ohmic environments,” *Phys. Rev. A* **97**, 012126 (2018).
110. Y.-D. Sha and W. Wu, “Continuous-variable quantum sensing of a dissipative reservoir,” *Phys. Rev. Res.* **4**, 023169 (2022).
111. H. Ather and A. Z. Chaudhry, “Improving the estimation of environment parameters via initial probe-environment correlations,” *Phys. Rev. A* **104**, 012211 (2021).
112. A. Peruzzo, J. McClean, P. Shadbolt, M.-H. Yung, X.-Q. Zhou, P. J. Love, A. Aspuru-Guzik, and J. L. O’Brien, “A variational eigenvalue solver on a quantum processor,” arXiv:1304.3061, 2013.
113. S. Zhou and L. Jiang, “An exact correspondence between the quantum Fisher information and the Bures metric,” arXiv:1910.08473, 2019
114. K. Tscharke, M. Wendlinger, A. Ahouzi, P. Bhardwaj, K. Amoi-Taleghani, M. Schrodl-Baumann, and P. Debus, "Quantum Autoencoder for Multivariate Time Series Anomaly Detection," arXiv:2504.17548 [quant-ph] (2025), last accessed 03/24/2026
115. R. Frehner and K. Stockinger, "Applying quantum autoencoders for time series anomaly detection," Quantum Machine Intelligence 7, 59 (2025). DOI: 10.1007/s42484-025-00285-1.
116. F. Shayeganfar, A. Ramazani, V. Sundararaghavan, and Y. Duan, "Quantum graph learning and algorithms applied in quantum computer sciences and image classification," Applied Physics Reviews 12, 021327 (2025). DOI: 10.1063/5.0237599.
117. H. Qiao, H. Tong, B. An, I. King, C. Aggarwal, and G. Pang, "Deep Graph Anomaly Detection: A Survey and New Perspectives," IEEE Transactions on Knowledge and Data Engineering 37, 5106-5126 (2025). DOI: 10.1109/TKDE.2025.3581578.
118. M. Zhong, M. Lin, C. Zhang, and Z. Xu, "A survey on graph neural networks for intrusion detection systems: Methods, trends and challenges," Computers & Security 141, 103821 (2024). DOI: 10.1016/j.cose.2024.103821.
119. D. Wierichs, J. Izaac, C. Wang, and C. Y.-Y. Lin, General Parameter-Shift Rules for Quantum Gradients, Quantum 6, 677 (2022).
120. S. Jiang, J.-a. Lu, J. Zhou, and Q. Dai, “Fiedler value: The cumulated dynamical contribution value of all edges in a complex network,” Phys. Rev. E 109, 054301 (2024). DOI: 10.1103/PhysRevE.109.054301.
121. H. Shakeri, N. Albin, F. Darabi Sahneh, P. Poggi-Corradini, and C. Scoglio, “Maximizing algebraic connectivity in interconnected networks,” Phys. Rev. E 93, 030301(R) (2016). DOI: 10.1103/PhysRevE.93.030301.
122. S. Wald and L. Böttcher, “From classical to quantum walks with stochastic resetting on networks,” Phys. Rev. E 103, 012122 (2021). DOI: 10.1103/PhysRevE.103.012122.
123. C. Nicolini, V. Vlasov, and A. Bifone, “Thermodynamics of network model fitting with spectral entropies,” Phys. Rev. E 98, 022322 (2018). DOI: 10.1103/PhysRevE.98.022322.
124. E. Gillman, F. Carollo, and I. Lesanovsky, “Using (1+1)D quantum cellular automata for exploring collective effects in large-scale quantum neural networks,” Phys. Rev. E 107, L022102 (2023). DOI: 10.1103/PhysRevE.107.L022102.